\documentclass[preprint,aip,jcp,citeautoscript,nofootinbib,superscriptaddress,amsmath,amssymb]{revtex4-1}
\usepackage{graphicx} 
\usepackage[mathlines]{lineno}
\usepackage[utf8]{inputenc} 
\usepackage[T1]{fontenc} 
\usepackage{mathptmx} 
\usepackage{newunicodechar}
\usepackage[section]{placeins}
\usepackage{etoolbox} 
\usepackage{color} 
\usepackage{mhchem}
\newunicodechar{−}{\ensuremath{-}}
\newunicodechar{₂}{$_2$}
\usepackage{booktabs}
\usepackage{newunicodechar} 
\newunicodechar{−}{\ensuremath{-}} 
\newunicodechar{₂}{$_2$} 
\usepackage{graphicx}
\usepackage{tikz}
\usetikzlibrary{positioning}

\makeatletter
\def\@email#1#2{%
 \endgroup
 \patchcmd{\titleblock@produce}
  {\frontmatter@RRAPformat}
  {\frontmatter@RRAPformat{\produce@RRAP{*#1\href{mailto:#2}{#2}}}\frontmatter@RRAPformat}
  {}{}
}

\makeatother
\begin{document}

%\preprint{AIP/123-QED}

\title{Non-Additive Ion Effects on the Coil–Globule Equilibrium of a Generic Polymer in Aqueous Salt Solutions}
\author{Kushagra Goel}
\affiliation{Department of Chemistry, Shiv Nadar Institution of Eminence, Gautam Buddh Nagar, Uttar Pradesh 201314, India}
\author{Monika Choudhary}
\affiliation{Department of Chemical Engineering, Shiv Nadar Institution of Eminence, Gautam Buddh Nagar, Uttar Pradesh 201314, India}
\author{Swaminath Bharadwaj}
\affiliation{Department of Chemical Engineering, Shiv Nadar Institution of Eminence, Gautam Buddh Nagar, Uttar Pradesh 201314, India}
\email{swaminath.bharadwaj@snu.edu.in}

\def\thefootnote{*}\footnotetext{Corresponding author}\def\thefootnote{\arabic{footnote}}

\date{\today}
\begin{abstract}
Mixtures of weakly and strongly hydrated anions induce non-additive changes in
the LCST of thermoresponsive polymers such as PNIPAM and PEO. Large-scale
atomistic simulations of PNIPAM--NaI--Na$_{2}$SO$_{4}$ mixtures have shown
that these effects arise from the interplay between favorable
PNIPAM--iodide interactions and depletion of strongly hydrated sulfate ions.
Here, we investigate whether chemically specific polymer--anion interactions
are necessary to reproduce such behavior. To this end, we study the
coil--globule transition of a generic uncharged linear polymer with
non-specific polymer--water and polymer--ion van der Waals interactions in
atomistic aqueous solutions of single and mixed salts. Simulations are
performed at fixed concentrations of the strongly hydrated salt,
Na$_{2}$SO$_{4}$, and increasing concentrations of the weakly hydrated salts,
NaSCN and NaI. The generic polymer qualitatively reproduces experimentally
observed trends in pure NaSCN and Na$_{2}$SO$_{4}$ solutions, as well as the
non-additive behavior in mixed salt solutions. In particular, the model
captures the mutually reinforcing preferential accumulation of the weakly
hydrated SCN$^{-}$ ions and depletion of the strongly hydrated
SO$_{4}^{2-}$ ions near the polymer that underlies the non-additive
behavior. This mutual enhancement correlates with partitioning of sodium ions
from the counterion cloud of SCN$^{-}$ ions to that of
SO$_{4}^{2-}$ ions and is consistent with atomistic simulations of PNIPAM
solutions. The model also reproduces the effects of background salt
concentration and weakly hydrated anion identity on the non-additive
behavior. These results demonstrate that non-specific polymer--ion and polymer-water
interactions are sufficient to reproduce non-additive salt effects,
suggesting a dominant role of bulk ion--ion and ion--water interactions.
\end{abstract}

\maketitle{
\textbf{}}
\section{\label{sec:level1}{Introduction}}
The relative ability of anions to salt-out proteins in aqueous solution has
been ranked according to the anionic Hofmeister
series.\cite{Hofmeister1888,Kunz2004} The typical salting-out order is given by
$\mathrm{CO_3^{2-}} > \mathrm{SO_4^{2-}} > \mathrm{S_2O_3^{2-}} >
\mathrm{H_2PO_4^{-}} > \mathrm{F^{-}} > \mathrm{Cl^{-}} > \mathrm{Br^{-}} >
\mathrm{NO_3^{-}} > \mathrm{I^{-}} > \mathrm{ClO_4^{-}} > \mathrm{SCN^{-}}$,
indicating a decreasing tendency to precipitate macromolecules. Strongly
hydrated anions such as $\mathrm{CO_3^{2-}}$ and $\mathrm{SO_4^{2-}}$, left end
of the series, tend to remain in the bulk leading to salting-out of
macromolecules. In contrast, weakly hydrated anions, on the right of the
series, such as SCN$^{-}$ and I$^{-}$ tend to partition to macromolecular
surfaces leading to salting-in at low salt
concentrations.\cite{Zang2005,Jungwirth2001,vandervegt:2019} Recently, there
has been growing interest in understanding specific ion effects on the phase
behavior of thermoresponsive polymers such as Poly(N-isopropylacrylamide) (PNIPAM) and Polyethylene oxide (PEO)
.\cite{cremert:2020,vandervegt:2019,Bruce2021,Bui2021,Zhao2023,Moghaddam2015} In
particular, the dependence of the lower critical solution temperature (LCST) on
salt concentration has been widely investigated through experimental and
theoretical methods in both single and mixed salt solutions containing strongly
and weakly hydrated anions and
cations.\cite{cremert:2020,vandervegt:2019,Bruce2021,Bui2021,Zhao2023,Jungwirth2014,Heyda2017,Zhao2022,Moghaddam2015} In
summary, these studies indicate that specific ion effects on the LCST arise
from a complex interplay between ion–ion, ion–water, and macromolecule–ion
interactions, suggesting that they cannot always be described as simple
additive contributions of individual ion effects.

A particularly notable phenomenon is the non-additive ion effects on the LCST
of thermoresponsive polymers in aqueous salt solutions containing mixtures of
weakly and strongly hydrated anions.\cite{cremert:2020,Bui2021} Bruce et
al.\cite{cremert:2020} experimentally investigated the LCST behavior of PNIPAM
in both single and mixed salt solutions. In aqueous solutions
of the weakly hydrated salt NaI, the LCST exhibits a non-monotonic dependence
on salt concentration, increasing initially and then decreasing with increasing
NaI concentration, indicating salting-in at low concentrations and salting-out
at higher concentrations. In contrast, in pure Na$_{2}$SO$_{4}$ solutions, the
LCST decreases monotonically with increasing salt concentration, indicating
progressive salting-out. These trends in single salt solutions are consistent
with the Hofmeister series. However, in mixed salt solutions containing a fixed
concentration of Na$_{2}$SO$_{4}$, the LCST shows a non-monotonic dependence on
NaI concentration with three distinct regimes. At low NaI concentrations
(region I), the LCST decreases with increasing NaI concentration, indicating
salting-out in the presence of the strongly hydrated background salt. At
intermediate concentrations (region II), the LCST increases with increasing NaI
concentration, corresponding to salting-in. At higher concentrations (region
III), the LCST again decreases, indicating reentrant salting-out behavior.
Notably, the trends observed in regions I and II differ qualitatively from
those in pure NaI solutions, highlighting the non-additive nature of ion
effects in mixed salt systems. Note that these observations are not limited to
PNIPAM.  Bruce et al.\cite{cremert:2020} have also shown this non-additive behavior in other
polymers such as Polyethylene oxide and Poly(N,N-dimethylacrylamide) in systems
containing different combinations of strongly and weakly hydrated anions.  To
gain mechanistic insight, Zhao et al.\cite{Zhao2022} performed molecular simulations of PNIPAM
in NaI–Na$_{2}$SO$_{4}$ solutions at low NaI concentrations (region I), and
were able to semi-quantitatively reproduce the experimentally observed changes
in LCST in both single and mixed salt solutions.  These simulations show that
iodide ions exhibit preferential accumulation near the polymer, while sulfate
ions are depleted from the polymer surface.  Importantly, these two effects are
coupled, such that sulfate depletion and iodide accumulation near the polymer
reinforce each other. Within this framework, the non-additive behavior arises
from the balance between these coupled contributions: at low NaI
concentrations, enhanced sulfate depletion dominates over the favorable
PNIPAM–iodide interactions, resulting in salting-out of the PNIPAM, whereas at
intermediate concentrations, increased PNIPAM–iodide interactions become
dominant, leading to salting-in and an increase in the LCST.  An important
question arising from this mechanism is whether chemically specific
polymer–anion interactions are necessary to reproduce the observed behavior,
and how their importance compares to that of bulk ion–ion and ion–water
interactions.

In this work, we study the coil--globule transition of a model 32-bead
hydrophobic polymer in pure and mixed salt solutions using molecular dynamics
simulations and advanced sampling techniques such as umbrella sampling. The
polymer interacts with water molecules and ions solely through non-specific van
der Waals interactions. We consider the mixed salt system
NaSCN--Na$_{2}$SO$_{4}$ at two different background salt concentrations and
also investigate NaI--Na$_{2}$SO$_{4}$ solutions to examine the effect of the
weakly hydrated anion identity on non-additive salt effects. Our results show
that the model polymer qualitatively reproduces the experimentally observed
trends in pure NaSCN and Na$_{2}$SO$_{4}$ solutions, as well as the
non-additive behavior observed in mixed salt solutions. We find that depletion
of SO$_{4}^{2-}$ ions and preferential accumulation of SCN$^{-}$ ions around
the polymer are mutually reinforcing. {This mutual enhancement correlates with
non-additive changes in bulk ion pairing and ion hydration, including (a)
partitioning of the cation from the counterion environment of the weakly
hydrated anion to that of the strongly hydrated anion, and (b) enhanced
hydration of both anions. }We further examine the effects of background salt
concentration and weakly hydrated anion identity. Our results show that
increasing the background salt concentration shifts the minimum in the polymer
collapse free energy in region I to higher NaSCN concentrations and
simultaneously deepens the minimum. {Upon replacing SCN$^{-}$ with I$^{-}$ at
fixed background salt concentration, the onset of region II shifts to
higher salt concentrations. This shift is associated with weaker preferential
accumulation of I$^{-}$ ions around the polymer due to which higher iodide concentrations are required to overcompensate the contribution from the depletion of sulfate ions.} These trends are consistent with experiments and atomistic
simulations on mixed salt solutions of PNIPAM. Overall, the ability of this generic polymer model to reproduce the observed
non-additive regimes, their dependence on background salt concentration, and
their sensitivity to weakly hydrated anion identity indicates that
non-specific polymer--ion interactions are sufficient to reproduce
non-additive salt effects. These results further suggest that bulk ion--ion
and ion--water interactions play a dominant role in governing the
coil--globule equilibrium in mixed salt solutions

\section{Methods}
\subsection{System and theoretical details}

In this study, we employ a generic hydrophobic polymer model originally proposed
by Zangi et al.\cite{Zangi2009}, which exhibits a two-state conformational
equilibrium between the coil (C) and globule (G) states, similar to that
observed in thermoresponsive polymers such as
PNIPAM.\cite{Tiktopulo:1995,Podewitz:2019,Palivec:2018} The simplified
coarse-grained representation of this model enables efficient sampling of
conformational transitions while avoiding the significant computational cost
associated with fully atomistic polymer simulations. The polymer is modeled as a linear chain of 32 uncharged Lennard-Jones beads,
each representing a united-atom CH$_2$ group with a diameter of 0.4 nm and an
interaction strength $\epsilon{\rm pp}$ of 1.0 kJ mol$^{-1}$. Adjacent beads are
connected by a harmonic bond potential with an equilibrium bond length of 0.153
nm and a force constant of $3.2 \times 10^{5}$ kJ mol$^{-1}$ nm$^{-2}$.  The
angle between adjacent bonds is modeled using a harmonic potential with an
equilibrium angle $\theta_0 = 111^\circ$ and a force constant of 100 kJ
mol$^{-1}$. Water is modeled using the SPC/E model. Non- polarizable force
fields were employed for all the ions. For Na$_{2}$SO$_{4}$, the optimized
forcefield developed by Bruce et al. was utilized.\cite{cremert:2020} Model I(4)
in the work of Fyta and Netz was used for NaI.\cite{Netz:2012} For SCN$^{-}$,
the force-field employed in K\v{r}\'{\i}\v{z}ek et al. was
used.\cite{Krizek2014} {At 1 bar and 300 K,  SPC/E water acts as a poor solvent for the above mentioned hydrophobic polymer with a negative polymer 
collapse free energy, $\Delta G^{C\rightarrow G}=-2.0$ kJ/mol. To change the quality of the solvent, the polymer bead-water Lennard-Jones parameter, $
\epsilon_{\rm pw}$ was tuned using a factor $\lambda_{\rm pw}$ in the following manner,
\begin{equation}
\epsilon_{\rm pw}=\lambda_{\rm pw}\sqrt{\epsilon_{\rm pp}\epsilon_{\rm ww}}.
\end{equation}
The polymer bead-water Lennard-Jones diameter, $\sigma_{\rm pw}=(\sigma_{\rm pp}+\sigma_{\rm ww})/2$, was kept constant. $\lambda_{\rm pw}$ was fixed at 1.075 to achieve good solvent conditions with a positive polymer collapse free energy, $\Delta G^{\rm C\rightarrow G}=2.42 \pm 0.04$ kJ/mol. See Sec.~S1 in the Supporting Information for non-bonded parameters.

We investigated the coil-to-globule transition of this 32-mer polymer in aqueous
solutions of single and mixed salts using a combination of molecular dynamics
simulations and umbrella sampling. In single salt solutions, simulations were
performed at varying concentrations of NaSCN and Na$_{2}$SO$_{4}$. In mixed salt
solutions, simulations were performed at varying NaSCN concentrations at two
fixed concentrations of Na$_{2}$SO$_{4}$ (0.5 and 1 m) to examine the effect of
background salt concentration. Simulations were also performed at varying NaI
concentrations in the presence of 0.5 m Na$_{2}$SO$_{4}$ to examine the effect
of different weakly hydrated anions. The number of water molecules was fixed at
32695 for all simulations. The number of ions at different concentrations is
listed in Sec.~S2 of the Supplementary information.

All molecular dynamics simulations were carried out using the \textsc{GROMACS}
(version 2022.1) software package\cite{Abraham:2015}.  The equations of motion
were integrated using the leap-frog algorithm with a time step of $2$~fs.  All
bonds were constrained using the LINCS algorithm. Long-range electrostatics were
computed using the Particle Mesh Ewald (PME) method\cite{PME}, with a real-space
cutoff of $1.4$~nm, a Fourier grid spacing of $0.12$~nm, and fourth-order
interpolation. van der Waals interactions were truncated at $1.4$~nm, with
long-range dispersion corrections applied to both energy and pressure. The
neighbor list was updated every $10$ integration steps ($20$~fs).  For each
system, energy minimization was performed using the steepest-descent algorithm,
followed by a $2$~ns equilibration run in the canonical (NVT) ensemble,
employing the v-rescale thermostat to maintain the temperature at $300$~K with a
time constant of $1$~ps. Subsequently, equilibration in the isothermal-isobaric
(NPT) ensemble was performed for $5$~ns using the v-rescale
thermostat\cite{Bussi2007} and the Berendsen barostat\cite{Berendson1984} to
maintain the pressure at $1$~bar. A production run of $400$~ns was then
performed in the NPT ensemble using the v-rescale thermostat and the
Parrinello–Rahman barostat\cite{Parrinello1981}, with a pressure coupling time
constant of $2$~ps. This run was used to compute monomer-ion and monomer-water
radial distribution functions (RDFs), and the final configuration was
subsequently used as the starting point for the umbrella sampling calculations.

\subsection{Umbrella sampling}

The following expression was used to compute the polymer collapse free energy,
$\Delta G^{\rm C\to G}$, from the potential of mean force (PMF) profiles,
$w(R_{\rm g})$:
{
\begin{equation}
    K_{\rm C\rightarrow G}=\frac{P_{\rm globule}}{P_{\rm coil}} = \frac{\int_{0}^{R_{\rm g}^\#} e^{-w(R_{\rm g})/RT} \, dR_{\rm g}}{\int_{R_{\rm g}^\#}^{\infty} e^{-w(R_{\rm g)}/RT}dR_{\rm g}}=e^{-\Delta G^{\rm C \rightarrow G} / RT}, 
    \label{umb_equation}
\end{equation}
where $K_{\rm C\rightarrow G}$ is the equilibrium constant, $P_{\rm globule}$ ($P_{\rm coil}$) is the probability for the globule (coil) state, $T$ is the temperature, $R$ is the gas constant, $R_{\rm g}$ represents the
radius of gyration, and $R_{\rm g}^{\#}$ is the threshold value used to distinguish
between the coil and globule states of the polymer.} The PMF profiles of the
polymer in aqueous salt solutions were obtained using umbrella sampling
simulations performed with \textsc{GROMACS} (version 2022.1)\cite{Abraham:2015}
interfaced with PLUMED (version 2.5.2)\cite{Plumed}, employing $R_{\rm g}$ as the
collective variable. A harmonic biasing potential, $V(R_{\rm g})=\frac{1}
{2}K_{\rm b}(R_{\rm g}-R_{\rm g}^0)^2$, with a force constant of $K_{\rm
b}=20000$ kJ mol$^{-1}$ nm$^{-2}$, was applied to restrain the polymer radius of
gyration. Sampling was performed over $R_g^0$ values ranging from $0.4$ nm to
$1.2$ nm, with a window spacing of $0.025$ nm. The coil and globule
conformations were identified from the two distinct minima in the PMF profiles,
positioned above and below the cutoff $R_g^{\#} = 0.7$ nm, respectively.  Each
umbrella sampling window was sampled for $110$ ns. Temperature and pressure were
maintained using the v-rescale thermostat (time constant $\tau_{\rm T}=0.5$ ps)
and the Parrinello–Rahman barostat (time constant $\tau_{\rm p}=1$ ps),
respectively. The unbiased PMF profiles were reconstructed using the weighted
histogram analysis method (WHAM) with the Grossfield WHAM
code\cite{grossfield_wham}. Details pertaining to the computation of the polymer
collapse free energy and associated errors can be found in Sec.~S3 of the
Supplementary Information. {For both pure and mixed salt systems, we define the quantity
$\Delta \Delta G^{\rm C\rightarrow G}$ as
\begin{equation}
\Delta \Delta G^{\rm C\rightarrow G}
=
\Delta G^{\rm C\rightarrow G}(c_{\rm salt})
-
\Delta G^{\rm C\rightarrow G}(c_{\rm salt}=0),
\end{equation}
where $\Delta G^{\rm C\rightarrow G}(c_{\rm salt}=0)$ corresponds to the
aqueous polymer solution for pure salt systems and to the corresponding
background salt solution for mixed salt systems.}
{
\subsection{Preferential binding coefficients}
The preferential accumulation of ions around the polymer, or depletion from its
vicinity, can be quantified using the preferential binding coefficient,
\begin{equation}\label{eq:prefbind}
\Gamma_{\rm polymer-ion} = \rho_{\rm ion}\left(G_{\rm polymer-ion} - G_{\rm polymer-water}\right),
\end{equation}
where $G_{\rm polymer-ion}$ and $G_{\rm polymer-water}$ are the polymer--ion and
polymer-- water KBIs, respectively. An ion preferentially accumulates around the
polymer when $\Gamma_{\rm polymer-ion} > 0$, and is depleted from its vicinity
when $\Gamma_{\rm polymer- ion} < 0$. 
%Note that this preferential binding coefficient is not specific to either the
%coil or globule state, but instead represents an average over the equilibrium
%polymer conformational ensemble. 
Note that the preferential binding coefficient is not computed separately for
the coil and globule states, but instead represents an average over all polymer
conformations sampled in equilibrium MD simulations. Consequently, this quantity
provides an overall measure of ion accumulation or depletion around the polymer
and cannot be directly related to the coil--globule collapse free energy, which
is expected to depend on the difference in preferential binding between the
coil and globule ensembles. Nevertheless, the trends in the preferential
binding coefficients can still be used to identify qualitative correlations
between ion accumulation/depletion and the coil--globule equilibrium.
The Kirkwood--Buff integrals, $G_{\rm ij}$, are defined as
\begin{equation}
G_{\rm{ij}} = \int_{0}^{\infty}\left[g_{\rm{ij}}(r)-1\right]4\pi r^2 dr,
\label{kbi}
\end{equation}
where $g_{\rm ij}(r)$ is the radial distribution function between components
\textit{i} and \textit{j}. The Ganguly correction was applied to account for
finite- size effects in the RDFs.\cite{Ganguly2013} Subsequently, the
extrapolation scheme proposed by Kr{\"u}ger \textit{et al.}\cite{kruger2013} was
used to obtain the KBIs from the corrected RDFs. This combination of RDF and KBI
corrections has been shown to improve KBI convergence.\cite{Milzetti2018}
The RDFs used for the KBI calculations were computed between the 
monomer beads and the atomic sites of the ions. For water molecules, the
center of mass was used as the reference point. The preferential binding
coefficients were obtained from the long-distance plateau of the KBIs.
Details of the polymer--water and polymer--anion RDFs, along with the associated
corrections, are provided in Sec.~S4 and Sec.~S5.1 of the Supplementary Information.

\subsection{Ion pairing and ion hydration}
Anion--cation pairing was characterized using the excess coordination number,
$\Delta N_{\rm anion,cation}$, defined as
\begin{equation}
\Delta N_{\rm anion,cation}=\frac{N_{\rm cation}}{V}\int^{r_{\rm out}}_{r_{\rm in}} \left(g_{\rm anion,cation}-1\right) 4\pi r^{2} {\rm d}r,
\end{equation}
where $g_{\rm anion,cation}$ is the anion--cation radial distribution function,
$N_{\rm cation}$ is the total number of cations, and $V$ is the average
simulation box volume. The quantity $\Delta N_{\rm anion,cation}$ can be
interpreted as the change in the number of cations within a spherical
observation region before and after placing an anion at its center. The choice
of $r_{\rm in}$ and $r_{\rm out}$ depends on the peaks in $g_{\rm
anion,cation}$, which can be assigned to different ion-pairing modes such as
contact ion pairing (CIP), solvent-shared ion pairing (SIP), and
solvent-separated ion pairing (SSIP). In the Supplementary Information, see Sec.~S5.2  for ion-ion rdfs and Sec.~S6 for individual excess coordination numbers from CIP, SIP and SSIP.

For SCN$^{-}$--Na$^{+}$, the first two peaks of $g_{\rm SCN^{-}-Na^{+}}$
correspond to CIP and SIP, respectively. In contrast, for
SO$_{4}^{2-}$--Na$^{+}$, the first two peaks of $g_{\rm SO_{4}^{2-}-Na^{+}}$
correspond to SIP and SSIP, respectively. This is consistent with experiments
showing that sulfate and sodium ions do not form contact ion pairs.\cite{Buchner1999} In the
present work, only the first two peaks of the anion--cation radial distribution
function were considered for the excess ion-pairing calculations. 

A similar expression can be defined for ion hydration,
\begin{equation}
\Delta N_{\rm anion,water}=\frac{N_{\rm water}}{V}\int^{r_{\rm out}}_{r_{\rm in}} \left(g_{\rm anion,water}-1\right) 4\pi r^{2} {\rm d}r,
\end{equation}
where $g_{\rm anion,water}$ is the anion--water radial distribution function and
$N_{\rm water}$ is the total number of water molecules. Similar to the
ion-pairing case, $\Delta N_{\rm anion,water}$ represents the change in the
number of water molecules within a spherical observation region before and after
placing an anion at its center. Here, the first two peaks of $g_{\rm
anion,water}$, corresponding to the first and second hydration shells, were
considered for calculating $\Delta N_{\rm anion,water}$. In the Supplementary Information, see Sec.~S5.2  for ion-water rdfs and Sec.~S6 for individual excess coordination numbers from the first and second hydration shells.
%%%%%
\section{Result and Discussion}

\subsection{\textbf{Polymer in aqueous solutions of a single salt}} 
 Figure~\ref{fig:pure_scn}(a) shows the non-monotonic dependence of the relative
polymer collapse free energy, $\Delta\Delta G^{\rm C\rightarrow G}$, on sodium
thiocyanate concentration in aqueous polymer solutions. The polymer chain
prefers the coil state in pure water ($\Delta G^{\rm C\rightarrow G}({\rm
pure~water}) > 0$). At low NaSCN concentrations ($c_{\rm NaSCN} < 0.2 \ \rm m$),
$\Delta\Delta G^{\rm C\rightarrow G}$ increases with increasing salt
concentration, indicating that salt addition shifts the coil--globule
equilibrium toward the coil state. In contrast, at higher salt concentrations,
$\Delta\Delta G^{\rm C\rightarrow G}$ decreases with increasing salt
concentration, indicating a shift of the equilibrium toward the globule state.
\begin{figure}[ht]
  \begin{center}
    {\includegraphics[width=0.45\textwidth]{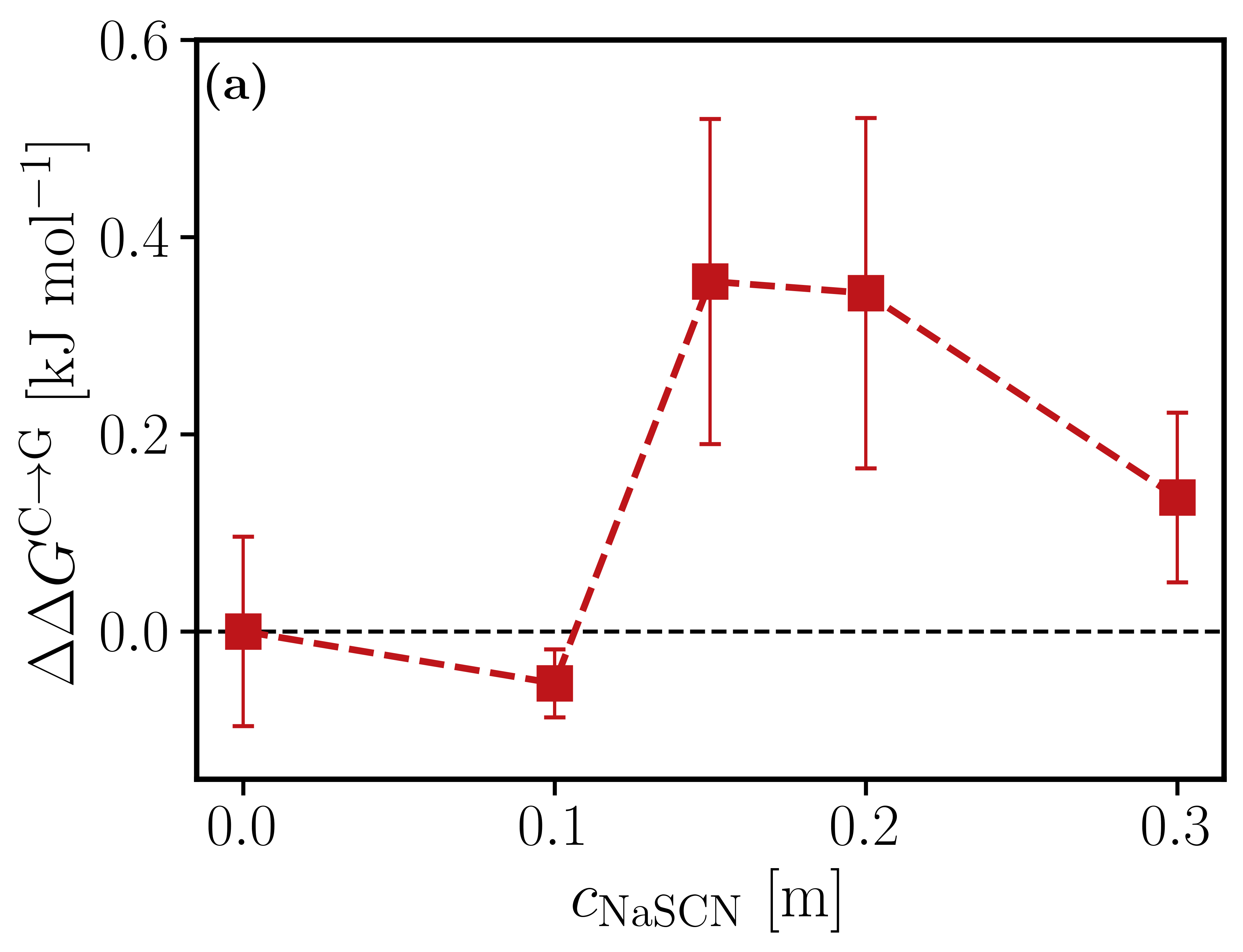}}
    {\includegraphics[width=0.45\textwidth]{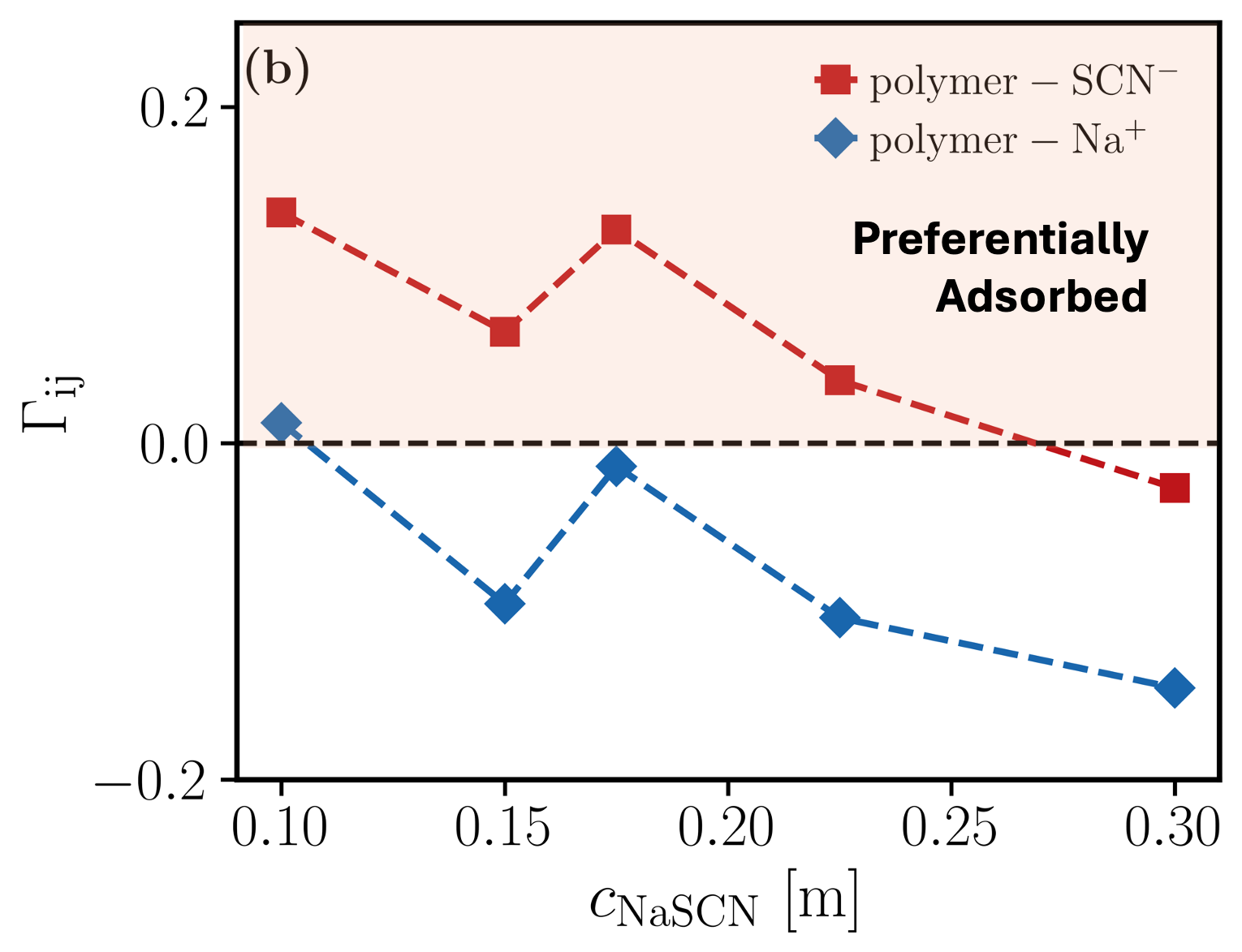}}
    \end{center}
\caption{Dependence of the (a) relative polymer collapse free energy, $\Delta\Delta G^{\rm C \to G}$ =$\Delta
G^{\rm C \to G}(c_{\rm NaSCN})$ - $\Delta G^{C \to G}(c_{\rm NaSCN}=0)$, and (b) polymer-SCN$^{-}$ and polymer-water KBIs on $c_{\rm NaSCN}$ in pure sodium thiocyanate solutions.}
\label{fig:pure_scn}
\end{figure}

{To rationalise these trends, we examine the polymer--SCN$^{-}$ and
polymer--Na$^{+}$ preferential binding coefficients (see Eq.~\ref{eq:prefbind}).
Figure~\ref{fig:pure_scn}(b) shows that $\Gamma_{\rm polymer-SCN^{-}}$ is
positive at all concentrations except 0.3~m, indicating preferential
accumulation of SCN$^{-}$ ions around the polymer. In contrast, $\Gamma_{\rm
polymer-Na^{+}}$ is negative at all salt concentrations, indicating depletion of
sodium ions from the polymer vicinity.  These observations suggest that
preferential accumulation of SCN$^{-}$ ions around the polymer is associated
with the salting-in behavior observed at low NaSCN concentrations. The radius of
gyration of the polymer exhibits similar qualitative trends, and its dependence
on salt concentration is shown in Sec.~S7 of the Supplementary Information.}

In contrast, $\Delta\Delta G^{\rm C\rightarrow G}$ decreases monotonically with
increasing sodium sulfate concentration (Fig.~\ref{fig:na2so4}(a)), indicating
that salt addition shifts the coil--globule equilibrium toward the globule
state. {Figure~\ref{fig:na2so4}(b) shows that both $\Gamma_{\rm polymer-Na^{+}}$
and $\Gamma_{\rm polymer-SO_{4}^{2-}}$ are negative and decrease monotonically
with increasing salt concentration, indicating depletion of both ions from the
vicinity of the polymer. }The trends observed in Fig.~\ref{fig:pure_scn} and
Fig.~\ref{fig:na2so4} are in qualitative agreement with experimental
observations for aqueous single-salt solutions of
Poly(N-isopropylacrylamide).\cite{cremert:2020} These results further show that
non-specific polymer-anion van der Waals interactions are sufficient to
reproduce salting-in behaviour at low concentrations of weakly hydrated anions.
\begin{figure}[ht]
  \begin{center}
    {\includegraphics[width=0.45\textwidth]{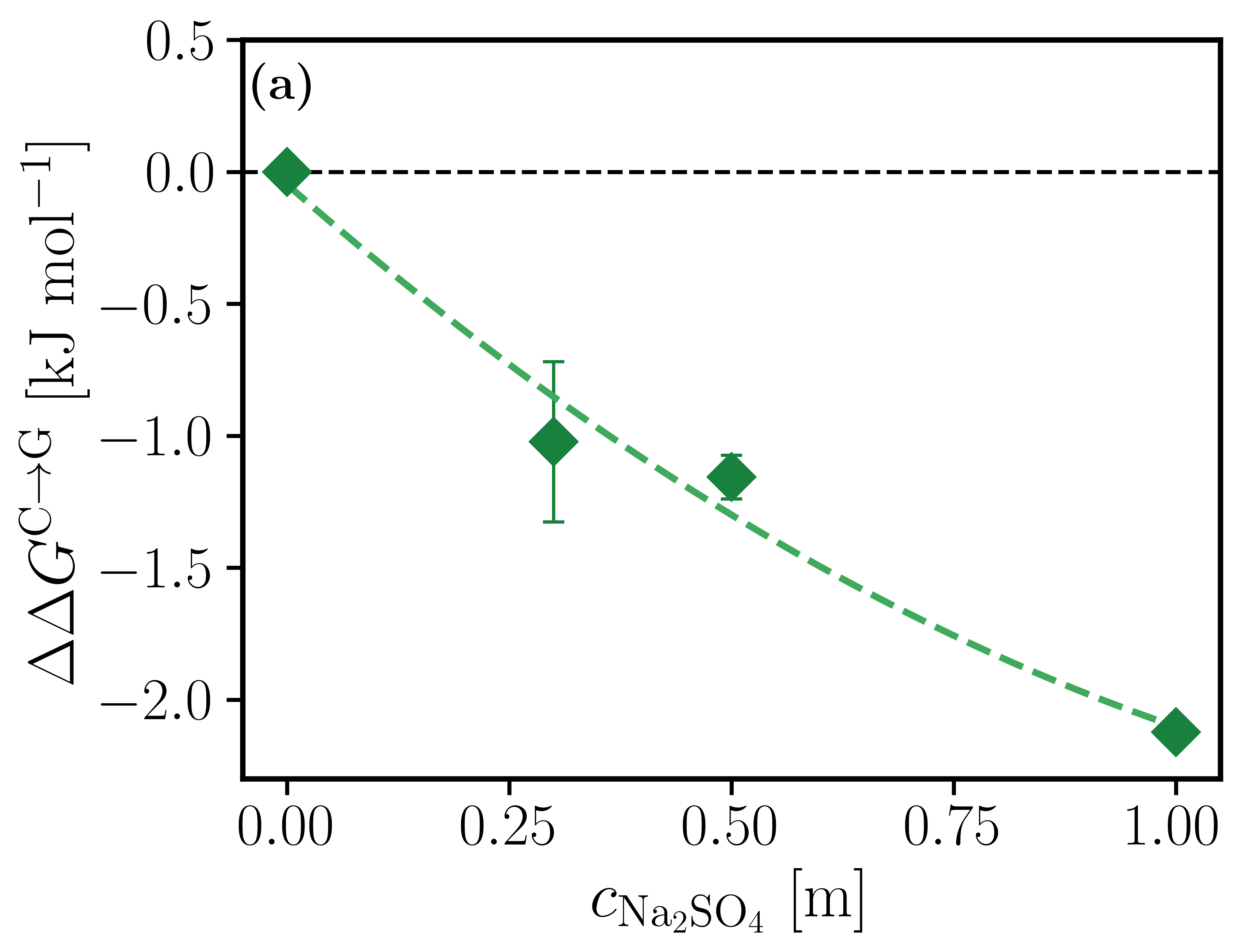}}
    {\includegraphics[width=0.45\textwidth]{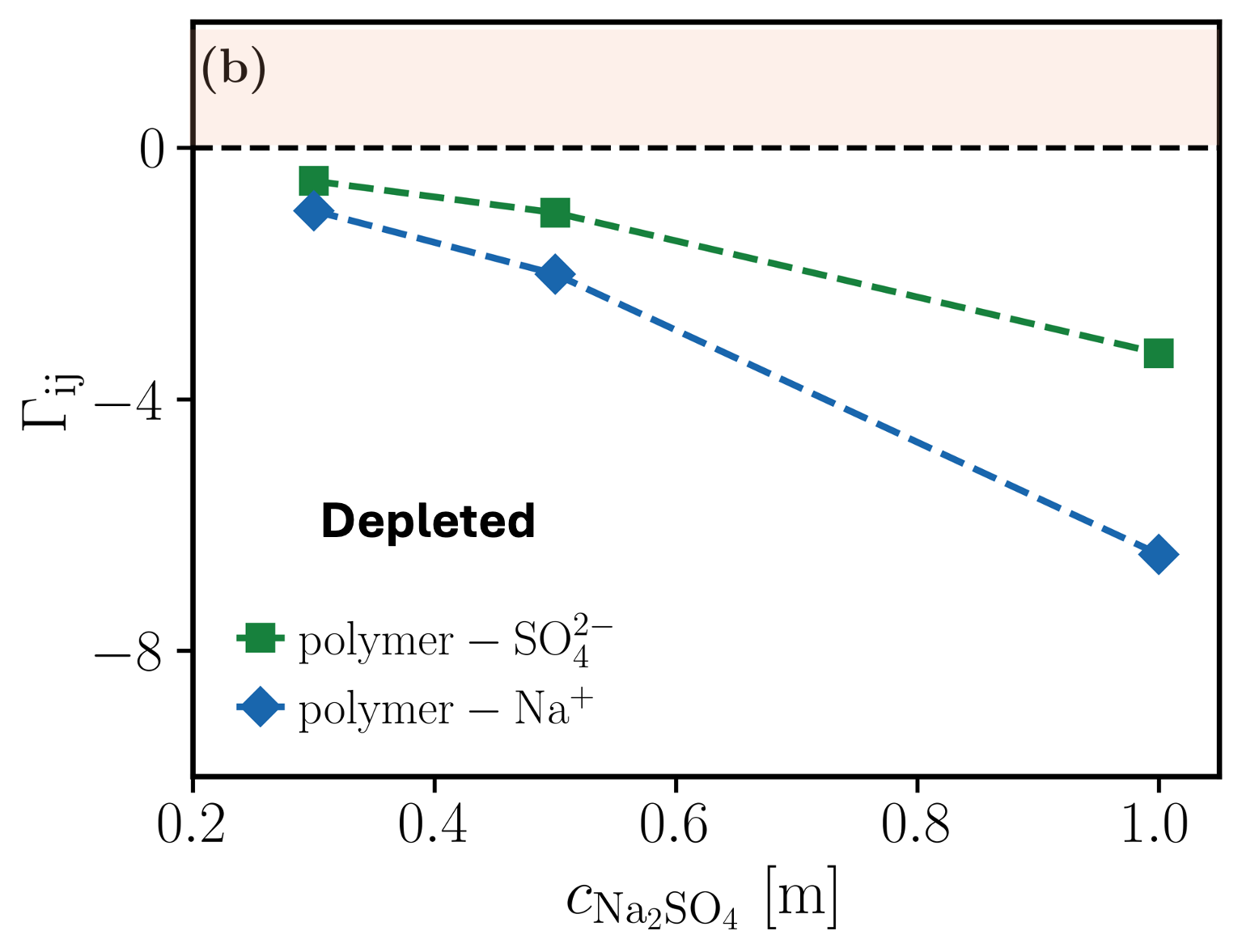}}
  \end{center}
\caption{Dependence of the (a) relative polymer collapse free energy,
$\Delta\Delta G^{\rm C \to G}$ =$\Delta G^{\rm C \to G}(c_{\rm Na_{2}SO_{4}})$ -
$\Delta G^{C \to G}(c_{\rm Na_{2}SO_{4}}=0)$, and (b) polymer-SO$_{4}^{2-}$ and
polymer-water KBIs on $c_{\rm Na_{2}SO_{4}}$ in pure sodium sulfate solutions.}
\label{fig:na2so4}
\end{figure}
\subsection{Polymer in aqueous solutions of mixed salts}
Figure~\ref{fig:mix_salt_scn}(a) shows the dependence of $\Delta\Delta G^{\rm
C\rightarrow G}$ on $c_{\rm NaSCN}$ at different concentrations of the strongly
hydrated background salt, Na$_{2}$SO$_{4}$. For a background salt concentration
of 1~m, $\Delta\Delta G^{\rm C\rightarrow G}$ exhibits a non-monotonic
dependence on $c_{\rm NaSCN}$, in qualitative agreement with experiments by
Bruce \textit{et al.}\cite{cremert:2020} At low NaSCN concentrations ($c_{\rm
NaSCN} < 0.15$~m), $\Delta\Delta G^{\rm C\rightarrow G}$ decreases with
increasing $c_{\rm NaSCN}$, indicating a shift of the coil--globule equilibrium
toward the globule state (region I). At higher concentrations,
$\Delta\Delta G^{\rm C\rightarrow G}$ increases with $c_{\rm NaSCN}$, favoring
the coil state (region II).  Thus, the model reproduces the experimentally
observed non-additive behaviour across region I and II. 
\cite{cremert:2020} {The uncertainty associated with the value of $\Delta\Delta G^{C\rightarrow G}$ at 0.225~m in
Fig.~\ref{fig:mix_salt_scn}(a) is larger than that at nearby concentrations,
making it difficult to conclusively identify the decrease in
$\Delta\Delta G^{C\rightarrow G}$ from 0.20~m to 0.225~m as the onset of region III. Since
the non-additive deviations from pure NaSCN solutions are primarily observed
in regions I and II, we focus our discussion on the trends up to 0.20~m.}

\begin{figure}[ht]
\centering
\includegraphics[width=0.45\textwidth]{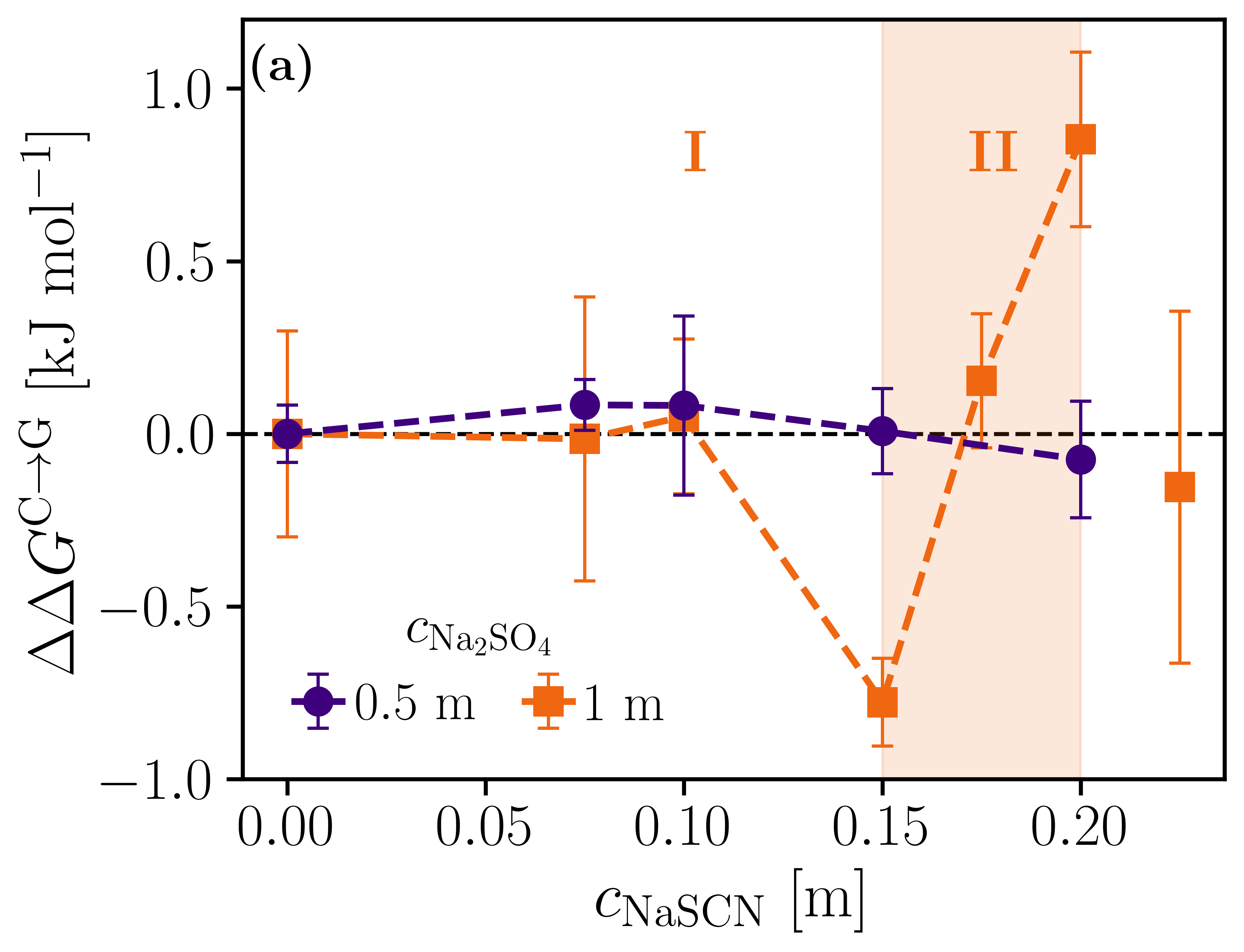}
\includegraphics[width=0.45\textwidth]{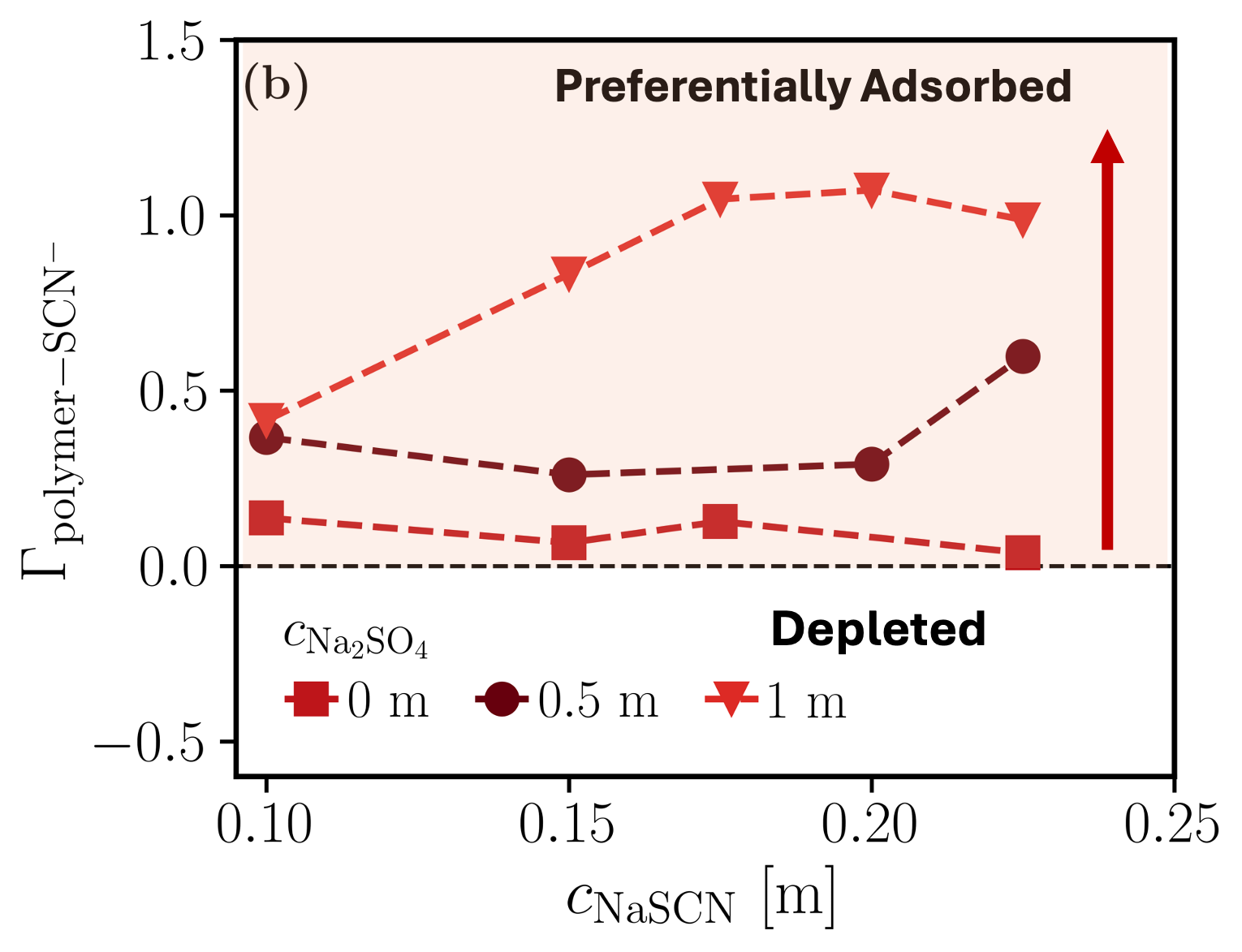}
\includegraphics[width=0.45\textwidth]{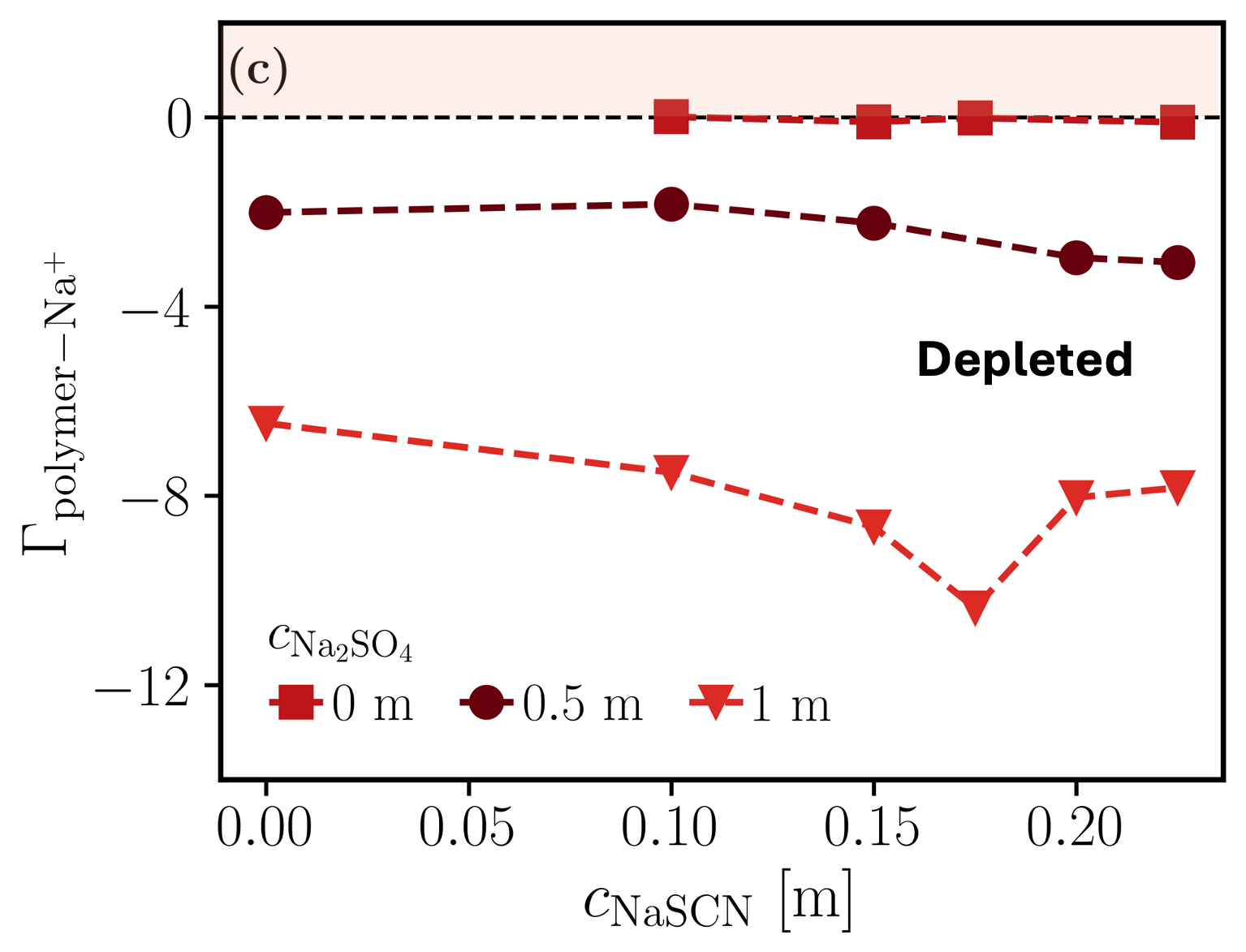} 
\includegraphics[width=0.45\textwidth]{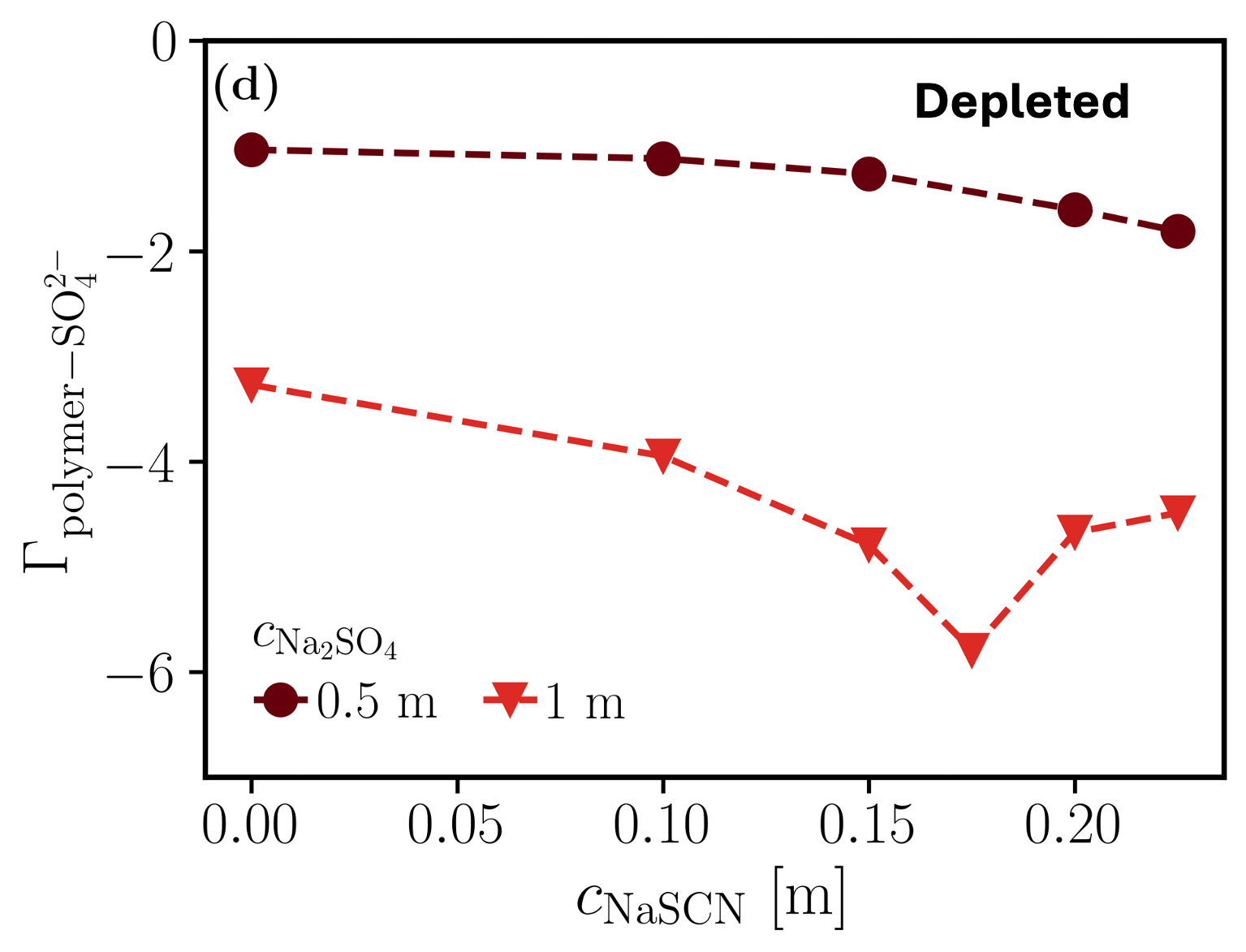}
 \caption{Dependence of (a) the relative polymer collapse free energy, $\Delta\Delta G^{\rm C \to G}$, and the preferential binding coefficients,(b) $\Gamma_{\rm polymer-SCN^{-}}$, (c) $\Gamma_{\rm polymer-Na^{+}}$, and (d) $\Gamma_{\rm polymer-SO_{4}^{2-}}$ on $c_{\rm NaSCN}$ at different background salt concentrations.} 
\label{fig:mix_salt_scn}
\end{figure}

{Figure~\ref{fig:mix_salt_scn}(b) shows the dependence of $\Gamma_{\rm
polymer-SCN^{-}}$ on $c_{\rm NaSCN}$ for different background salt
concentrations. For a fixed $c_{\rm NaSCN}$, $\Gamma_{\rm polymer-SCN^{-}}$ is
positive and increases with increasing $c_{\rm Na_{2}SO_{4}}$, indicating
enhanced preferential accumulation of SCN$^{-}$ ions around the polymer in the
presence of the background salt. For a background salt concentration of 1~m,
$\Gamma_{\rm polymer-SCN^{-}}$ increases monotonically with increasing $c_{\rm
NaSCN}$ up to 0.2~m NaSCN. This increase in $\Gamma_{\rm polymer-SCN^{-}}$
correlates with the increase in $\Delta\Delta G^{\rm C\rightarrow G}$ observed
between 0.15~m and 0.2~m NaSCN. In contrast, Figs.~\ref{fig:mix_salt_scn}(c) and
(d) show that both $\Gamma_{\rm polymer-Na^{+}}$ and $\Gamma_{\rm
polymer-SO_{4}^{2-}}$ decrease monotonically with increasing $c_{\rm NaSCN}$ up
to 0.175~m NaSCN for a background salt concentration of 1~m. This indicates
enhanced depletion of sulfate ions upon addition of NaSCN. Furthermore, the
decrease in $\Gamma_{\rm polymer-Na^{+}}$ and $\Gamma_{\rm polymer-SO_{4}^{2-}}$
correlates with the decrease in $\Delta\Delta G^{\rm C\rightarrow G}$ observed
up to 0.15~m NaSCN. Taken together, Figs.~\ref{fig:mix_salt_scn}(b)--(d)
indicate a mutual enhancement between preferential accumulation of the weakly
hydrated SCN$^{-}$ ions and depletion of the strongly hydrated SO$_{4}^{2-}$
ions. These trends are consistent with atomistic simulations of
PNIPAM--NaI--Na$_{2}$SO$_{4}$ systems by Zhao et al.}\cite{Zhao2022}

{To understand this mutual enhancement, we examine the trends in anion--cation
pairing and anion hydration. Figures~\ref{fig:ion_pair_hydration}(a) and (b)
show the excess thiocyanate--sodium and sulfate--sodium ion pairing,
respectively. The excess SCN$^{-}$--Na$^{+}$ ion pairing increases with
increasing NaSCN concentration. However, for a fixed NaSCN concentration,
increasing the background Na$_{2}$SO$_{4}$ concentration suppresses
SCN$^{-}$--Na$^{+}$ pairing. At the same time, the excess
SO$_{4}^{2-}$--Na$^{+}$ pairing increases with increasing NaSCN concentration.
Together, these observations suggest that sodium ions partition from the
counterion environment of the weakly hydrated SCN$^{-}$ ions to that of the
strongly hydrated SO$_{4}^{2-}$ ions. This cation partitioning is accompanied by
increased hydration of both anions. Figure~\ref{fig:ion_pair_hydration}(c) shows
that the SCN$^{-}$--water affinity progressively increases with increasing
background salt concentration. Similarly, the SO$_{4}^{2-}$--water affinity
(Fig.~\ref{fig:ion_pair_hydration}(d)) increases with increasing NaSCN
concentration. These trends in ion pairing and ion hydration are consistent with
atomistic simulations of PNIPAM--NaI--Na$_{2}$SO$_{4}$ solutions reported by
Zhao \textit{et al.}}\cite{Zhao2022}

\begin{figure}[ht]
\centering
\includegraphics[width=0.45\textwidth]{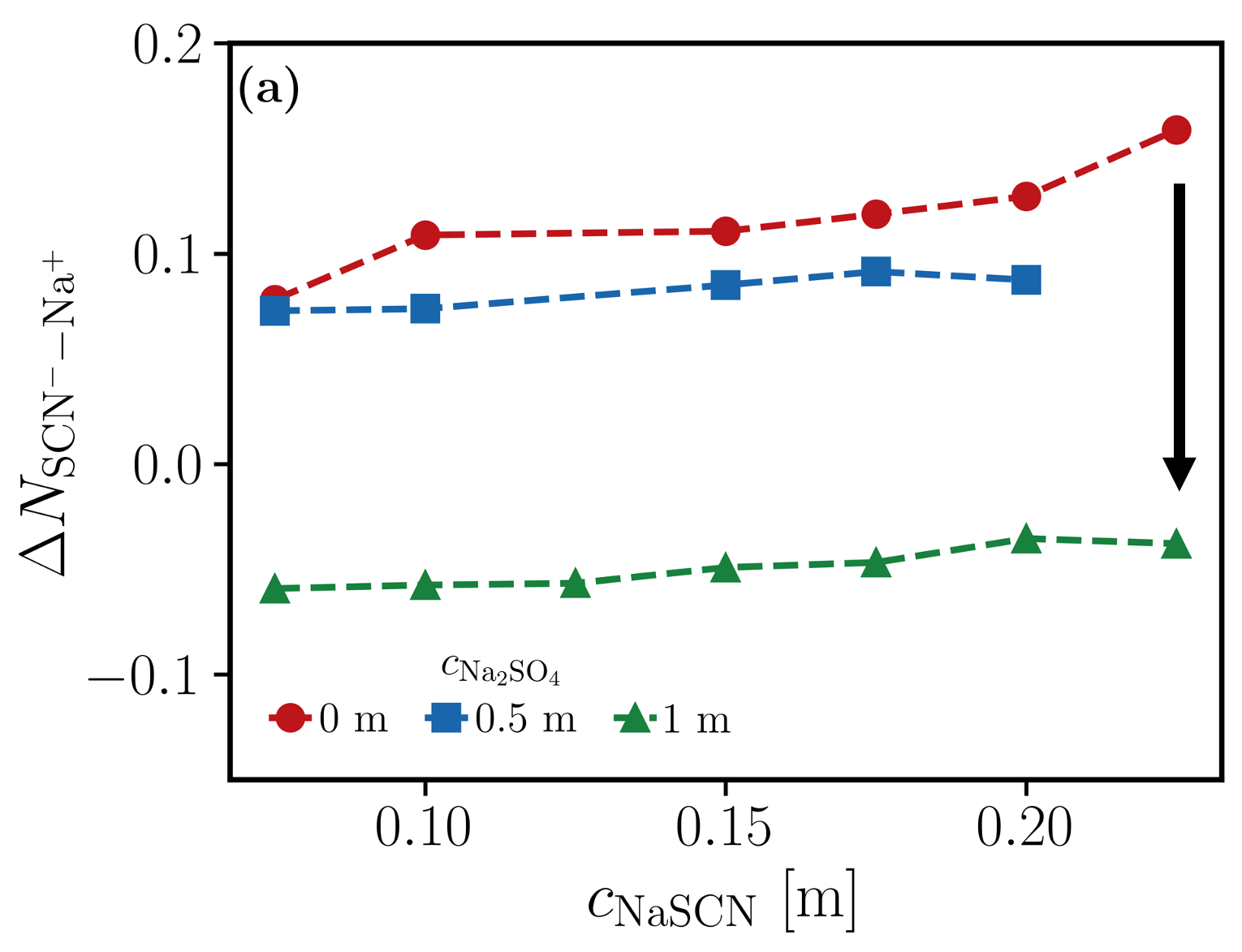}
\includegraphics[width=0.45\textwidth]{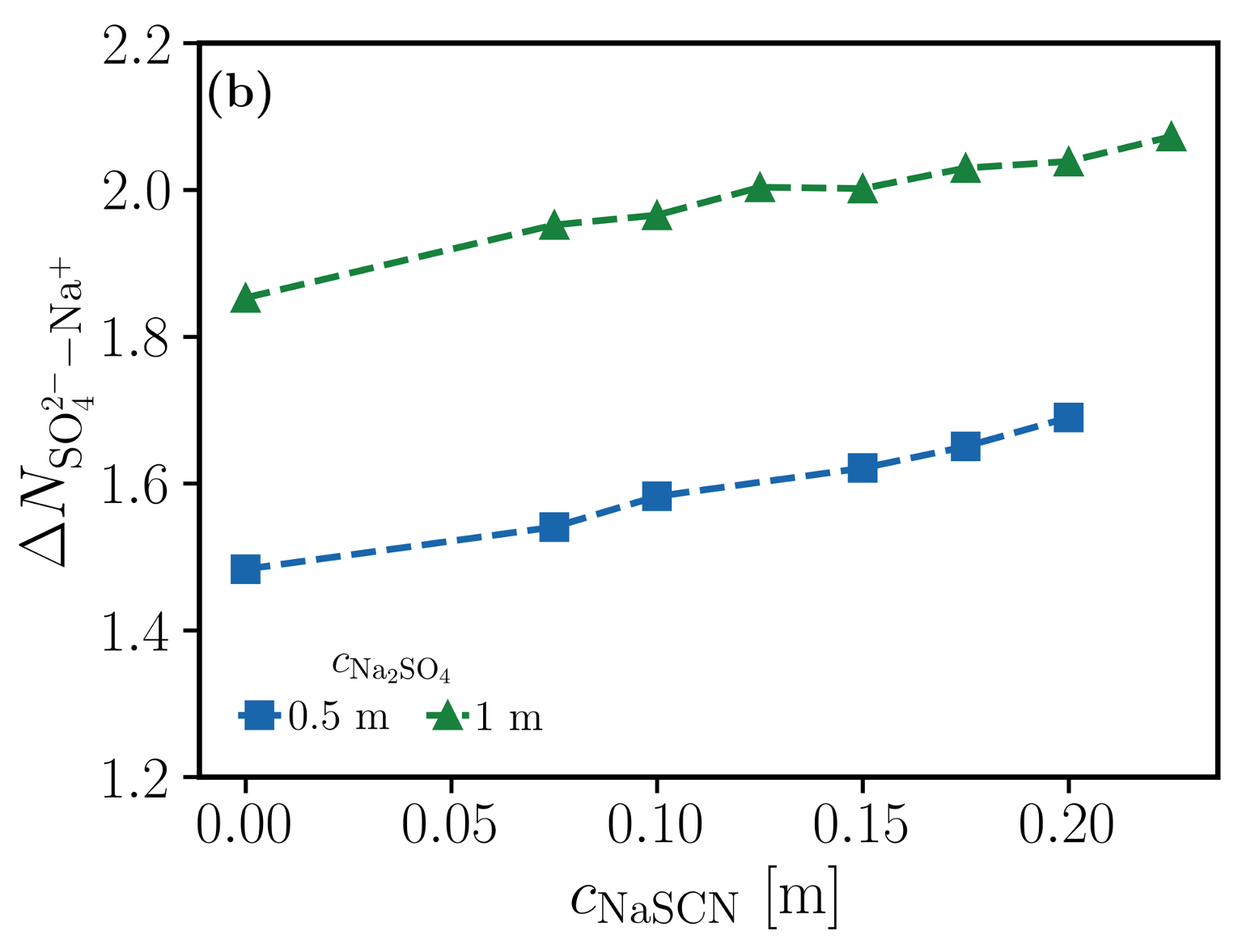} 
\includegraphics[width=0.45\textwidth]{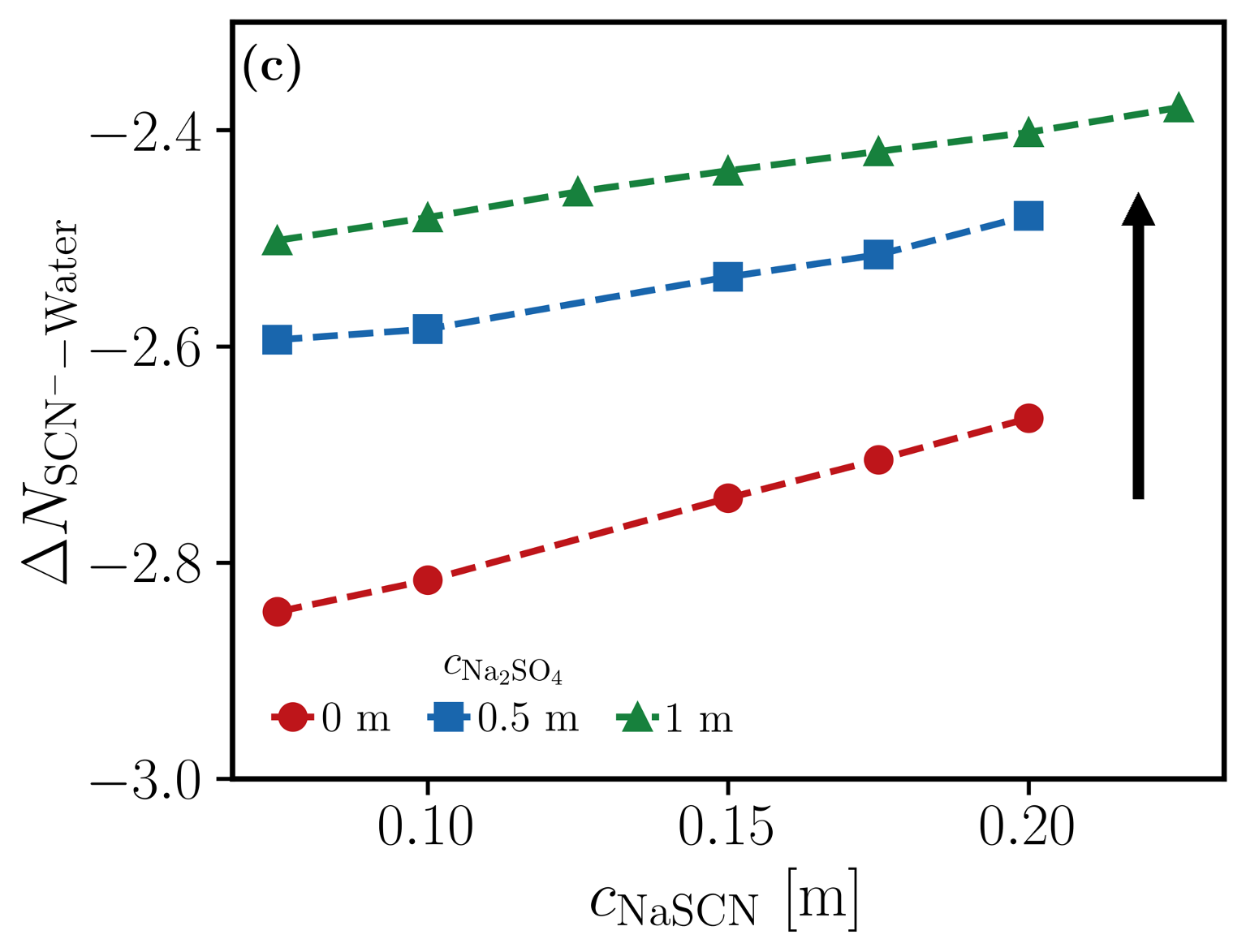}
\includegraphics[width=0.45\textwidth]{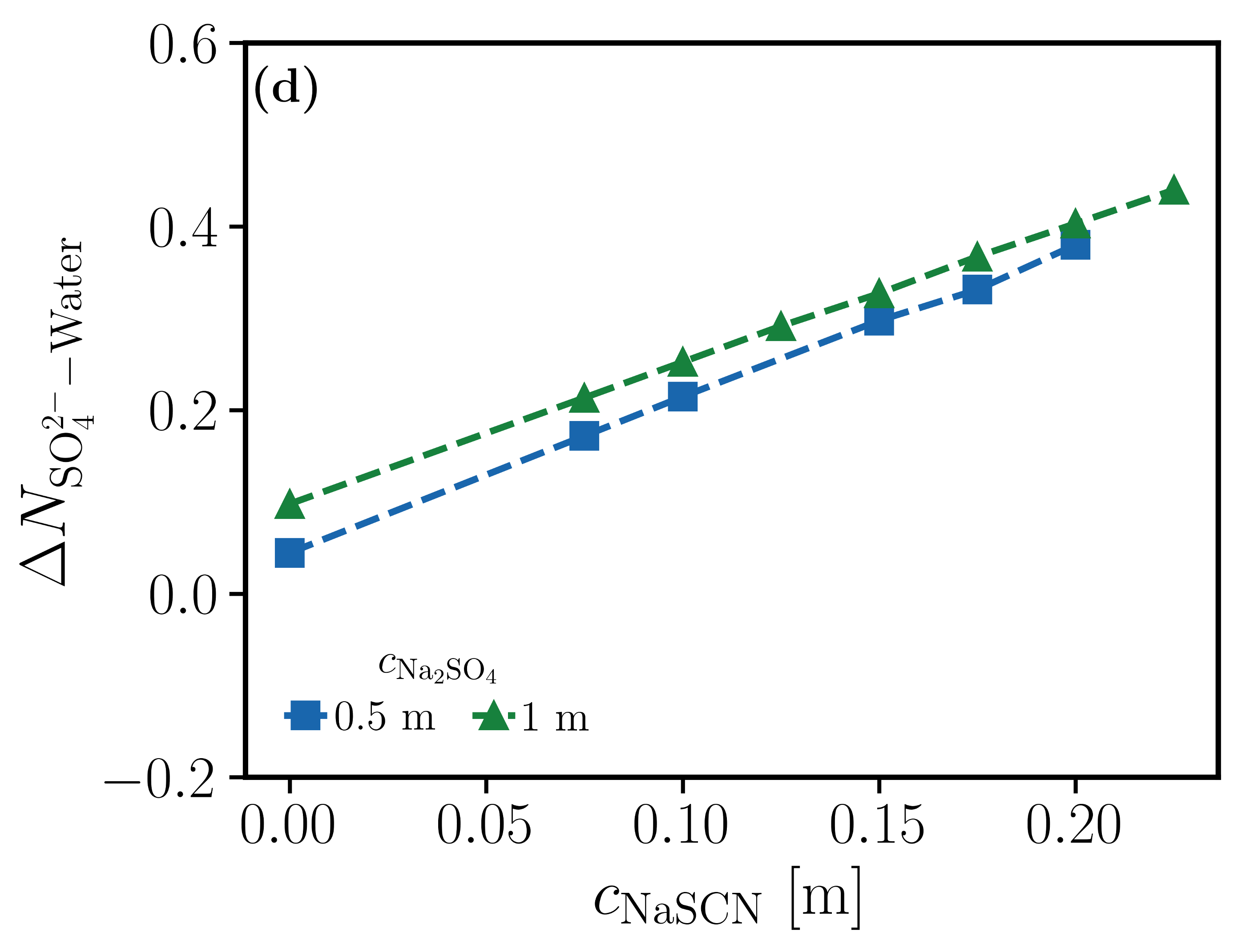}
 \caption{Dependence of (a) $\Delta N_{\rm SCN^{-}-Na^{+}}$, (b) $\Delta N_{\rm
SO_{4}^{2-}-Na^{+}}$, (c) $\Delta N_{\rm SCN^{-}-Water}$ and (d) $\Delta N_{\rm
SO_{4}^{2-}-Water}$ on $c_{\rm NaSCN}$ for different background salt
concentrations}
 \label{fig:ion_pair_hydration}
\end{figure}

{As in the case of I$^{-}$ ions in PNIPAM-NaI-Na$_{2}$SO$_{4}$
solutions,\cite{Zhao2022} the preferential accumulation of SCN$^{-}$ ions
increases even though they become more hydrated in the bulk. This suggests that
SCN$^{-}$ ions can readily shed part of their hydration shell and accumulate
near the polymer chain. Furthermore, the increased accumulation of SCN$^{-}$
ions correlates with the partitioning of sodium ions toward the counterion
environment of the sulfate ions. In contrast, enhanced hydration of sulfate
ions, together with electrostatic repulsion from the adsorbed SCN$^{-}$ ions, is
associated with increased depletion of sulfate ions from the vicinity of the
polymer chain. These results support the hypothesis that the coil--globule
equilibrium is governed by the competition between polymer--weakly hydrated
anion interactions, which favor swelling, and depletion of strongly hydrated
ions, which favors collapse. Importantly, both contributions are mutually
enhanced through non-additive changes in bulk ion pairing and ion hydration. In
region I, depletion of sulfate ions dominates, leading to polymer collapse. With
increasing $c_{\rm NaSCN}$, polymer--thiocyanate interactions become dominant,
stabilizing the coil state in region II. At higher concentrations, increasing
salting-out effects shift the balance again, leading to reentrant collapse in
region III. For 0.5~m Na$_{2}$SO$_{4}$, these competing effects seem to nearly
compensate each other, resulting in an approximately constant $\Delta\Delta
G^{\rm C\rightarrow G}$. Within the studied concentration range, these trends
suggest that the system may lie near the transition between regions II and III,
while region I occurs below 0.075~m NaSCN. This interpretation is consistent with
experimental observations showing that the minimum in the LCST shifts to lower
salt concentrations with decreasing background salt concentration.}

\begin{figure}[ht]
\begin{center}
\includegraphics[width=0.45\textwidth]{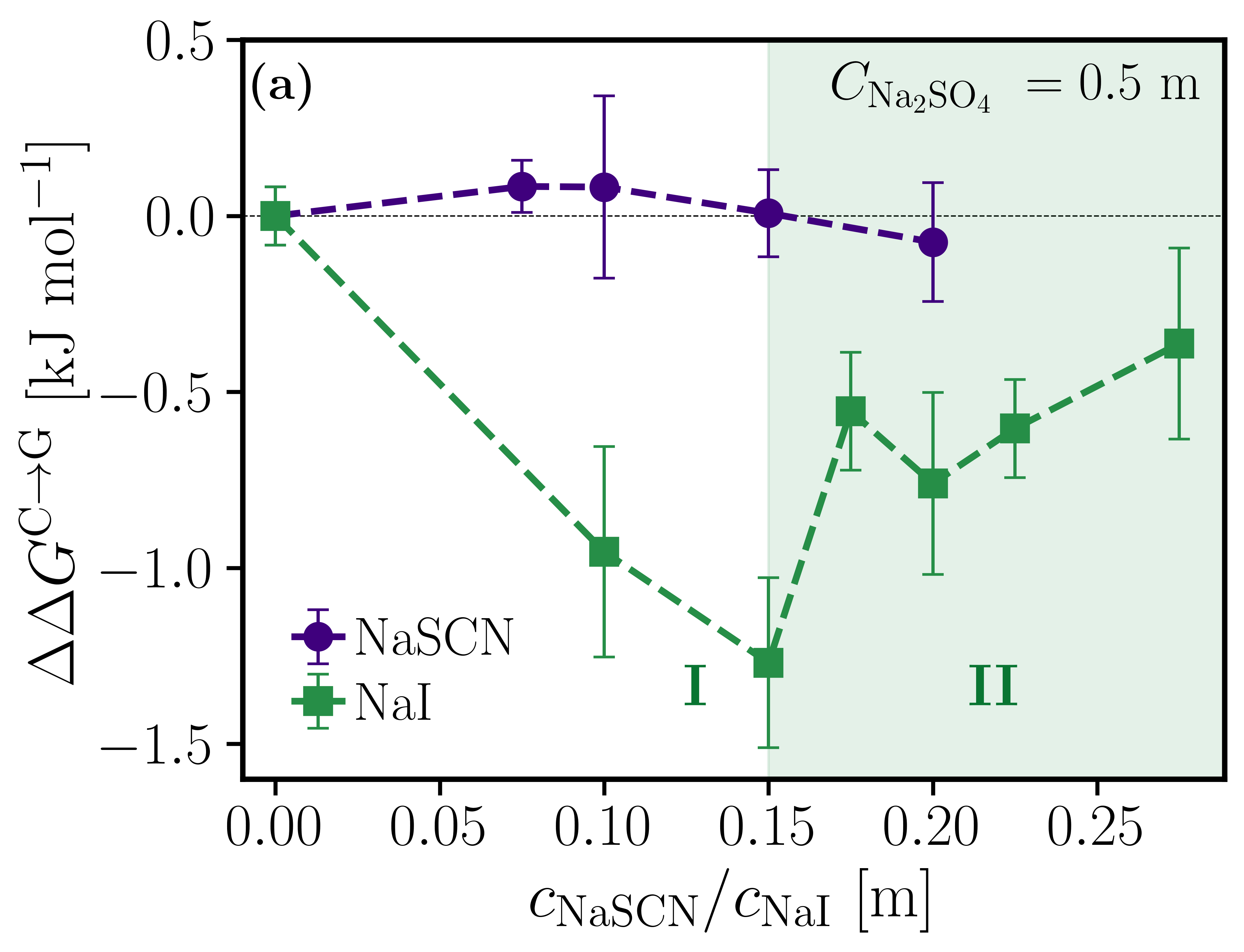}
\includegraphics[width=0.45\textwidth]{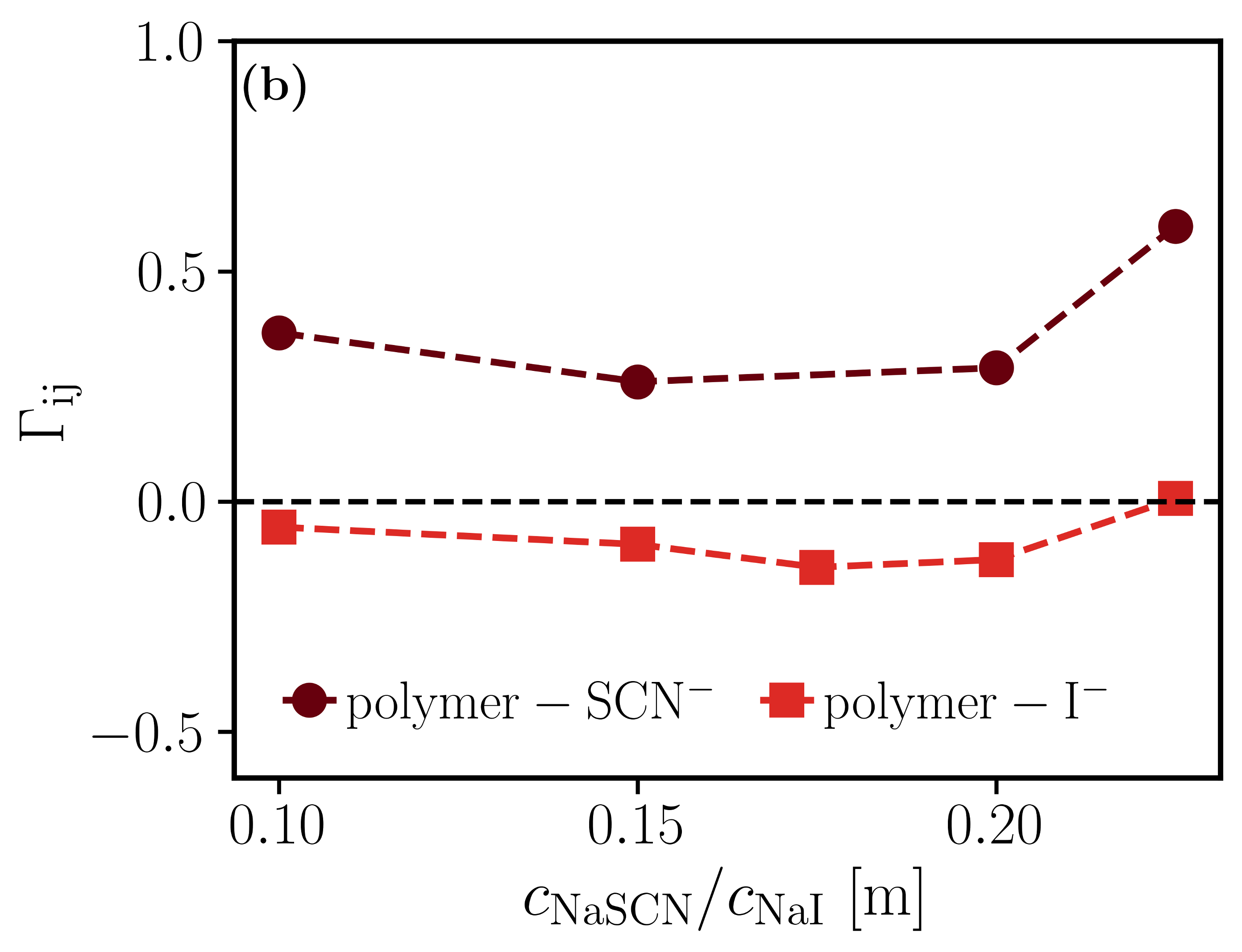}
\includegraphics[width=0.45\textwidth]{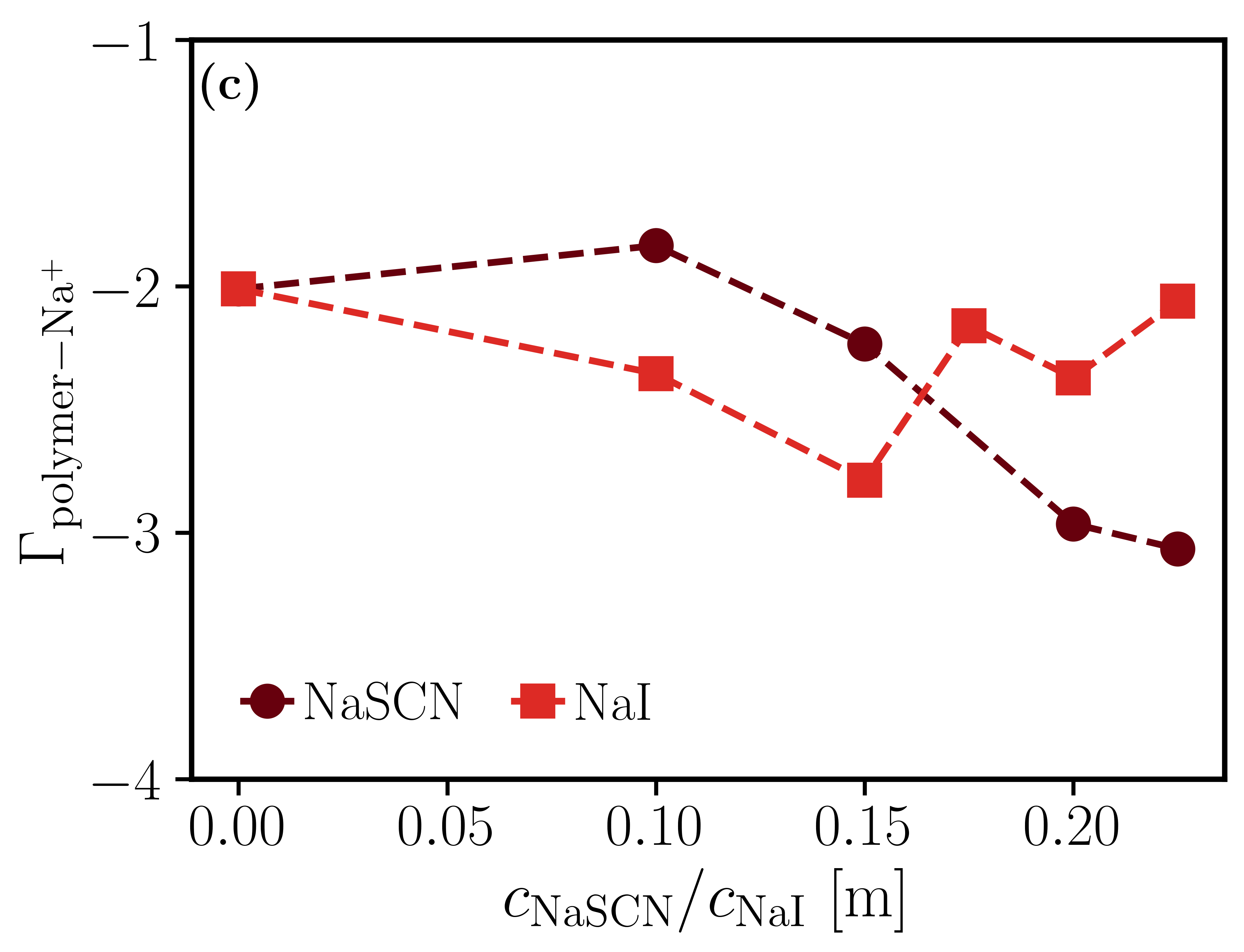}
\includegraphics[width=0.45\textwidth]{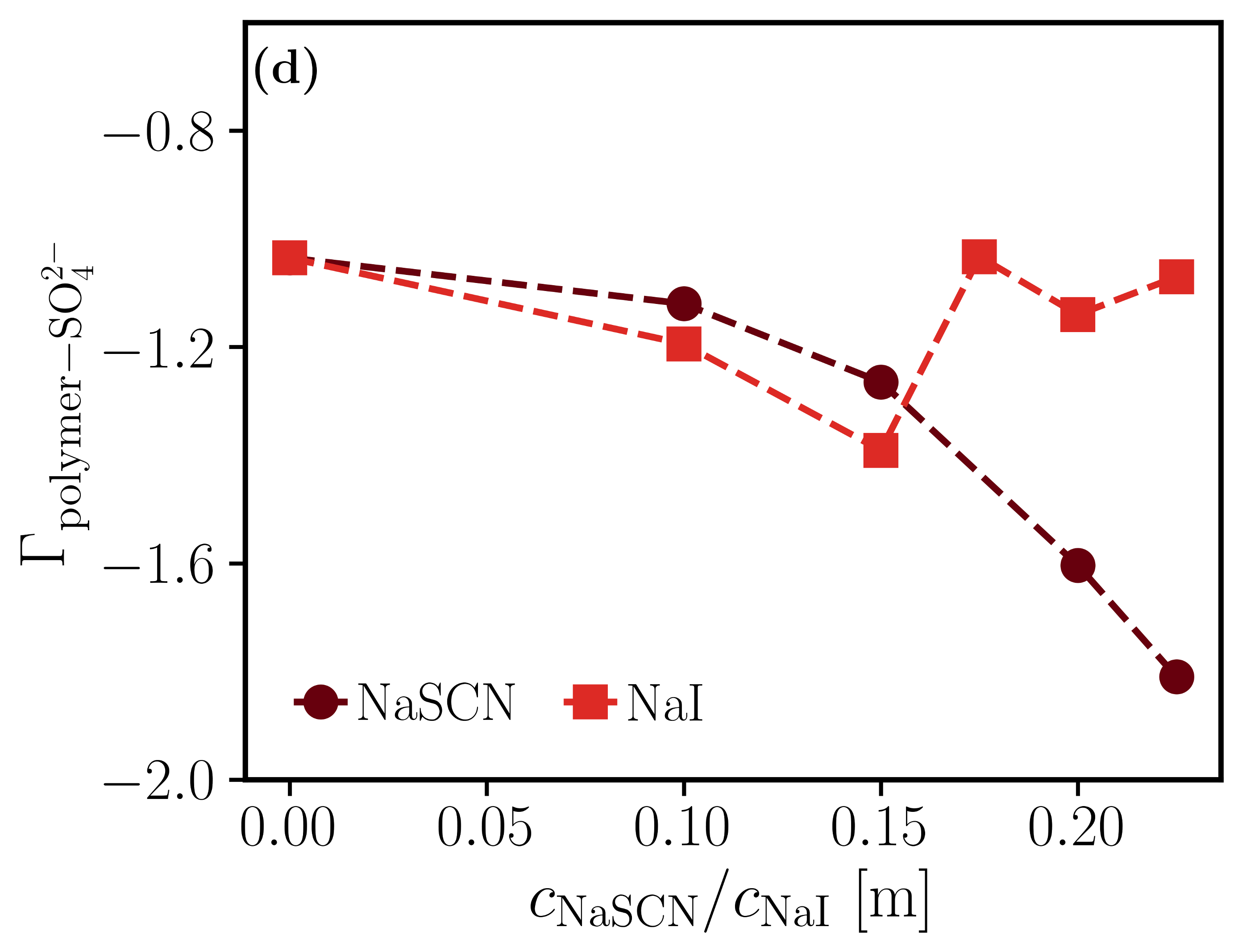}
\end{center}
 \caption{Dependence of (a) the relative polymer collapse free energy, $\Delta\Delta G^{\rm C \to G}$, and the preferential binding coefficients, (b) $\Gamma_{\rm polymer-SCN^{-}}/\Gamma_{\rm polymer-I^{-}}$, (c) $\Gamma_{\rm polymer-Na^{+}}$, and (d) $\Gamma_{\rm polymer-SO_{4}^{2-}}$ on $c_{\rm NaSCN}/c_{\rm NaI}$ at a background salt concentration of 0.5 m.} 
% \caption{Dependence of (a) the relative polymer collapse free energy, $\Delta\Delta G^{\rm C \to G}$, and (b) the polymer–SCN$^{-}$ and polymer–I$^{-}$ KBIs on $c_{\rm NaSCN}$ and $c_{\rm NaI}$, respectively, at a background salt concentration of 0.5 m.} 
\label{fig:mix_salt_scn_iodide}
\end{figure}

{Figure~\ref{fig:mix_salt_scn_iodide}(a) shows the dependence of $\Delta\Delta
G^{\rm C\rightarrow G}$ on the concentrations of two weakly hydrated salts,
NaSCN and NaI, in the presence of 0.5~m Na$_{2}$SO$_{4}$. In
NaI--Na$_{2}$SO$_{4}$ solutions, $\Delta\Delta G^{\rm C\rightarrow G}$ exhibits
a non-monotonic dependence on salt concentration, decreasing initially and
subsequently increasing with increasing NaI concentration.
Figure~\ref{fig:mix_salt_scn_iodide}(b) shows that preferential accumulation of
iodide ions around the polymer chain is weaker than that of thiocyanate ions. In
contrast, depletion of sodium and sulfate ions from the polymer vicinity is
comparable in both cases (Figs.~\ref{fig:mix_salt_scn_iodide}(c) and (d)).
Consequently, the weaker polymer--iodide interactions lead to a deeper minimum
in $\Delta\Delta G^{\rm C\rightarrow G}$ and shift the minimum to higher NaI
concentrations, as larger iodide concentrations are required to overcome the
collapse-favoring depletion of sulfate ions.}

{The concentration ranges over which the different non-additive regimes (regions
I and II) are observed span approximately 0.20--0.25~m, which is comparable to
experimentally reported systems such as PDMA in NaI--Na$_{2}$SO$_{4}$ solutions
with a 0.5~M background salt concentration.\cite{cremert:2020} An interesting
observation is that, for NaSCN--Na$_{2}$SO$_{4}$ solutions with a 1~m background
salt concentration, the $\Delta\Delta G^{\rm C\rightarrow G}$ in the re-entrant
regime exceeds the corresponding value in pure background salt solutions. In
contrast, for NaI--Na$_{2}$SO$_{4}$ solutions, the $\Delta\Delta G^{\rm
C\rightarrow G}$ in the re-entrant regime remains lower than the corresponding
value in pure background salt solutions. This is consistent with experiments
showing that the LCST in the re-entrant regime can either exceed or remain below
the LCST in pure background salt solutions depending on the polymer system.
Together, these results show that non-specific polymer--anion and polymer--water
van der Waals interactions are sufficient to reproduce the observed non-additive
behavior in mixed salt solutions, suggesting a dominant role of bulk ion--ion
and ion--water interactions in governing the coil--globule equilibrium.}

\section{Conclusion}
We have studied the coil--globule equilibrium of an uncharged linear polymer
consisting of 32 Lennard--Jones beads in aqueous solutions of the weakly
hydrated salts NaSCN and NaI, in the presence and absence of the strongly
hydrated background salt Na$_{2}$SO$_{4}$, using atomistic molecular dynamics
simulations. In pure NaSCN solutions, the polymer initially swells and
subsequently collapses with increasing NaSCN concentration. The swelling at
low salt concentrations is correlated with preferential accumulation of
SCN$^{-}$ ions near the polymer arising from favorable polymer--SCN$^{-}$
interactions. In contrast, the polymer collapses monotonically with
increasing salt concentration in aqueous Na$_{2}$SO$_{4}$ solutions due to
the progressive depletion of SO$_{4}^{2-}$ ions from the polymer vicinity.
In mixed salt solutions with a fixed concentration of Na$_{2}$SO$_{4}$, the
polymer initially collapses and subsequently swells with increasing
concentration of the weakly hydrated salt. Furthermore, these non-additive
salt effects become more pronounced with increasing background salt
concentration.

In agreement with previous atomistic simulations on
PNIPAM--NaI--Na$_{2}$SO$_{4}$ solutions, we observe that these non-additive
ion effects are governed by the interplay between polymer--weakly hydrated
anion interactions, which favor swelling, and depletion of the strongly
hydrated anions, which favors collapse. We further observe a mutual
enhancement between preferential accumulation of the weakly hydrated anions
and depletion of the strongly hydrated anions. {This mutual enhancement
correlates with partitioning of sodium ions from the counterion environment
of the weakly hydrated anions to that of the strongly hydrated anions and is
accompanied by enhanced hydration of both anions. Consequently,
non-additive changes in bulk ion pairing and ion hydration regulate both the
depletion of sulfate ions and the preferential accumulation of
SCN$^{-}$/I$^{-}$ ions near the polymer.}

{Upon replacing SCN$^{-}$ with I$^{-}$, a higher NaI concentration is
required to shift the equilibrium toward the coil state. This shift is
associated with weaker preferential accumulation of I$^{-}$ ions near the
polymer, leading to an expansion of region I before the onset of the
re-entrant regime.} Overall, our results show that non-specific
polymer--ion and polymer--water van der Waals interactions are sufficient to
qualitatively reproduce experimentally observed trends in pure and mixed salt
solutions. These findings suggest that bulk ion--ion and ion--water
interactions play a dominant role in governing non-additive salt effects,
without requiring chemically specific polymer interactions. Given that
generic polymer simulations are computationally inexpensive and compatible
with advanced sampling methods, this framework provides a route for
systematic computational investigations of other aspects of mixed salt
systems. The preferential accumulation of weakly hydrated anions and depletion of strongly hydrated anions predicted here could, in principle, be tested experimentally through equilibrium dialysis measurements, which provide ensemble-averaged measures of preferential accumulation and depletion around macromolecules. Such measurements may also provide insight into the thermodynamic coupling underlying the mutual enhancement of these competing effects.\cite{Lin1994,Smith2006}

\section{Supplementary Material}
See the Supplementary Material for more details on system composition, PMF
profiles and error estimates for the polymer collapse free energy, KBI
calculations and polymer-anion and polymer-water RDF profiles in single and
mixed solutions, and dependence of $R_{\rm g}$ on the salt concentration in
single and mixed salt solutions.
\section{Data Availability Statement}
The data that support the findings of this study are available from the corresponding author upon reasonable request.

\section{Acknowledgments}SB acknowledges financial assistance provided by the
Ministry of Education, India through SERB, India (SRG/2023/000202). KG
acknowledges the financial assistance provided by Ministry of Education, India
through SERB, India (SRG/2023/000202) under the "Scientific Social Responsibility" section.
We acknowledge the super-computing facility of Shiv Nadar Institution of
Eminence, Magus02, for providing the computational support.

\bibliography{aipsupp,reference}% Produces the bibliography via BibTeX

\section*{Supplementary Information}
\setcounter{section}{0}
\setcounter{figure}{0}
\setcounter{table}{0}
\renewcommand{\thesection}{S\arabic{section}}
\renewcommand{\thefigure}{S\arabic{figure}}
\renewcommand{\thetable}{S\arabic{table}}
\section{Non-bonded interaction Parameters}
\label{sec:interaction_parameters}
The generic hydrophobic polymer model originally proposed by Zangi et
al.\cite{Zangi2009} is employed. Water is modeled using the SPC/E model. Non-
polarizable force fields were employed for all the ions. For Na$_{2}$SO$_{4}$,
the optimized forcefield developed by Bruce et al. was
utilized.\cite{cremert:2020} Model I(4) in the work of Fyta and Netz was used
for NaI.\cite{Netz:2012} For SCN$^{-}$, the force-field employed in
K\v{r}\'{\i}\v{z}ek et al. was used.\cite{Krizek2014} The partial charges and
self-interaction Lennard-Jones parameters are listed in
Table~\ref{tab:SI_nonbonded}.

\begin{table}[tbp]
    \centering
    \caption{Partial charges and self-interaction Lennard-Jones parameters}
    \label{tab:SI_nonbonded}
    \begin{tabular}{l  c c c c }
        \toprule
        \textbf{Name}  & \textbf{Charge}& \textbf{$\sigma_{\rm ii}$ (nm)} &
\textbf{$\epsilon_{\rm ii}$ (kJ/mol)} \\\hline
        CT1  & 0.0000& 0.40000 & 1.00000  \\\hline
        OW  & -0.8476& 0.31656 & 0.65019 \\
        HW  & 0.4238 & 0.00000 & 0.00000 \\\hline
        NA   & 1.0000 & 0.25830 & 0.41860  \\\hline
        I   & -1.0000& 0.53310 & 0.15700 \\\hline
        S    & 2.0000 & 0.35500 & 1.04600 \\
        OS   &  -1 & 0.315000 & 0.836800  \\\hline
        S-P  & -0.573 & 0.38308 & 1.52256 \\
        CSP &0.483& 0.334977  & 0.425093 \\
        NSP &-0.910& 0.369722  & 0.310034\\\hline
        \bottomrule
    \end{tabular}
\end{table}

 \begin{table}[tbp]
    \centering
    \caption{$\lambda_{\rm \epsilon,ij}$ and $\lambda_{\rm,\sigma,ij}$ values
for cross-interactions utilizing LB rule}
    \label{tab:SI_LB}
    \begin{tabular}{l l c c}
        \toprule
        i & j & $\lambda_{\rm \epsilon,ij}$ & $\lambda_{\rm \sigma,ij}$ \\
        \midrule
        OW  & NA &1 & 1 \\
        NA  & I  &0.9 &1  \\
        NA  & S  &1 &1.7  \\
        NA  & OS &1 &1.7  \\
        CT1 & OW &1.075 &1   \\
        \bottomrule
    \end{tabular}
\end{table}

For certain cross-interactions, the Lorentz-Berthelot (LB) rule along with
scaling factors was employed,
\begin{equation}
\begin{split}
\epsilon_{\rm ij}=\lambda_{\rm \epsilon,ij}\sqrt{\epsilon_{\rm ii}\epsilon_{\rm
jj}},~~\sigma_{\rm ij}=\lambda_{\rm \sigma,ij}\left(\frac{\sigma_{\rm ii}+\sigma_{\rm
jj}}{2}\right).
\end{split}
\end{equation}
Table~\ref{tab:SI_LB} lists the $\lambda_{\rm \epsilon,ij}$ and
$\lambda_{\rm,\sigma ij}$ values for the cross-interactions utilizing LB rule.
All other cross-interaction parameters were computed using standard geometric
mixing rules.

%%%%%%%%%%%%%%%%%%%%%%%%%%%%%%%%%%%%%%%%%%%%%%%%%%%%%%%%%%%%%%%%%%%%%%%%%%%%
\section{System composition in single and mixed salt solutions}
Table~\ref{tab:ion_numbers} lists the number of ions used in the simulations for pure NaSCN and
Na$_2$SO$_4$ solutions, as well as for mixed-salt systems containing NaSCN or
NaI in the presence of Na$_2$SO$_4$ as the background salt at concentrations of
0.5 and 1 m.

\begin{table*}[tbp]
\centering
\caption{Number of ions in pure and mixed salt solutions at different concentrations}
\label{tab:ion_numbers}

\begin{tabular}{c|cccc|cccc}
\hline
\multicolumn{5}{c|}{\textbf{(a) Pure NaSCN}} &
\multicolumn{3}{c}{\textbf{(b) Pure Na$_2$SO$_4$}} \\
\hline

$c_{\mathrm{NaSCN}}$ (m)
& 0.10 & 0.15 & 0.20 & 0.30
&
$c_{\mathrm{Na_2SO_4}}$ (m)
& 0.30 & 0.50 & 1.00 \\
\hline

Na$^+$
& 60 & 90 & 120 & 177
&
Na$^+$
& 354 & 590 & 1180 \\

SCN$^-$
& 60 & 90 & 120 & 177
&
SO$_4^{2-}$
& 177 & 295 & 590 \\
\hline
\end{tabular}

\vspace{0.8cm}

\begin{tabular}{c|cccccc}
\hline
\multicolumn{7}{c}{\textbf{(c) Mixed salt:}
$c_{\mathrm{Na_2SO_4}} = 0.5~\mathrm{m}$} \\
\hline

$c_{\mathrm{NaI/NaSCN}}$ (m)
& 0.10 & 0.15 & 0.175 & 0.20 & 0.225 & 0.275 \\
\hline

Na$^+$
& 650 & 680 & 695 & 710 & 725 & 752 \\

I$^-$/SCN$^-$
& 60 & 90 & 105 & 120 & 135 & 162 \\

SO$_4^{2-}$
& 295 & 295 & 295 & 295 & 295 & 295 \\
\hline
\end{tabular}

\vspace{0.8cm}

\begin{tabular}{c|cccccc}
\hline
\multicolumn{7}{c}{\textbf{(d) Mixed salt:}
$c_{\mathrm{Na_2SO_4}} = 1~\mathrm{m}$} \\
\hline

$c_{\mathrm{NaSCN}}$ (m)
& 0.075 & 0.10 & 0.15 & 0.175 & 0.20 & 0.225 \\
\hline

Na$^+$
& 1225 & 1240 & 1270 & 1285 & 1300 & 1315 \\

SCN$^-$
& 45 & 60 & 90 & 105 & 120 & 135 \\

SO$_4^{2-}$
& 590 & 590 & 590 & 590 & 590 & 590 \\
\hline
\end{tabular}

\end{table*}
\section{Polymer collapse free energy}

The polymer collapse free energy can be determined from the potential of mean force (PMF) profiles, $w(R_g)$, using

\begin{equation}
e^{-\Delta G^{C \rightarrow G}/RT}
=
\frac{\int_{0}^{R_g^\#} e^{-w(R_g)/RT}\, dR_g}
{\int_{R_g^\#}^{\infty} e^{-w(R_g)/RT}\, dR_g}
\label{umb_equation}
\end{equation}
\begin{figure}[tbp]
\centering
\includegraphics[scale=0.4]{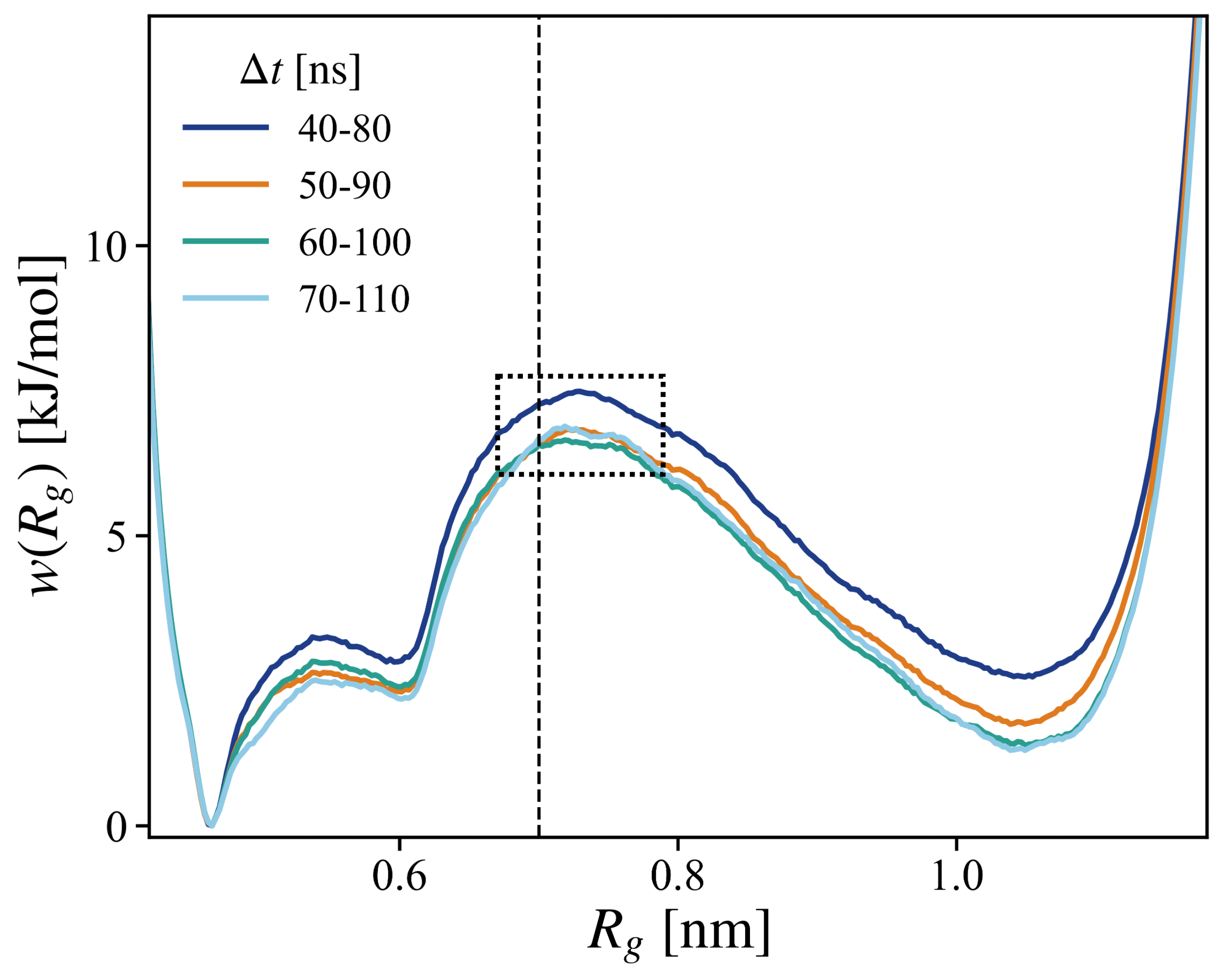}
\caption{PMF profiles, $w(R_g)$, for a mixed salt solution containing
0.1~m NaSCN and 1~m {Na$_{2}$SO$_{4}$}, computed over 40~ns averaging intervals
using different starting times.} 
\label{pmffinal}
\end{figure}
 The unbiased potential of mean force (PMF) profiles was reconstructed using the weighted histogram analysis method (WHAM) implemented by the code developed by the Grossfield laboratory.\cite{grossfield_wham} The WHAM calculation was carried out using a command that requires several input parameters: the minimum value of the histogram range (\texttt{hist\_min}), the maximum value of the histogram range (\texttt{hist\_max}), the total number of histogram bins (\texttt{num\_bins}), the convergence tolerance (\texttt{tol}), the simulation temperature (\texttt{temperature}), and the number of padding points added to the histogram (\texttt{numpad}). %These parameters collectively define the resolution, numerical stability, and convergence criteria of the PMF reconstruction.

Umbrella sampling simulations were carried out using 33 sampling windows. Each
umbrella window was simulated for 110 ns to ensure adequate sampling of
conformational space.  The PMF profile exhibits noticeable drift during the
initial stage of the simulation and does not reach equilibrium within the first
50 ns, as shown in Fig.~\ref{pmffinal}. Therefore, the initial portion of each
trajectory, up to 50 ns, was discarded to remove non-equilibrated data and
ensure that only equilibrated configurations contributed to the analysis. To
assess the convergence of the free-energy calculations, the potential of mean
force (PMF), denoted as $w(R_g)$, was computed over multiple time intervals
from the equilibrated segments of the trajectories. The resulting
time-dependent PMF profiles were compared to evaluate the stability of the
free-energy landscape, as illustrated in Fig.~\ref{pmffinal}. After confirming
convergence, the collapse free-energy difference between the coil and globule
states, $\Delta G^{C\rightarrow G}$, was calculated separately from each
equilibrated PMF profile. The final reported value of $\Delta G^{C\rightarrow
G}$ corresponds to the mean of the values obtained from the different
equilibrated profiles, while the associated uncertainty was estimated as the
standard deviation of these independently calculated $\Delta G^{C\rightarrow
G}$ values.
\begin{figure}[tbp]
\centering
\includegraphics[width=0.32\textwidth]{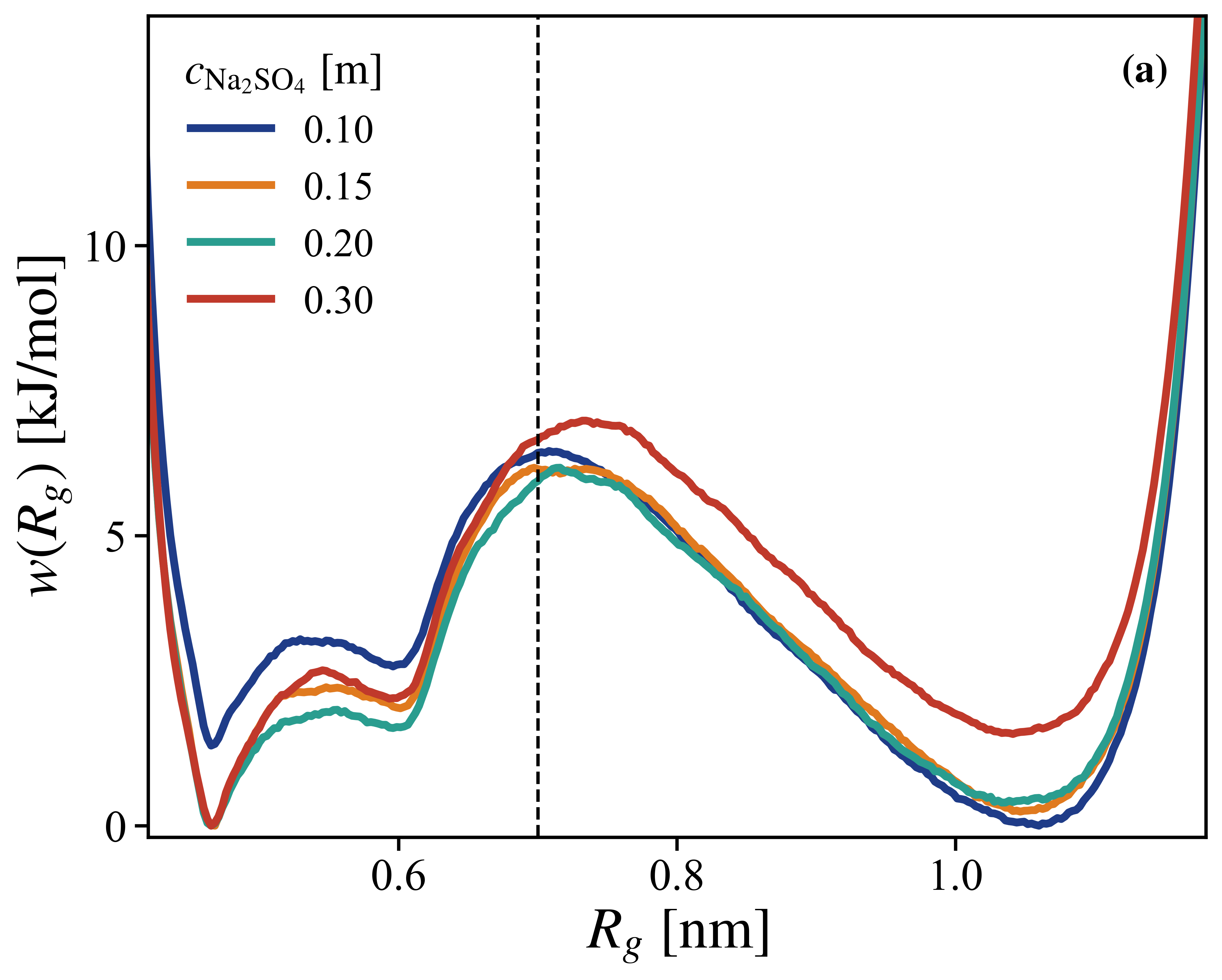}
\includegraphics[width=0.32\textwidth]{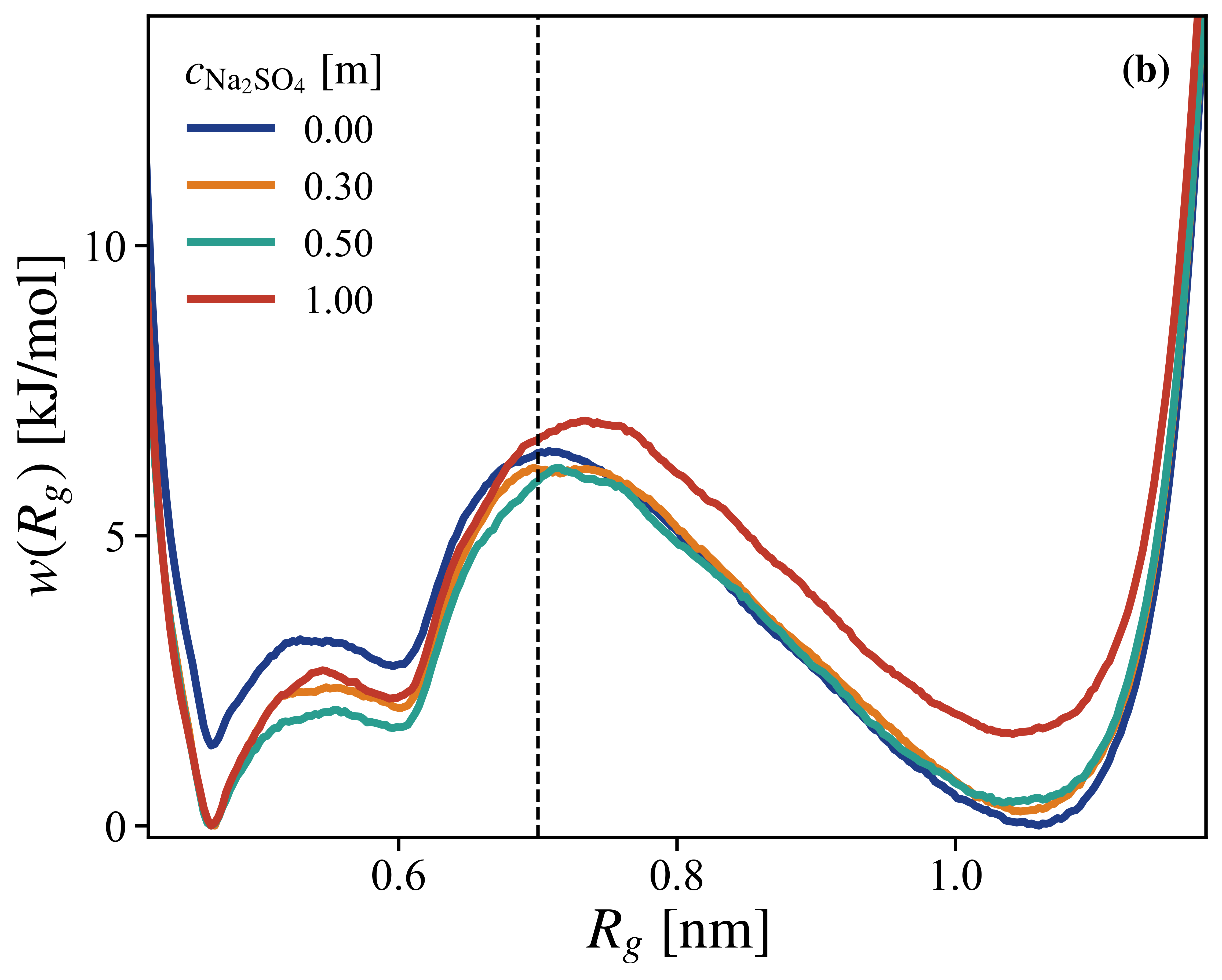}
\includegraphics[width=0.32\textwidth]{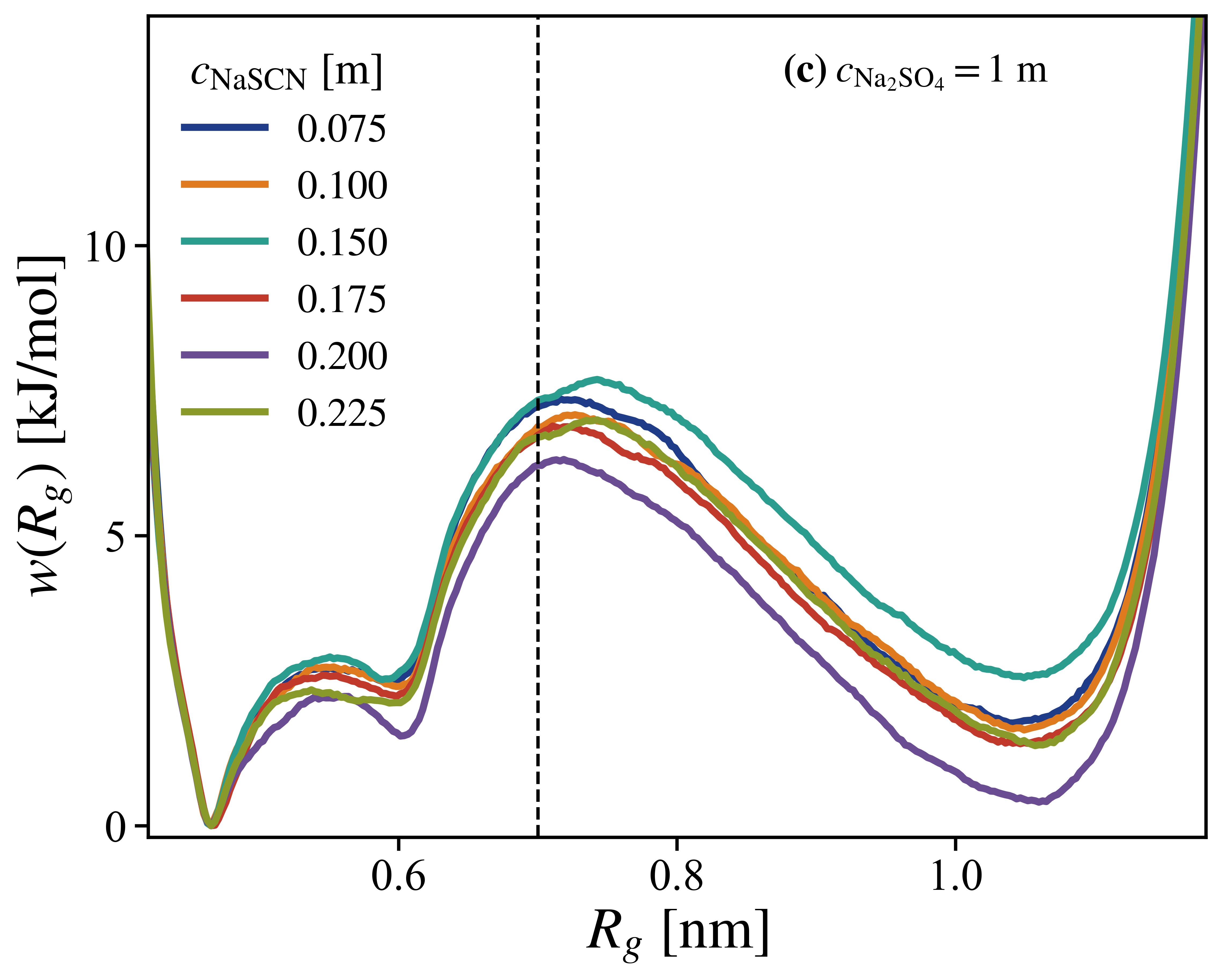}
\includegraphics[width=0.32\textwidth]{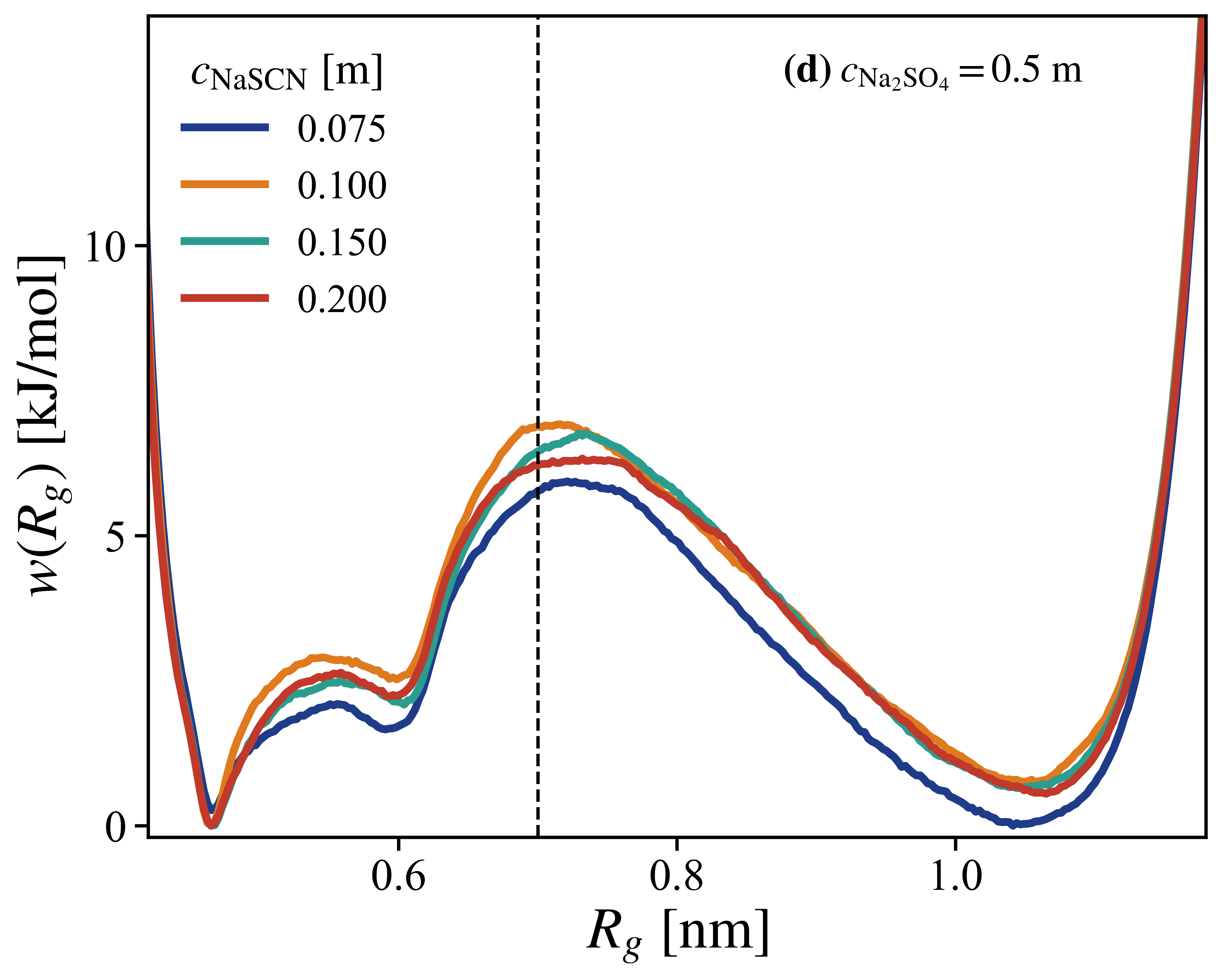}
\includegraphics[width=0.32\textwidth]{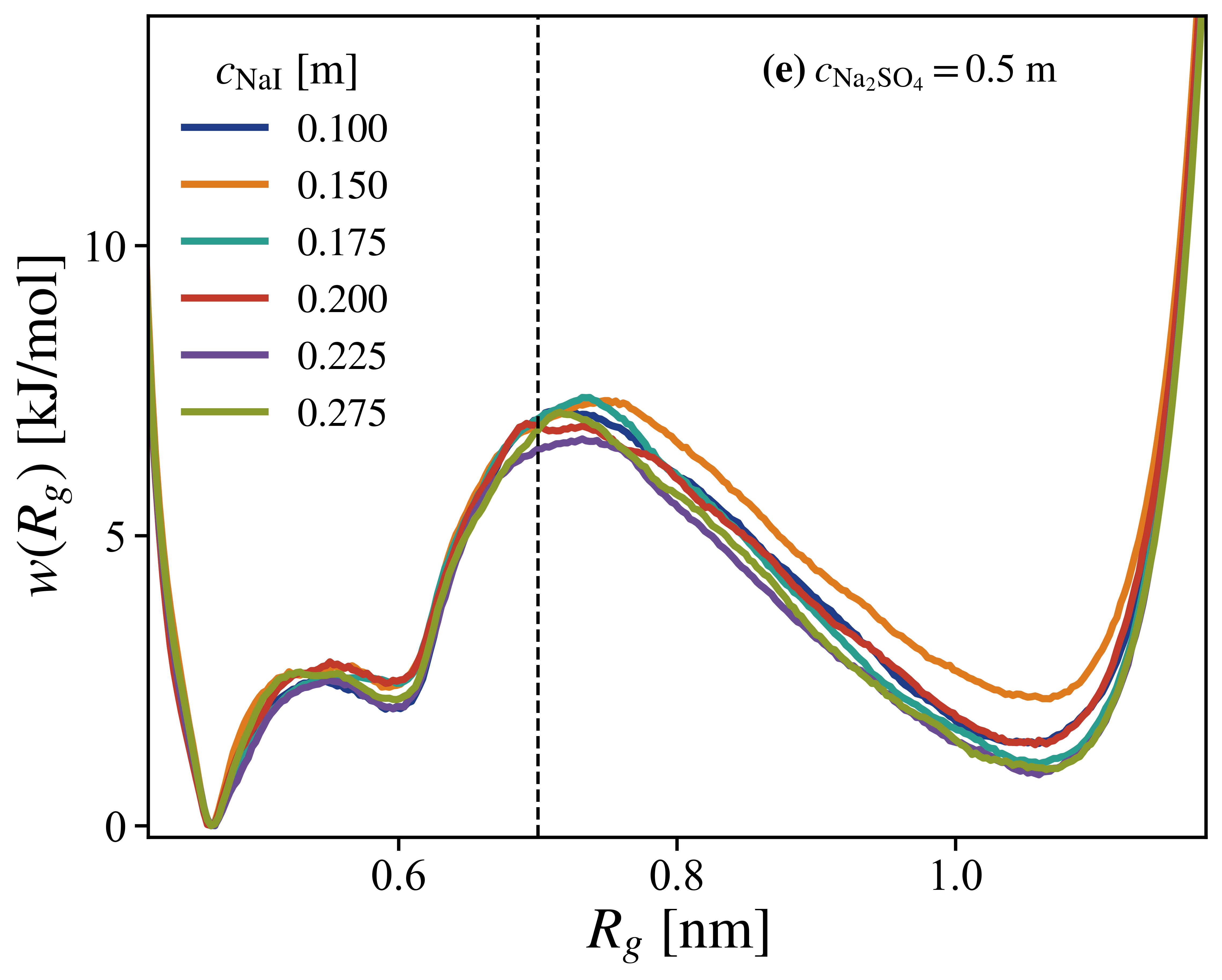}
\caption{The PMF profiles used to compute the change in collapse free energy for pure
NaSCN and Na$_2$SO$_4$ solutions are shown in panels (a) and (b),
respectively. Panels (c)--(e) show the corresponding PMF profiles for mixed
salt systems containing 0.5~m and 1~m Na$_2$SO$_4$ as the background salt,
together with varying concentrations of the weakly hydrated salts NaSCN or
NaI.}
\label{S2}
\end{figure}
\section{Kirkwood--Buff integrals}

The Kirkwood–Buff integral (KBI) provides a connection between microscopic structural information obtained from radial distribution functions (RDFs) and macroscopic thermodynamic properties, such as preferential binding.

\begin{equation}
G_{\rm{ij}} =
\int_{0}^{\infty}
\left[g_{\rm{ij}}(r)-1\right]4\pi r^2 dr
\label{kbi}
\end{equation}

where $G_{\rm{ij}}$ is the Kirkwood–Buff integral and $g_{\rm{ij}}(r)$ is the
radial distribution function between species $\rm{i}$ and $\rm{j}$ (here, the
polymer and counterion). The RDFs were calculated using the \texttt{rdf}
routine implemented in GROMACS. Finite-size effects arising from the limited
simulation box were corrected using the Ganguly correction.\cite{ganguly} The
corrected RDF is

\begin{equation}
g^{\mathrm{corrected}}_{\mathrm{ij}}(r)
=
g_{\mathrm{ij}}(r)
\,
\frac{N_{\mathrm{j}}\left(1-\frac{4\pi r^{3}}{3V}\right)}
{N_{\mathrm{j}}\left(1-\frac{4\pi r^{3}}{3V}\right) - \Delta N_{\mathrm{ij}}(r) - \delta_{\mathrm{ij}}}
\end{equation}

where $N_{\mathrm{j}}$ is the number of particles of type $\mathrm{j}$, $V$ is the system volume, $\Delta N_{\mathrm{ij}}(r)$ is the excess number of particles of type $\mathrm{j}$ within a radius $r$, and $\delta_{\mathrm{ij}}$ is the Kronecker delta. To improve the convergence of the KBI, the extrapolation scheme proposed by Krüger \textit{et al.} were applied to the corrected RDFs.\cite{kruger} The finite-size KBIs can then be be computed in the following way,
\begin{equation}
\begin{split}
G_{\rm ij}(R)
&=
\int_0^{R}
\omega(r,R)
\left[g^{\mathrm{corrected}}_{\rm {ij}}(r)-1\right] dr, ~~~~~\omega(r,R)
=
4\pi r^2
\left[
1-\left(\frac{r}{R}\right)^3
\right]
\label{eq:kbi_finite}
\end{split}
\end{equation}
The Kirkwood--Buff integrals were computed using different correction
schemes proposed by Milzetti \textit{et al.}~\cite{Milzetti2018}. These
combinations of KBI and RDF corrections have been shown to improve KBI
convergence in both ideal and nonideal aqueous mixtures. Figure~\ref{fig:kbi}
shows the polymer--SCN$^{-}$ KBIs computed using three different 
schemes in both pure and mixed salt solutions. It can be seen that the KBIs
obtained from the expression used in this work exhibit satisfactory
convergence in all cases.\begin{figure}[tbp]
\centering
\includegraphics[scale=0.35]{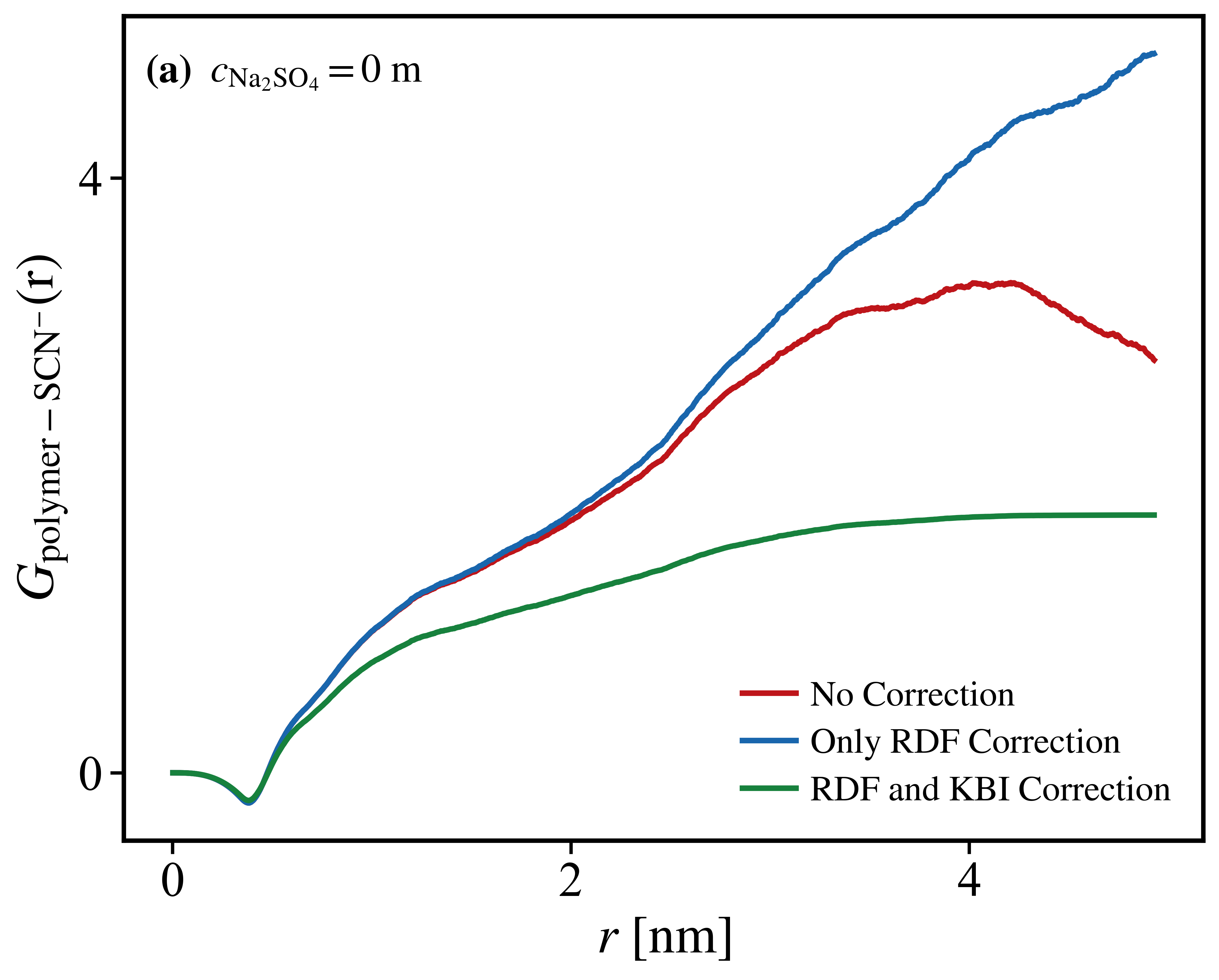}
\includegraphics[scale=0.35]{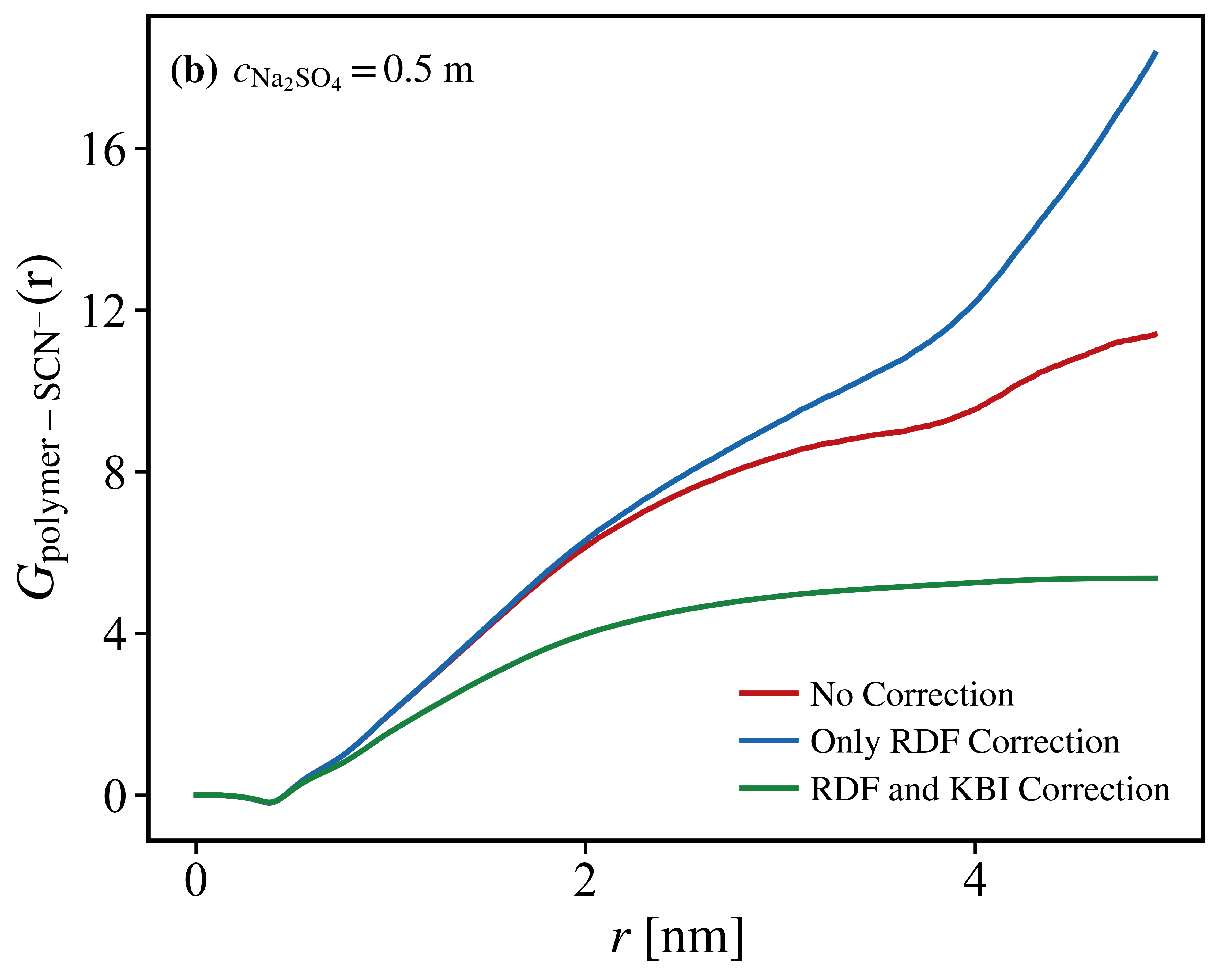}
\includegraphics[scale=0.35]{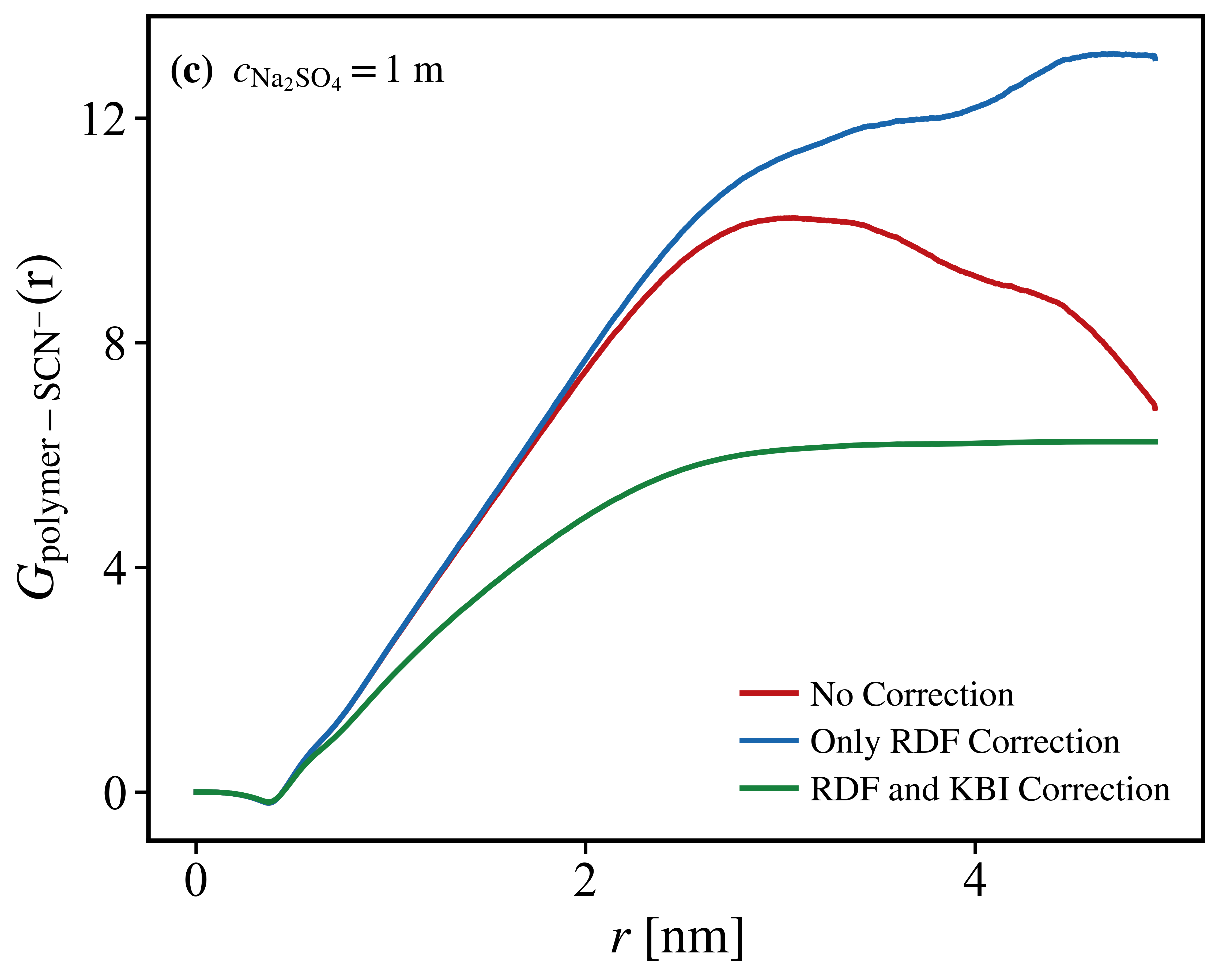}
\caption{$\rm G_{polymer\text{-}SCN^{-}}(r)$ functions computed using different
correction schemes, shown as a function of $r$ for 0.1 m NaSCN
at background salt concentrations of
(a) 0, (b) 0.5, and (c) 1~m \ce{Na2SO4}.}
\label{fig:kbi}
\end{figure}
The resulting plateau value is defined as the corrected $G_{\rm{ij}}$. The RDF without correction is provided in Fig.~\ref{rdfa}, 
\ref{rdfb}, \ref{rdfc}, \ref{rdfd} in Sec.~S5.1
\section{Radial Distribution function}
The radial distribution functions (RDFs) were computed using the \texttt{rdf} routine implemented in the GROMACS simulation package\cite{gromacs2015}. 
\subsection{Monomer-Ions RDFs}
\label{sec:monomerrdf}
For this analysis, each polymer bead was used as a reference site for computing
the RDFs. Figures~\ref{rdfa}(a--c) show the RDFs for pure sodium thiocyanate
solutions, while Figures~\ref{rdfb}(a--c) present the RDFs for pure sodium
sulfate solutions. Figure~\ref{rdfc} shows the monomer--SO$_4^{2-}$ RDFs at
background salt concentrations of (a) 0.5 and (b) 1~m sodium sulfate with
varying \ce{NaSCN} concentrations. Figures~\ref{rdfd}(a--c) show the RDFs
between the monomer and weakly hydrated anions (\ce{SCN^-}/\ce{I^-}) in mixed
salt solutions at fixed \ce{Na_2SO_4} concentration with varying weakly
hydrated salt (\ce{NaI}/\ce{NaSCN}) concentrations. Figures~\ref{rdfe}(a--c)
present the RDFs between the monomer and \ce{Na^+} ions in mixed salt solutions
at fixed \ce{Na_2SO_4} concentration with varying weakly hydrated salt
(\ce{NaI}/\ce{NaSCN}) concentrations.
\begin{figure}
\centering
\includegraphics[scale=0.35]{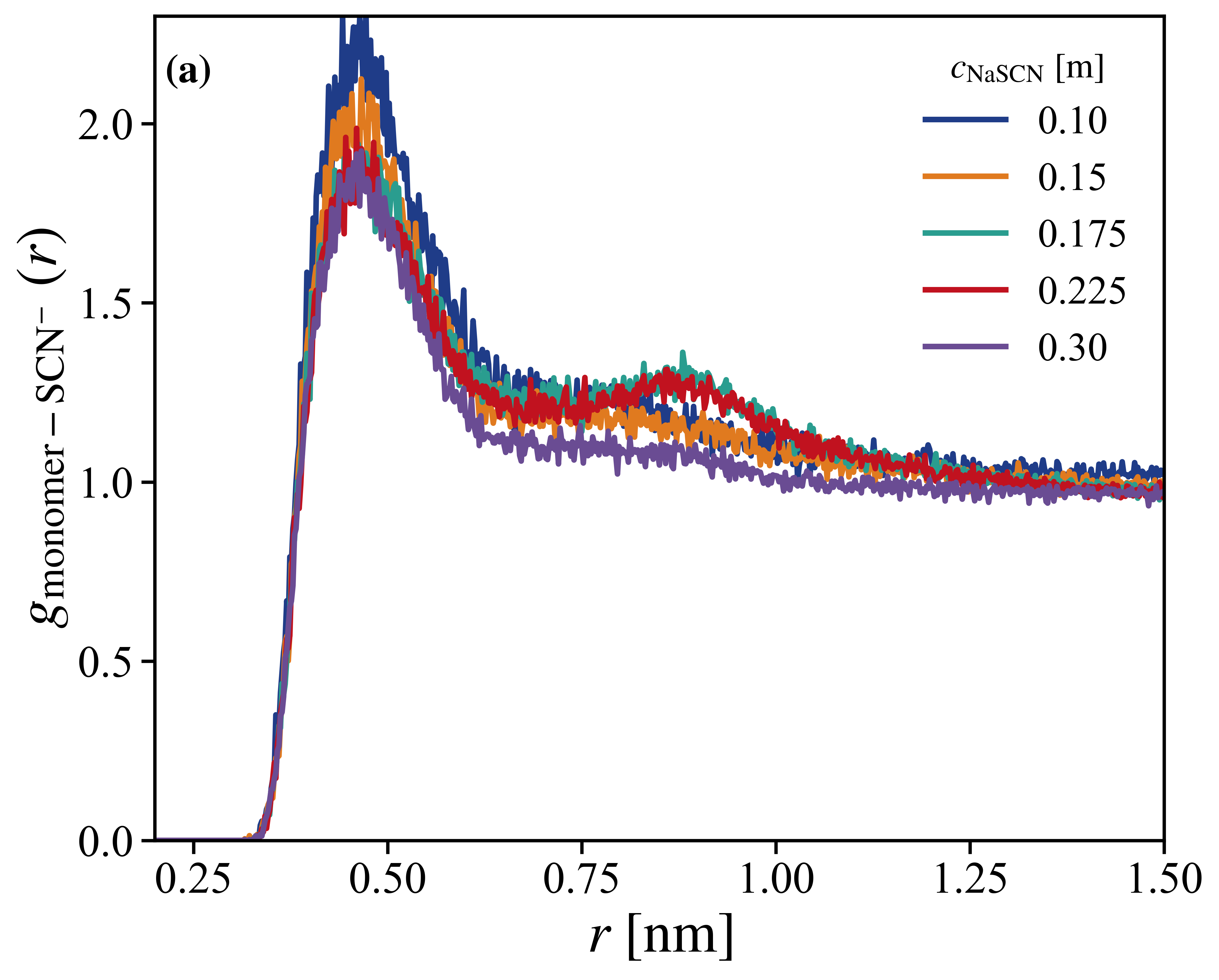}
\includegraphics[scale=0.35]{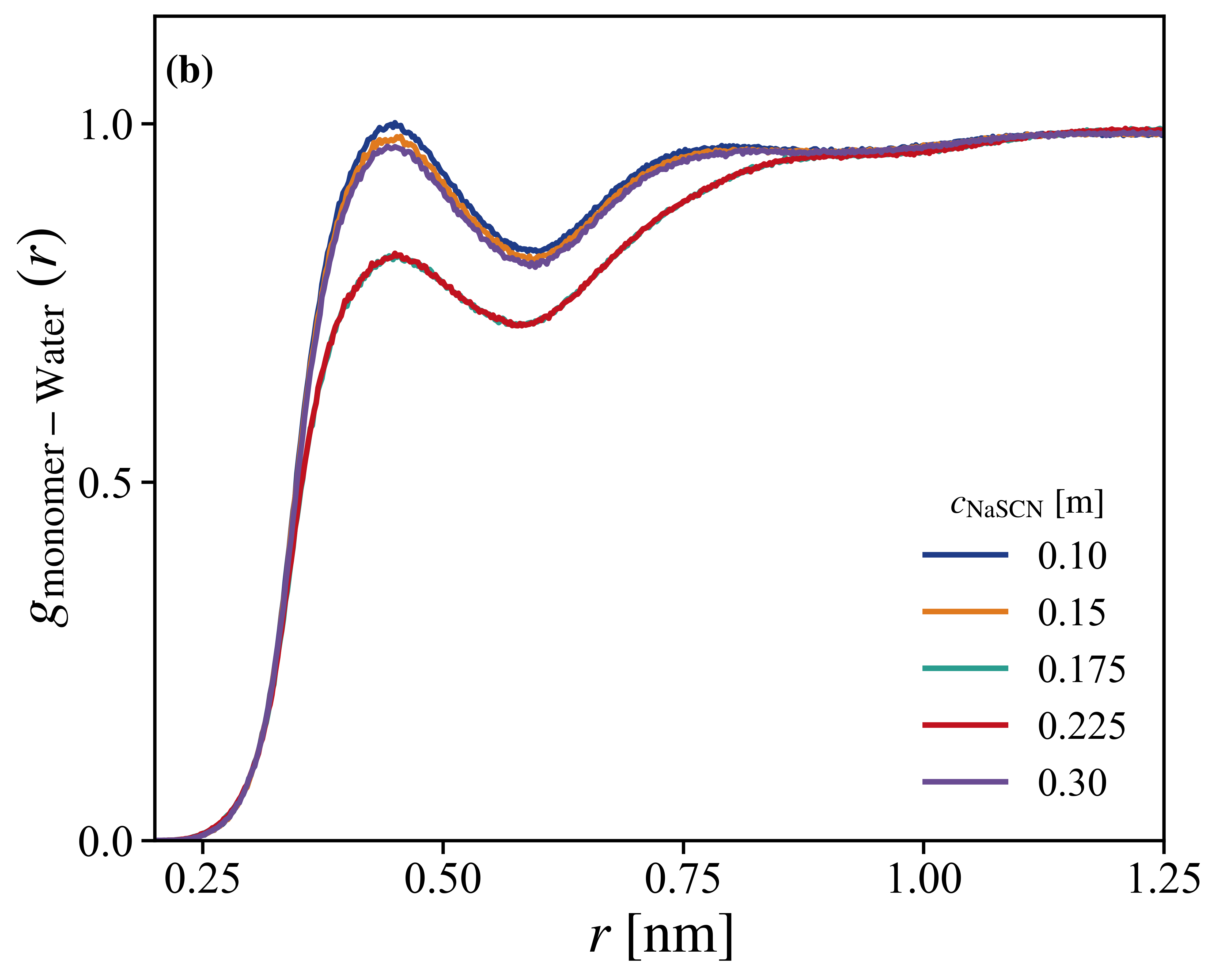}  
\includegraphics[scale=0.35]{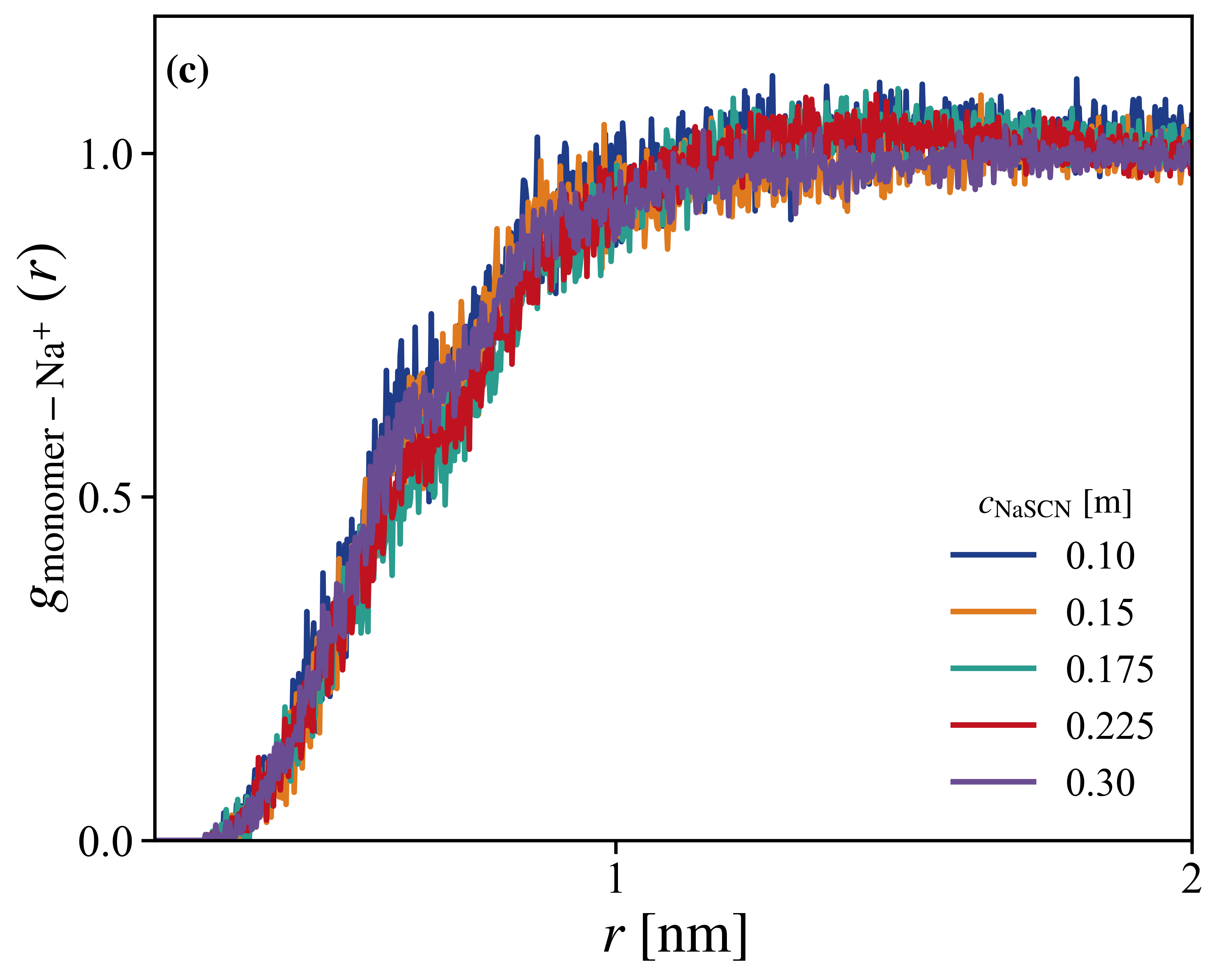}
\caption{Radial distribution functions for pure sodium thiocyanate solutions at
varying NaSCN concentrations: (a) monomer--SCN$^{-}$,
(b) monomer--water, and (c) monomer--Na$^{+}$ RDFs. }
\label{rdfa}
\end{figure}

\begin{figure}
\centering
\includegraphics[width=0.32\textwidth]{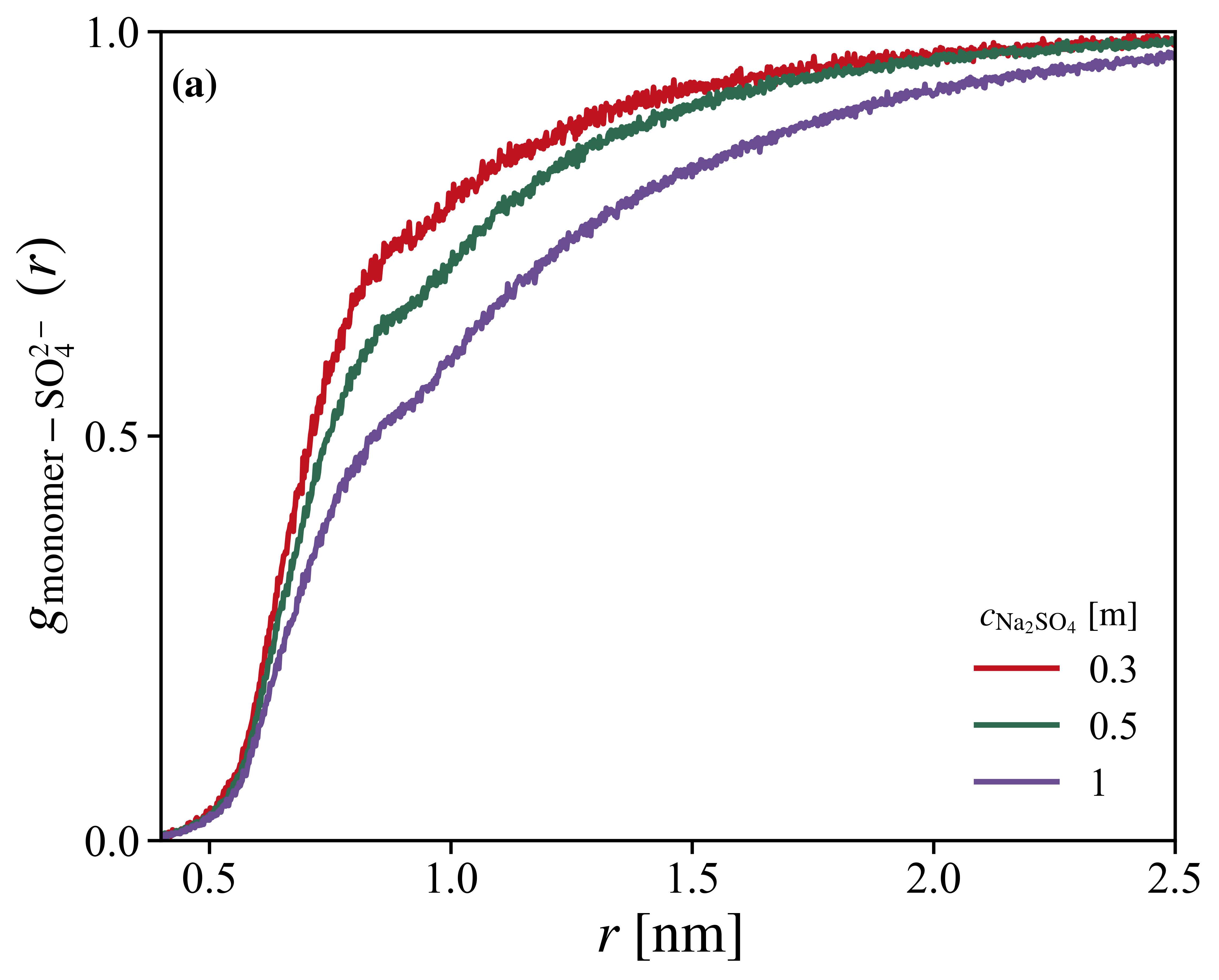}
\includegraphics[width=0.32\textwidth]{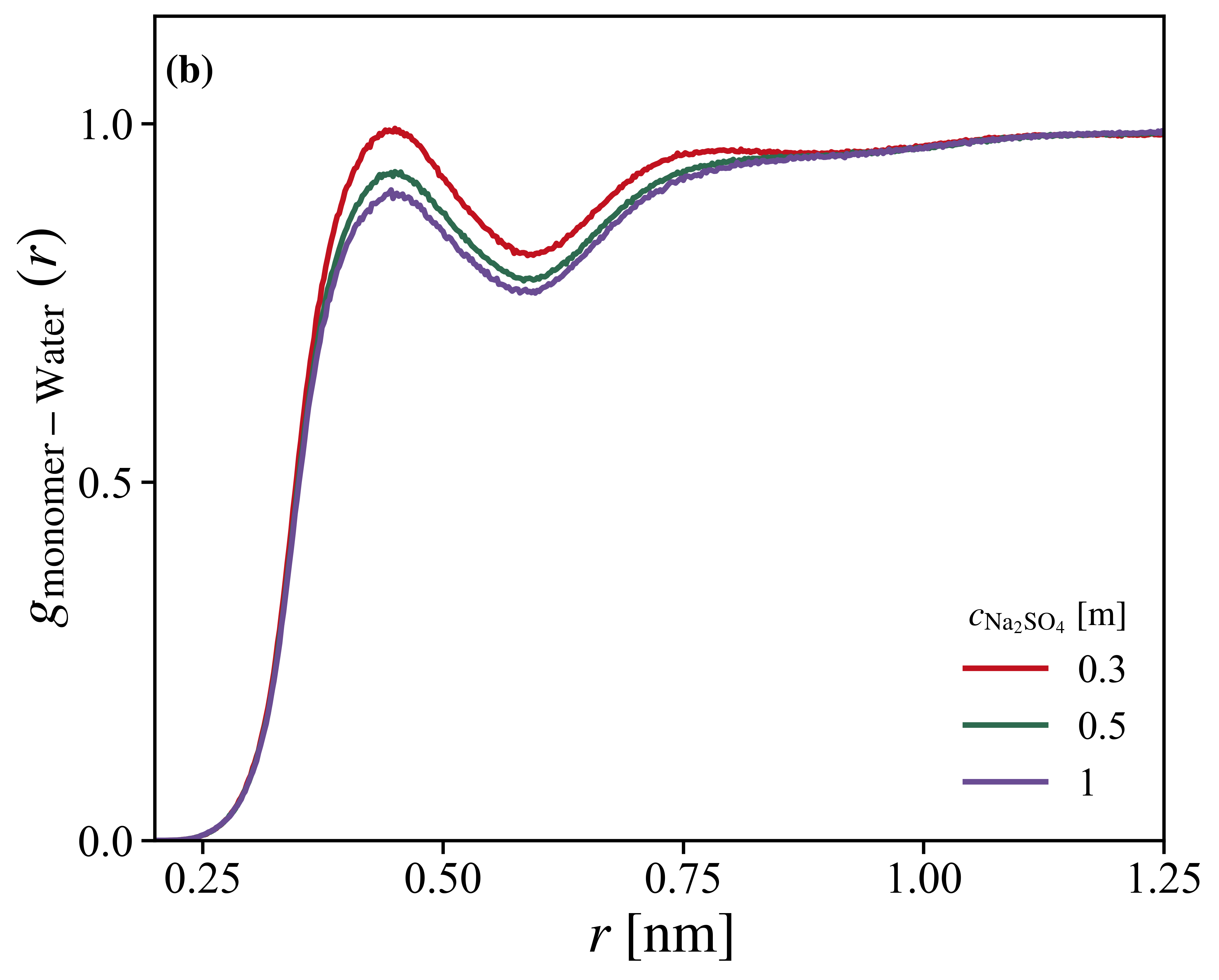} 
\includegraphics[width=0.32\textwidth]{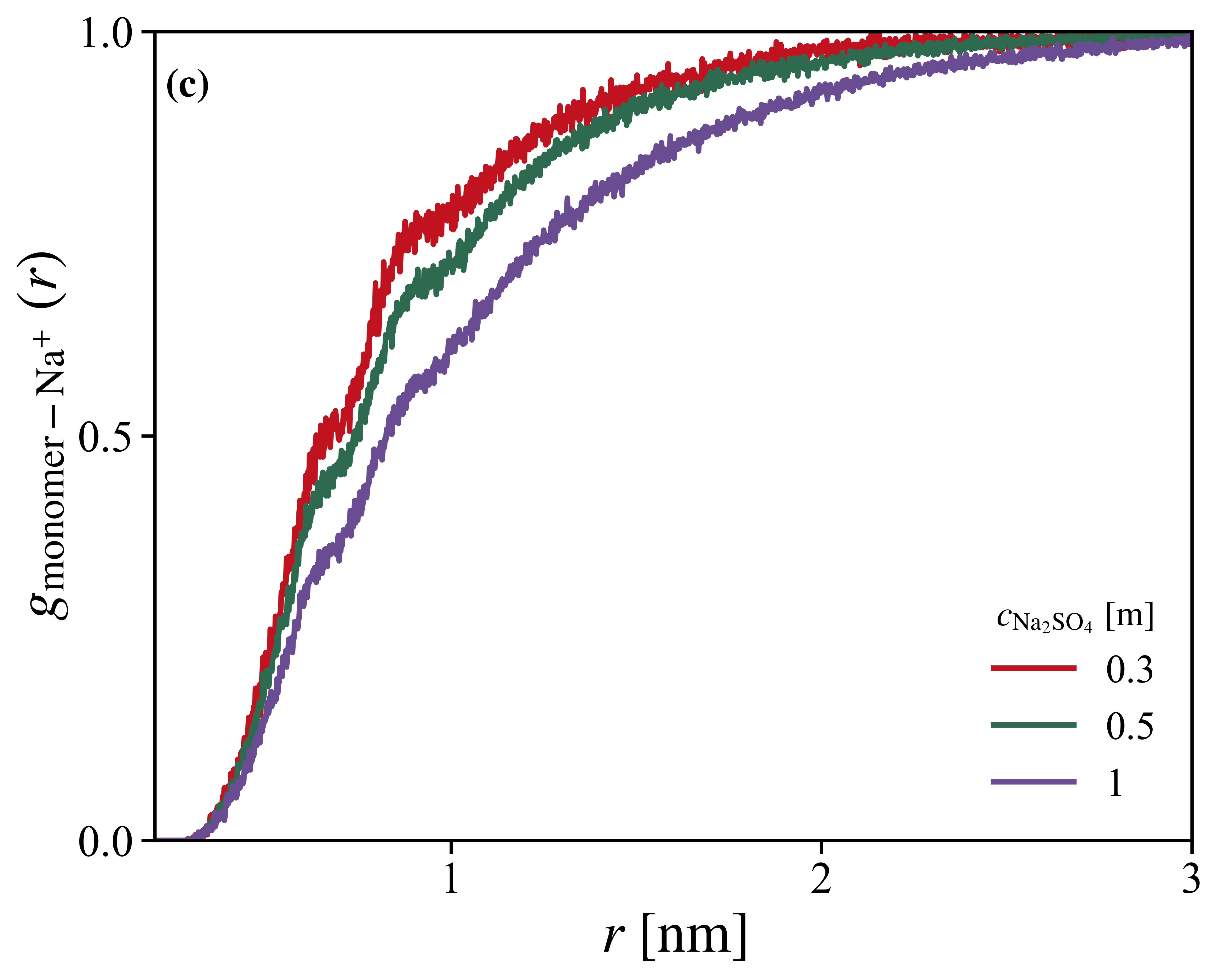}
\caption{Radial distribution functions for pure sodium sulfate solutions at varying
Na$_2$SO$_4$ concentrations: (a) monomer--SO$_4^{2-}$,
(b) monomer--water, and (c) monomer--Na$^{+}$ RDFs.}
\label{rdfb}
\end{figure}
\begin{figure}
\centering
\includegraphics[scale=0.35]{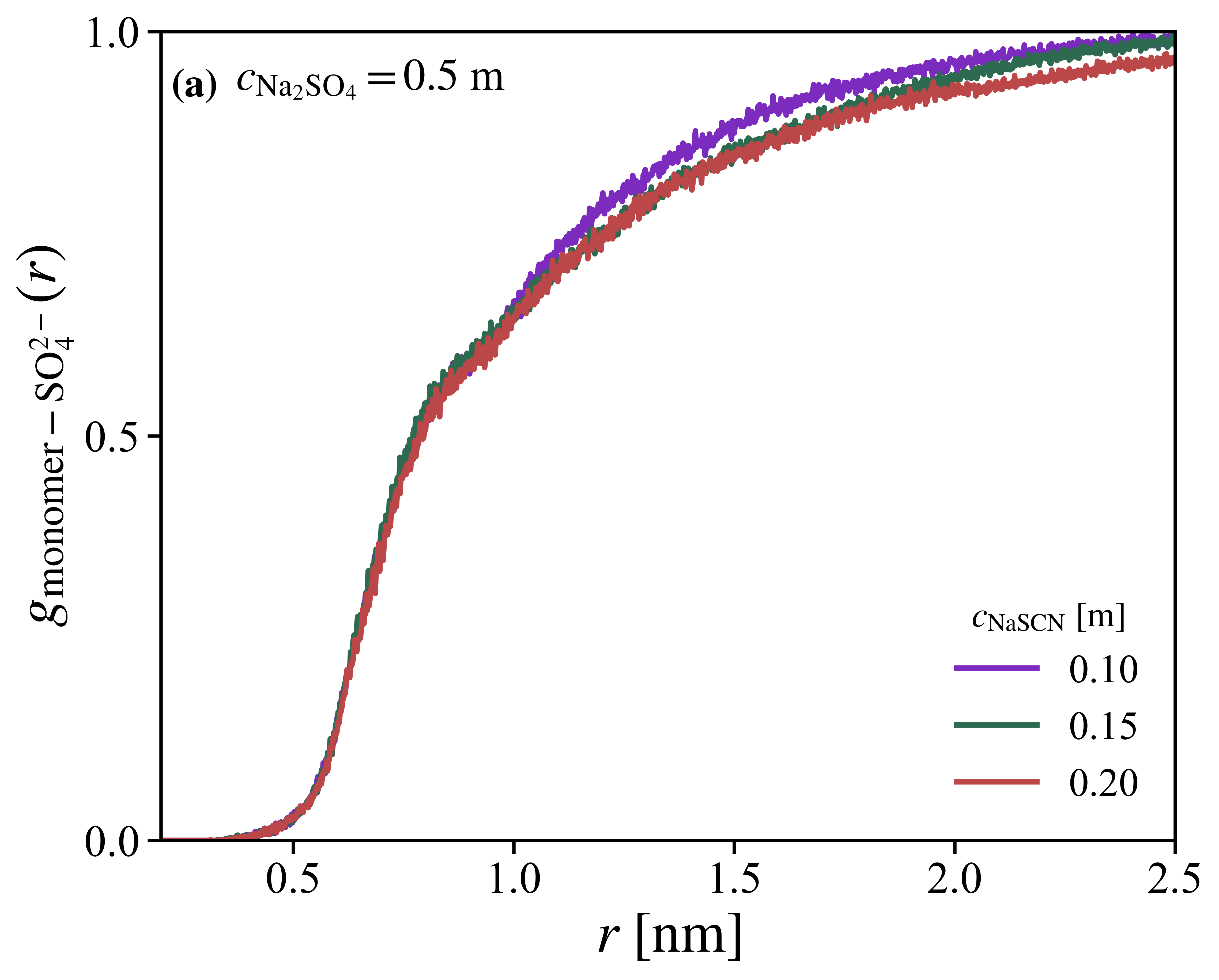}
\includegraphics[scale=0.35]{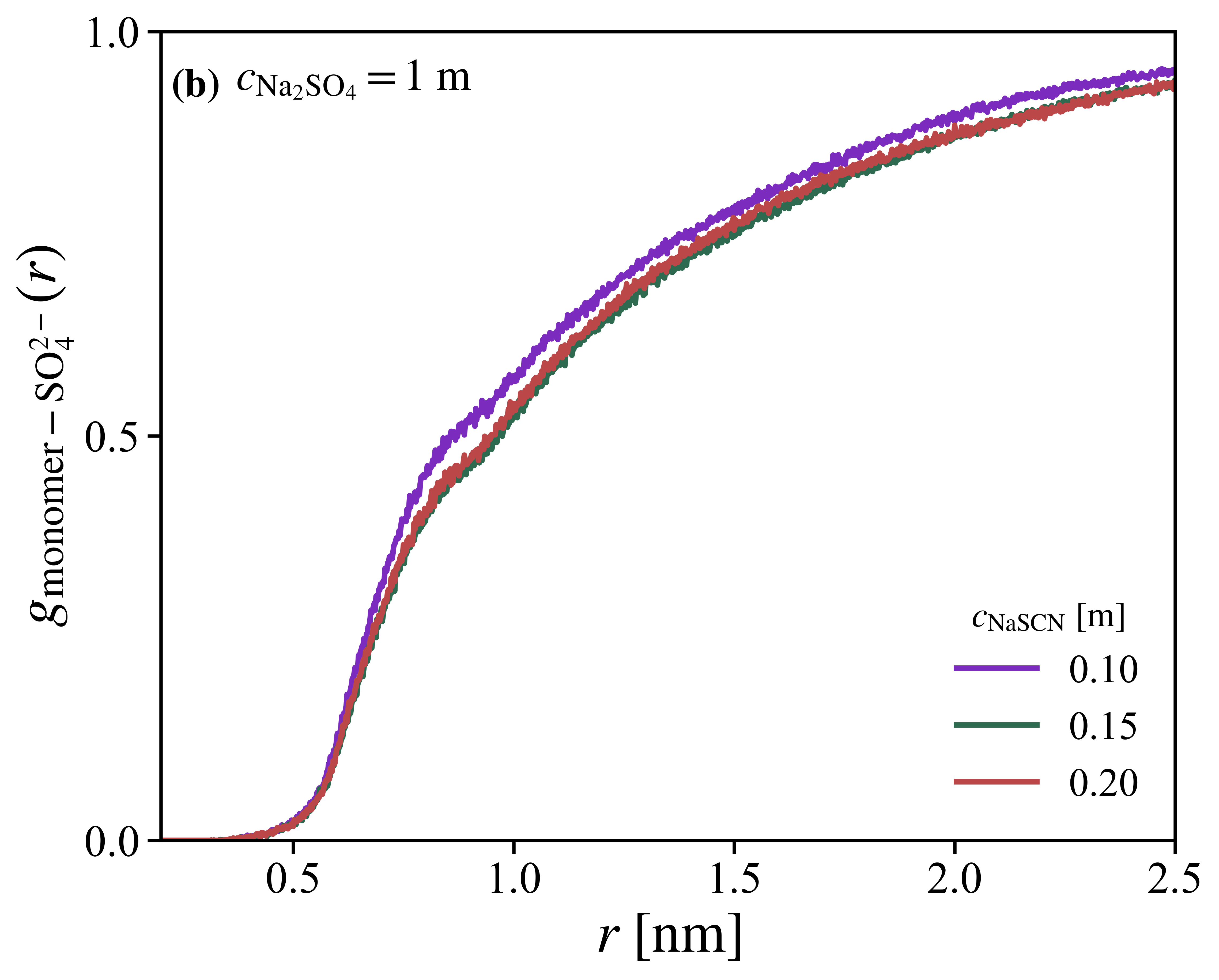}  
\caption{
Monomer--SO$_4^{2-}$ radial distribution functions for mixed salt solutions
with varying NaSCN concentrations at different Na$_2$SO$_4$ background salt
concentrations: (a) 0.5~m and (b) 1~m.}
\label{rdfc}
\end{figure}

\begin{figure}
\centering
\includegraphics[scale=0.35]{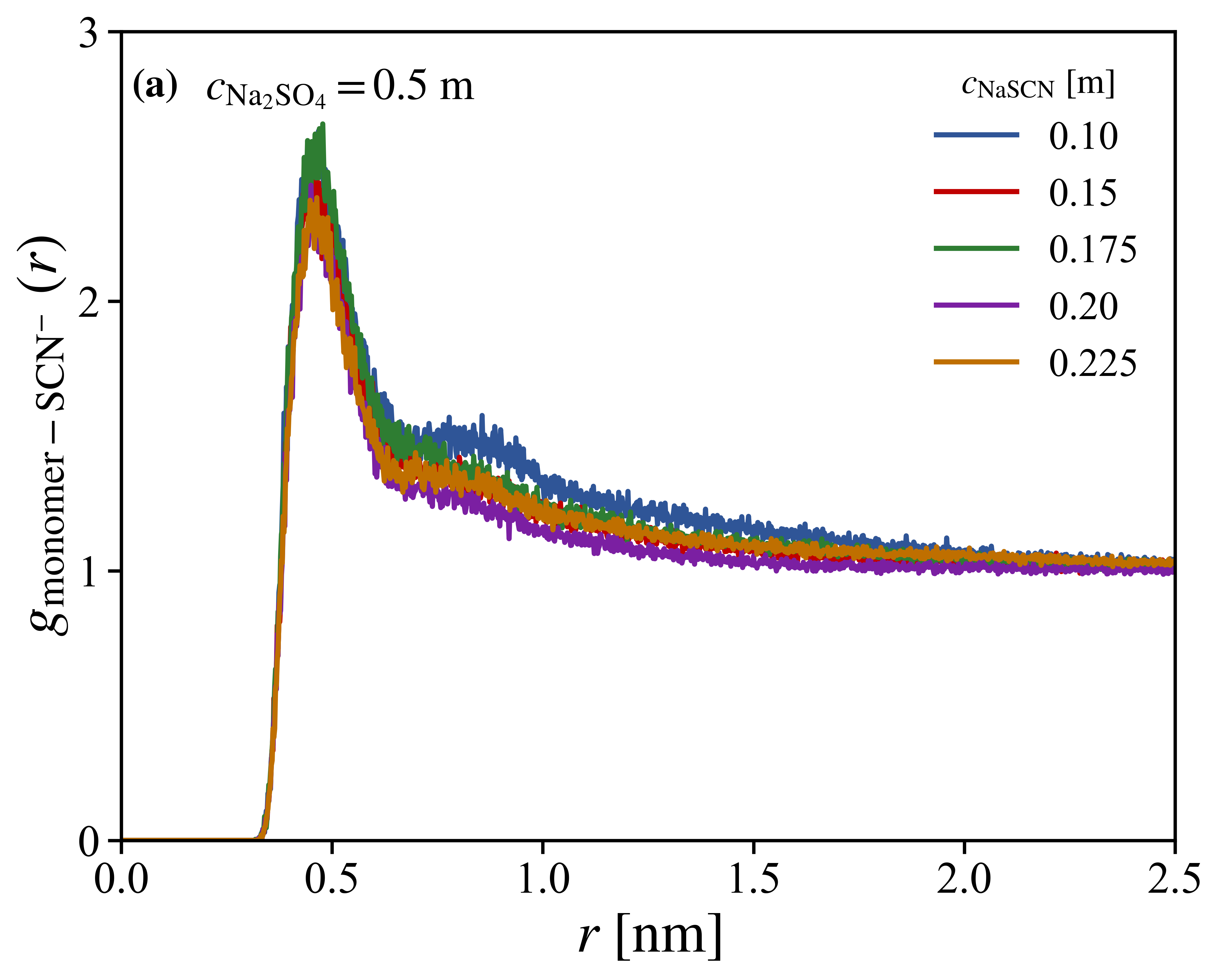}
\includegraphics[scale=0.35]{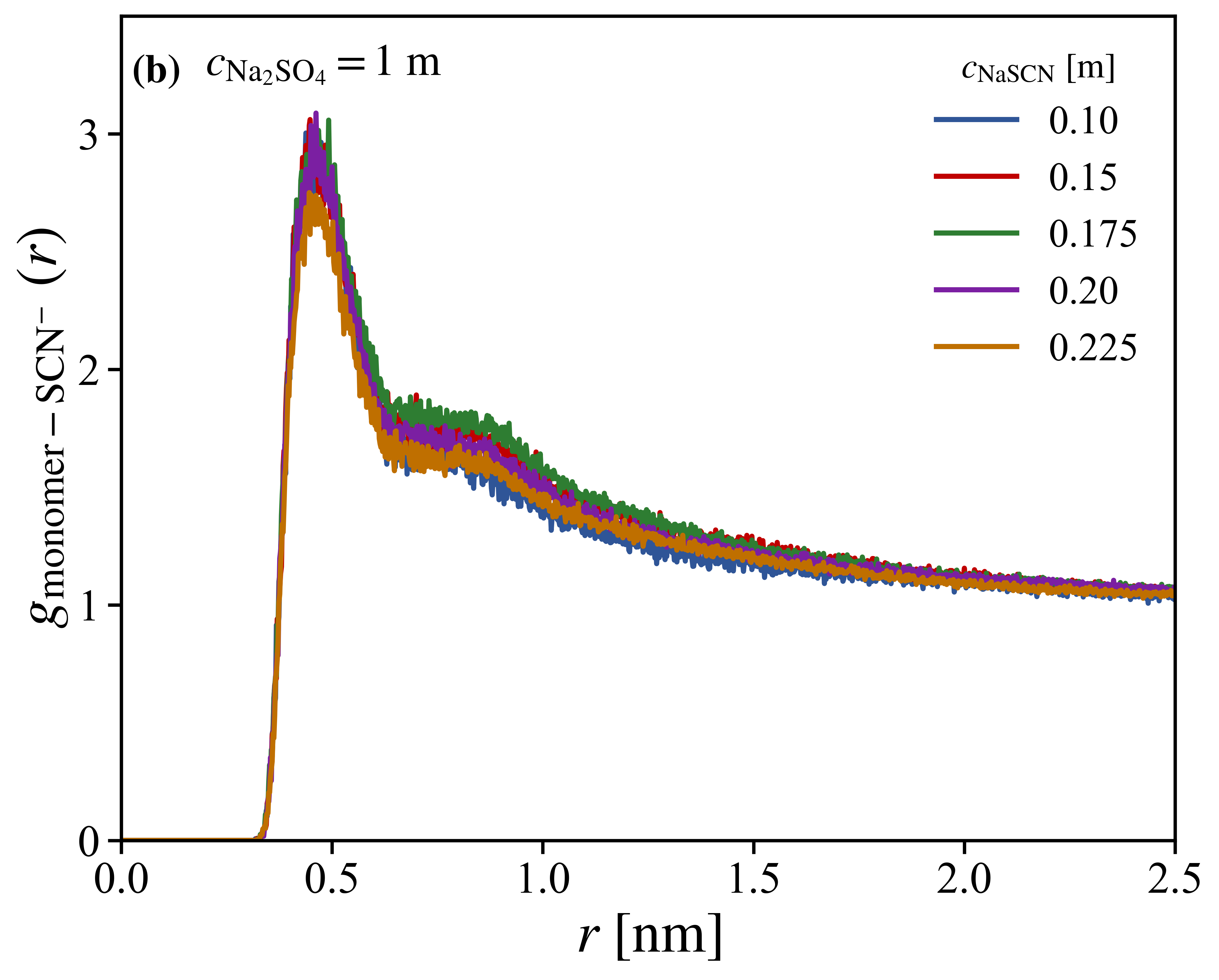} 
\includegraphics[scale=0.35]{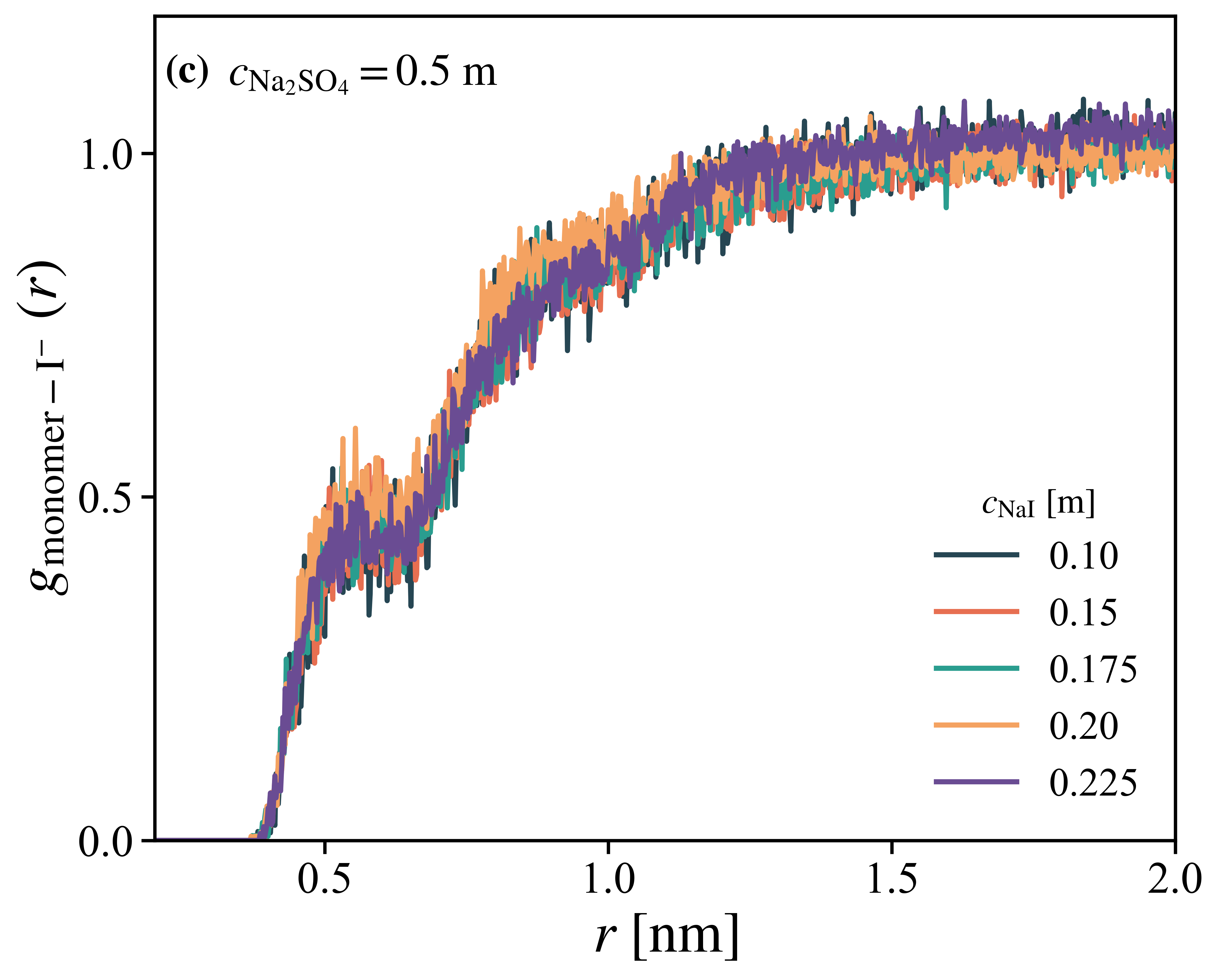}  
\caption{
Monomer--anion radial distribution functions for mixed salt solutions with
varying weakly hydrated salt concentrations:
(a) monomer--SCN$^{-}$ RDFs at a Na$_2$SO$_4$ background salt concentration
of 0.5~m,
(b) monomer--SCN$^{-}$ RDFs at a Na$_2$SO$_4$ background salt concentration
of 1~m, and
(c) monomer--I$^{-}$ RDFs at a Na$_2$SO$_4$ background salt concentration
of 0.5~m.}
\label{rdfd}
\end{figure}
\begin{figure}
\centering
\includegraphics[width=0.32\textwidth]{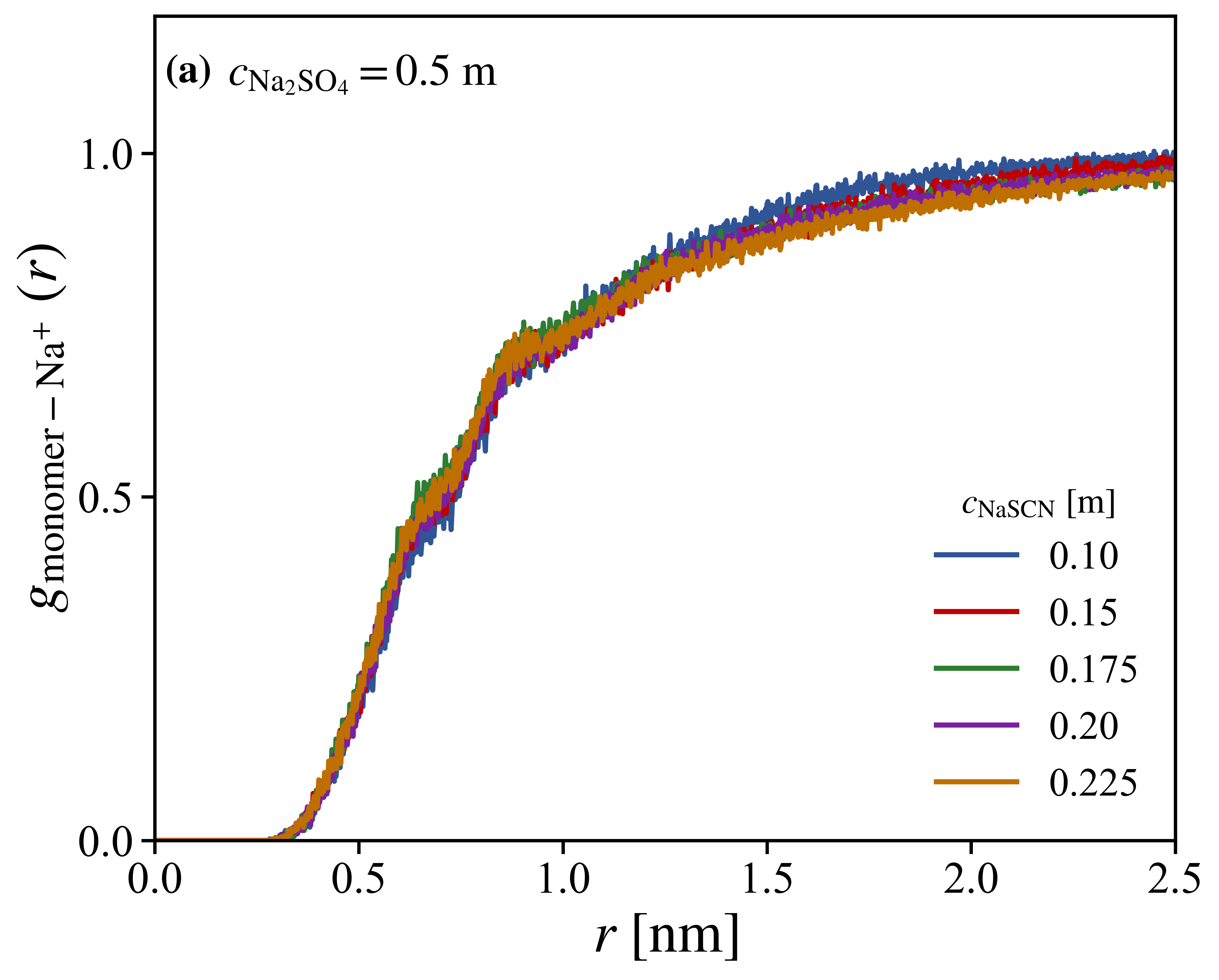}
\includegraphics[width=0.32\textwidth]{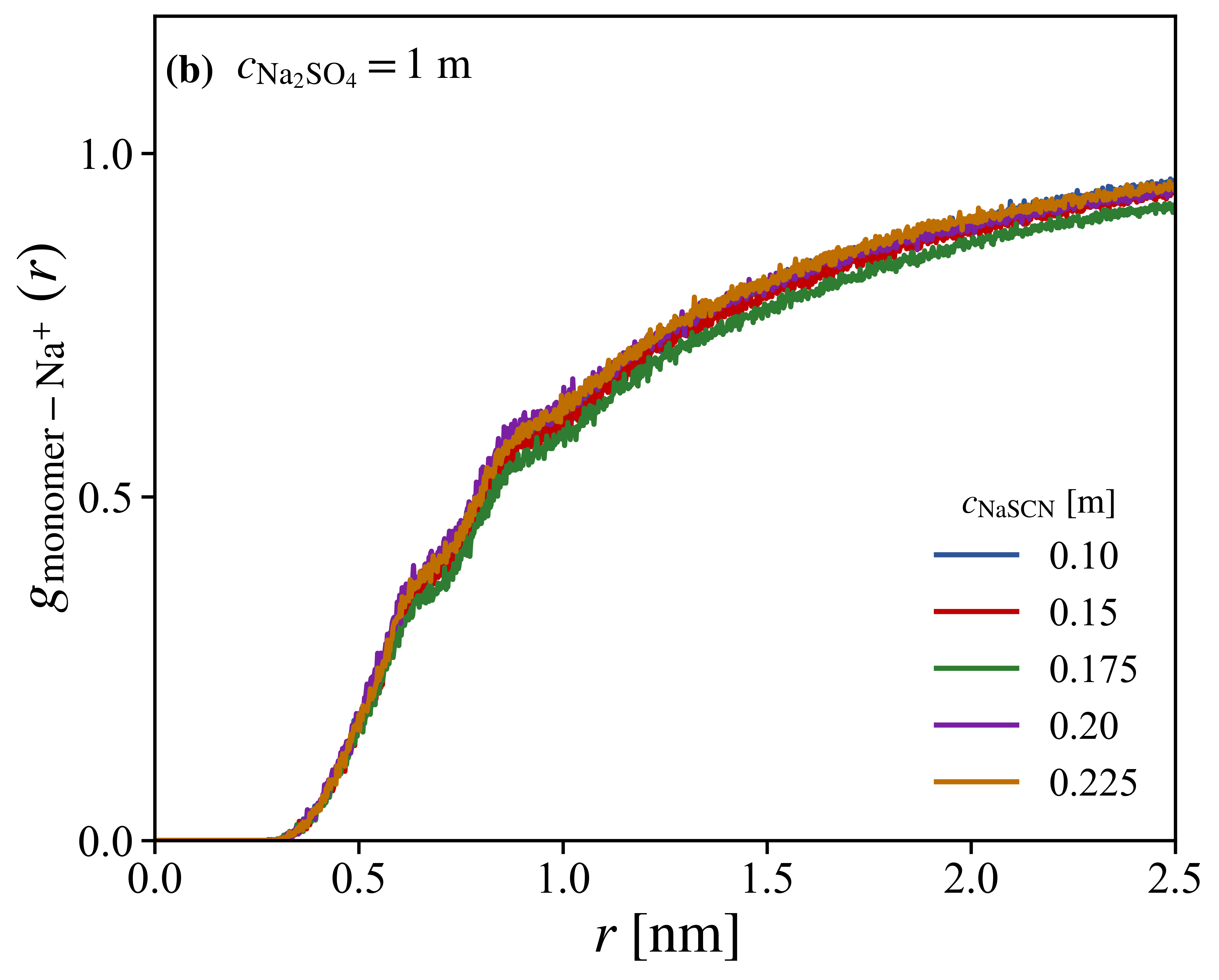} 
\includegraphics[width=0.32\textwidth]{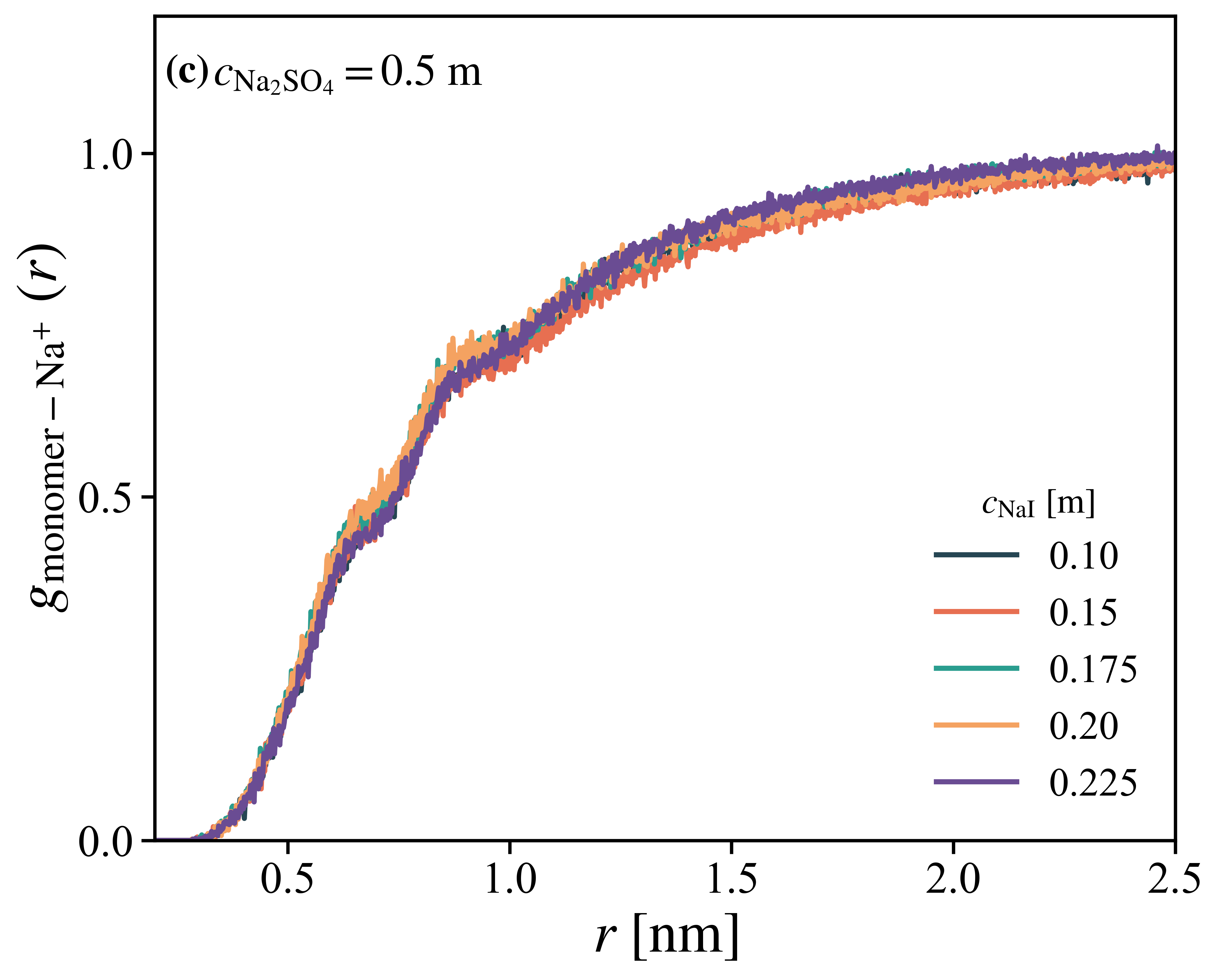}  
\caption{Monomer--Na$^{+}$ radial distribution functions for mixed salt solutions
with varying weakly hydrated salt concentrations:
(a,b) NaSCN--Na$_2$SO$_4$ solutions at Na$_2$SO$_4$ background salt
concentrations of 0.5~m and 1~m, respectively, and
(c) NaI--Na$_2$SO$_4$ solutions at a Na$_2$SO$_4$ background salt
concentration of 0.5~m.}
\label{rdfe}
\end{figure}
\newpage	
\subsection{Ion-ion and ion-water RDFs}
The ion--ion RDFs for the pure \ce{NaSCN} solution are shown in Figures~\ref{rdff}(a,b), where the thiocyanate--water and thiocyanate--sodium RDFs are presented. For the mixed salt solutions containing varying concentrations of \ce{NaSCN} in the presence of 0.5~m sodium sulfate (Na$_2$SO$_4$), the RDFs corresponding to \ce{SCN$^-$}--water, \ce{SCN$^-$}--\ce{Na$^+$}, \rm SO$_4^{2-}$--\ce{Na$^+$}, and \rm SO$_4^{2-}$--water are shown in Figures~\ref{rdfg}(a--d). Similarly, for the mixed salt solutions containing varying concentrations of \ce{NaSCN} with 1~m sodium sulfate (Na$_2$SO$_4$), the RDFs for \ce{SCN$^-$}--water, \ce{SCN$^-$}--\ce{Na$^+$}, \rm SO$_4^{2-}$--\ce{Na$^+$}, and \rm SO$_4^{2-}$--water are presented in Figures~\ref{rdfg2}(a--d). Table~\ref{tab:ion_pairs} lists the $r_{\rm in}$ and $r_{\rm out}$ for different types of pairing and hydration shells.

\begin{figure}
\centering
\includegraphics[scale=0.35]{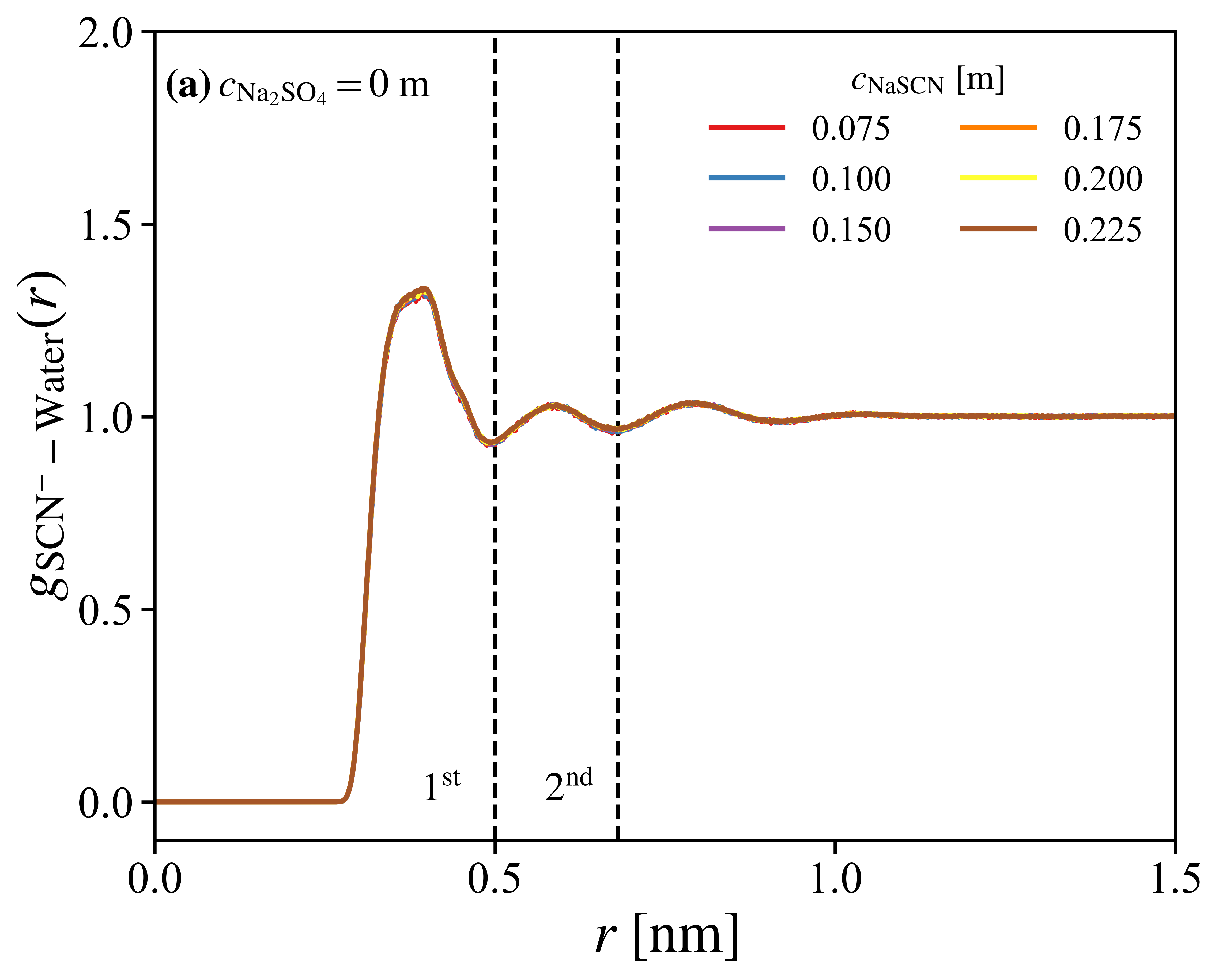}
\includegraphics[scale=0.35]{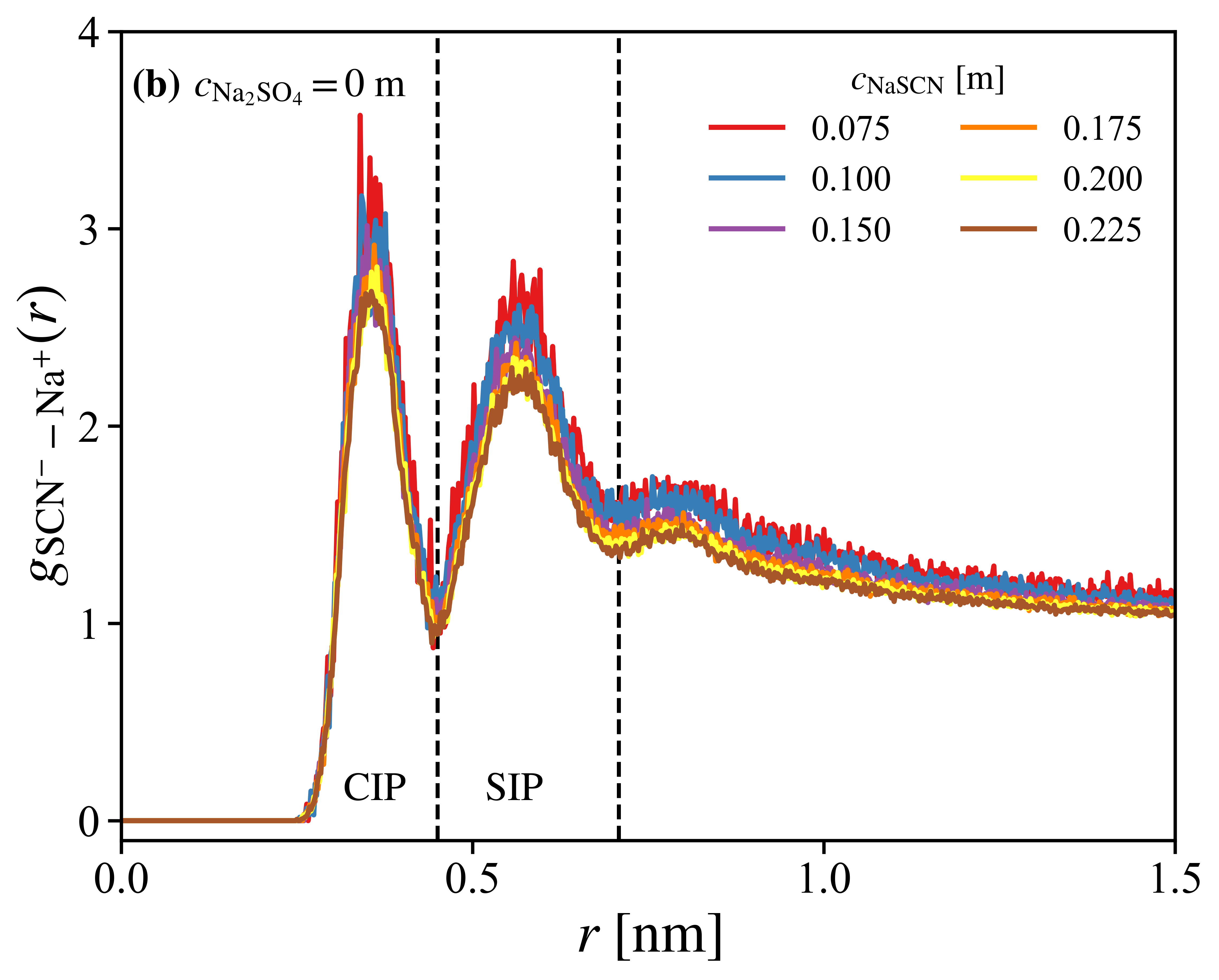} 
\caption{
Radial distribution functions for pure NaSCN solutions at varying NaSCN
concentrations: (a) SCN$^{-}$--water and (b) SCN$^{-}$--Na$^{+}$ RDFs. The
vertical lines in panel (a) indicate the regions used to compute the
thiocyanate--water affinity for the first and second hydration shells. The
vertical lines in panel (b) indicate the regions used to compute the excess
ion pairing corresponding to contact ion pairs (CIP) and solvent-shared ion
pairs (SIP).}
\label{rdff}
\end{figure}

\begin{figure}
\centering
\includegraphics[scale=0.35]{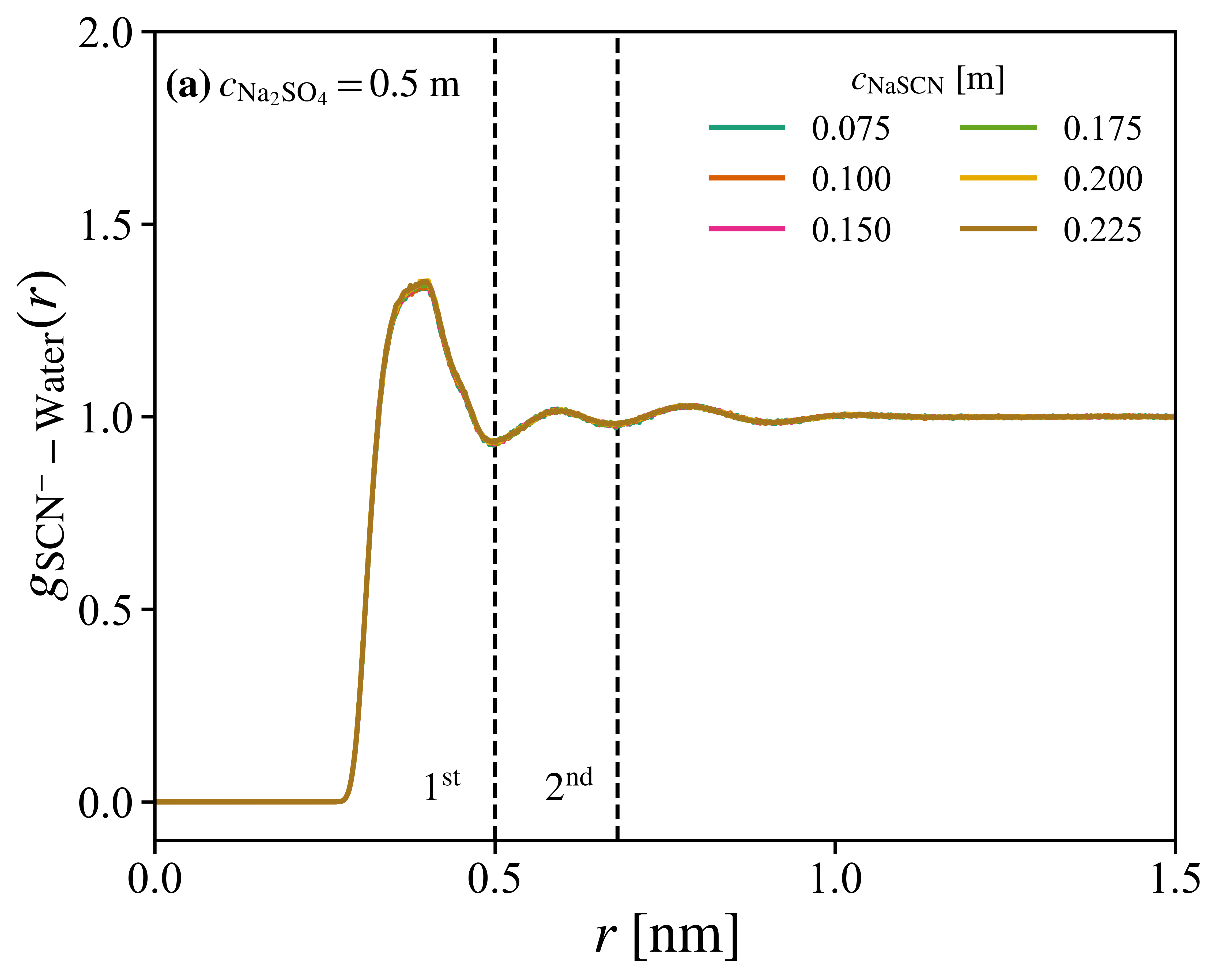}
\includegraphics[scale=0.35]{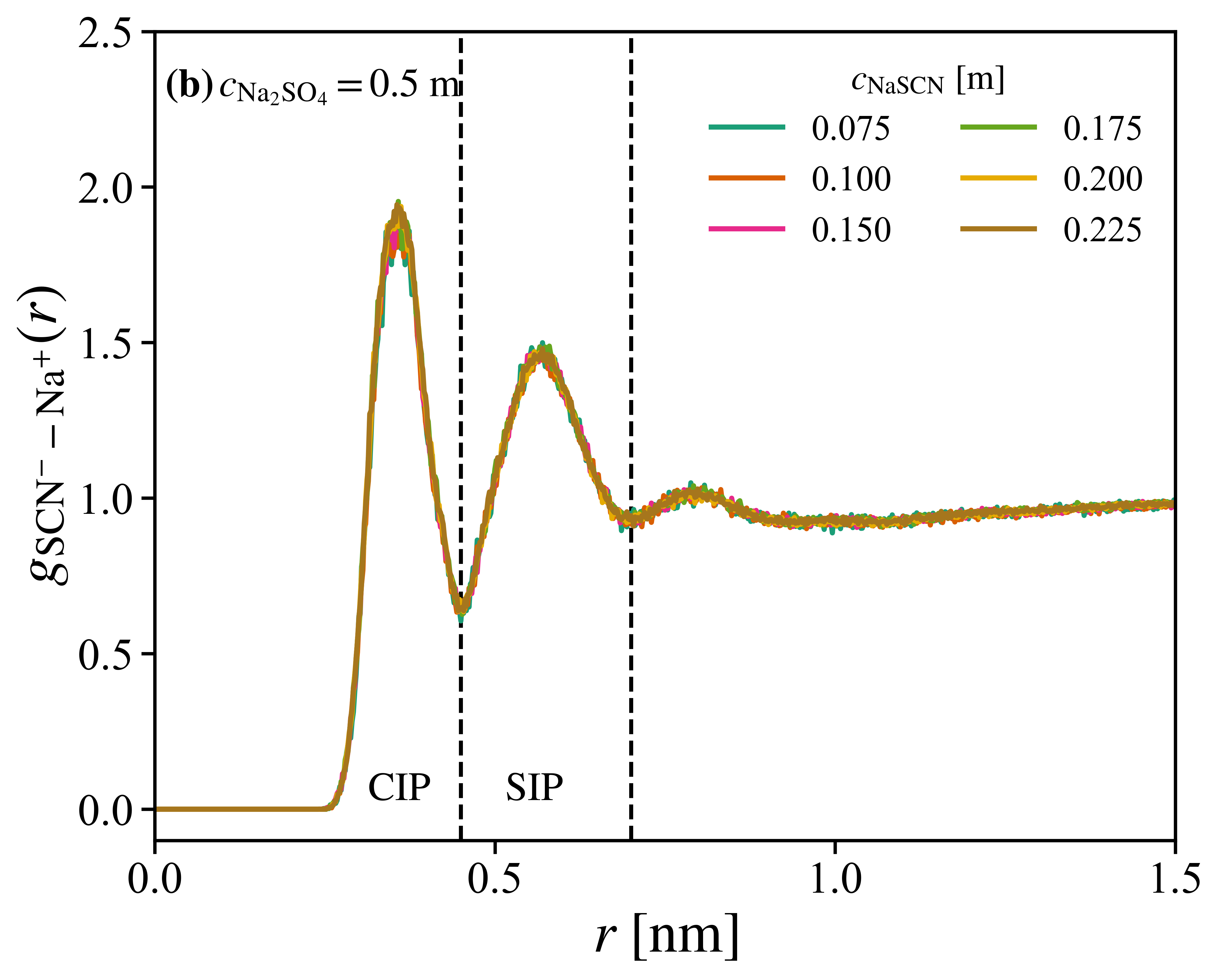} 
\includegraphics[scale=0.35]{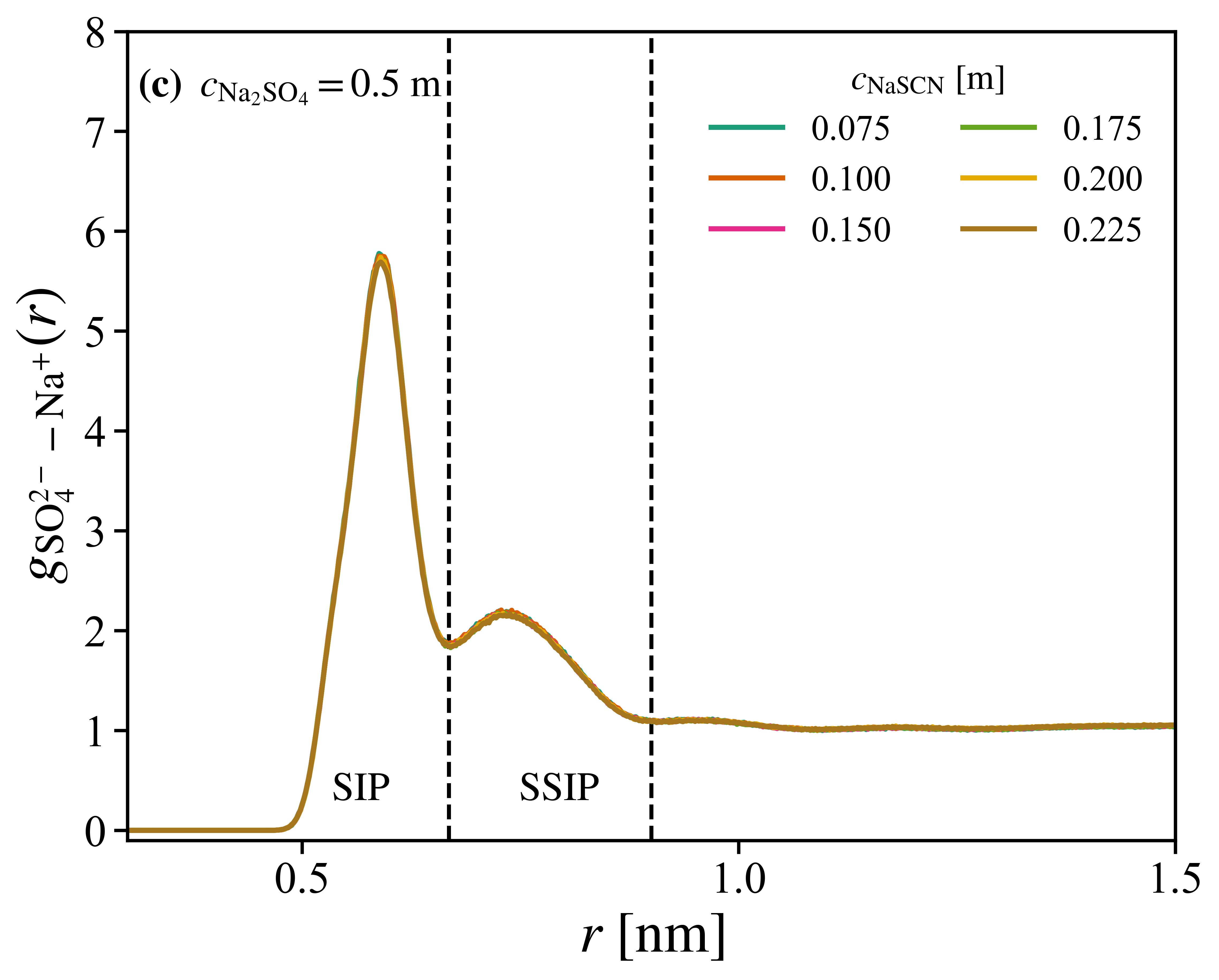}
\includegraphics[scale=0.35]{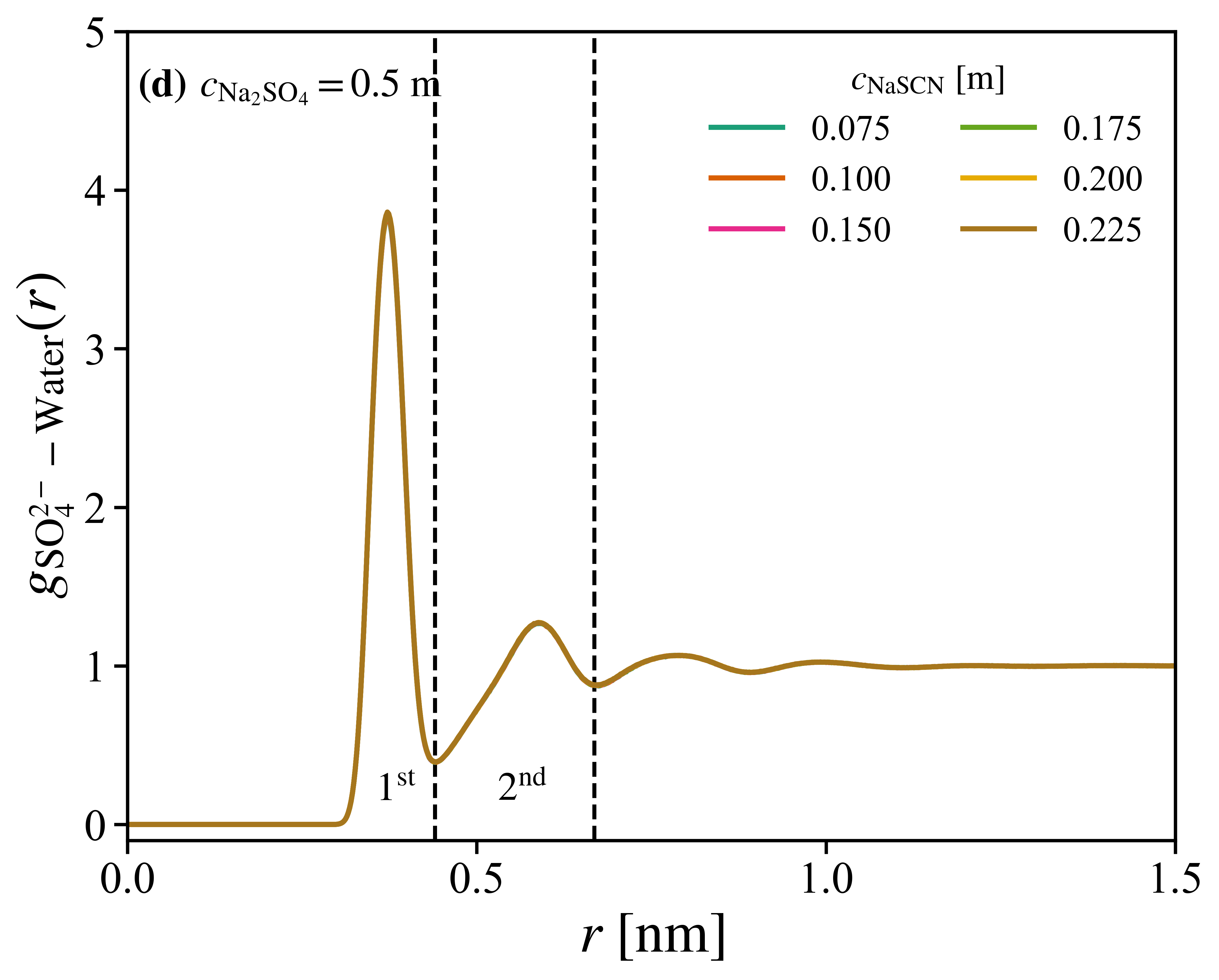}
\caption{
Radial distribution functions for mixed salt solutions at a 0.5~m Na$_2$SO$_4$
background salt concentration with varying weakly hydrated salt concentrations:
(a) SCN$^{-}$--water, (b) SCN$^{-}$--Na$^{+}$, (c) SO$_4^{2-}$--Na$^{+}$, and
(d) SO$_4^{2-}$--water RDFs. The vertical lines in panels (a) and (d) indicate
the regions used to compute the thiocyanate--water and sulfate--water
affinities for the first and second hydration shells. The vertical lines in
panels (b) and (c) indicate the regions used to compute the excess ion pairing
corresponding to contact ion pairs (CIP), solvent-shared ion pairs (SIP), and
solvent-separated ion pairs (SSIP).}
\label{rdfg}
\end{figure}

\begin{figure}
\centering
\includegraphics[scale=0.35]{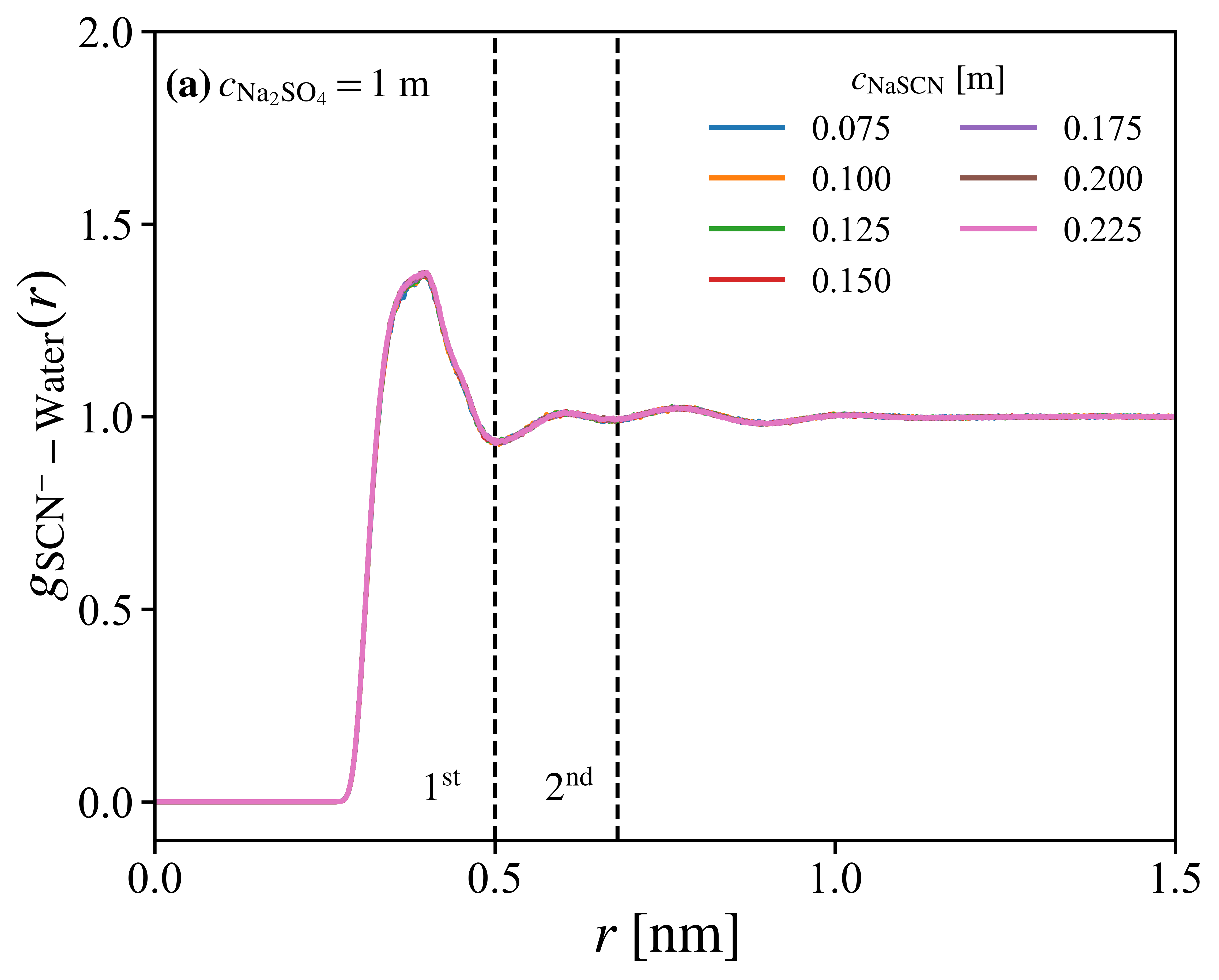}
\includegraphics[scale=0.35]{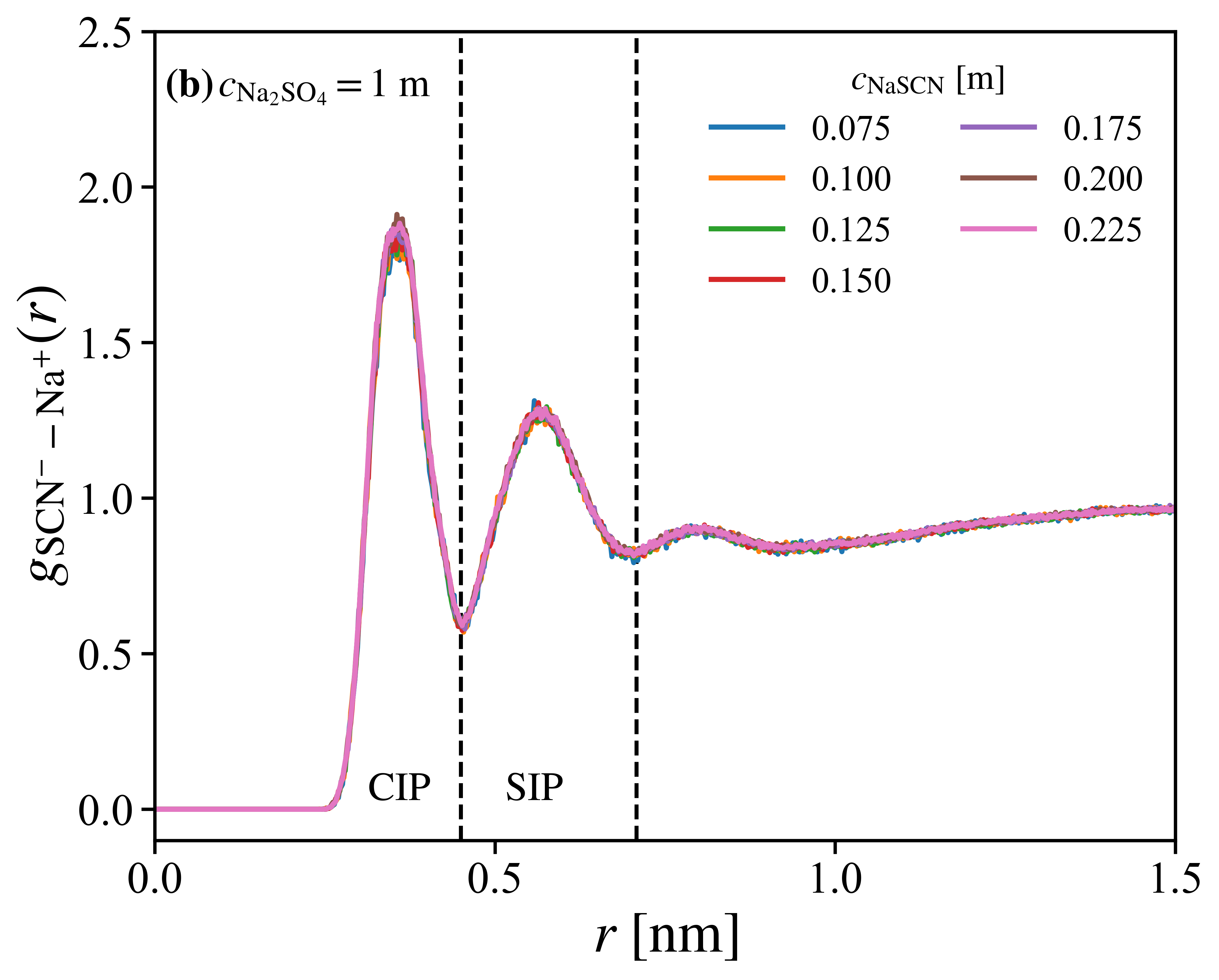}  
\includegraphics[scale=0.35]{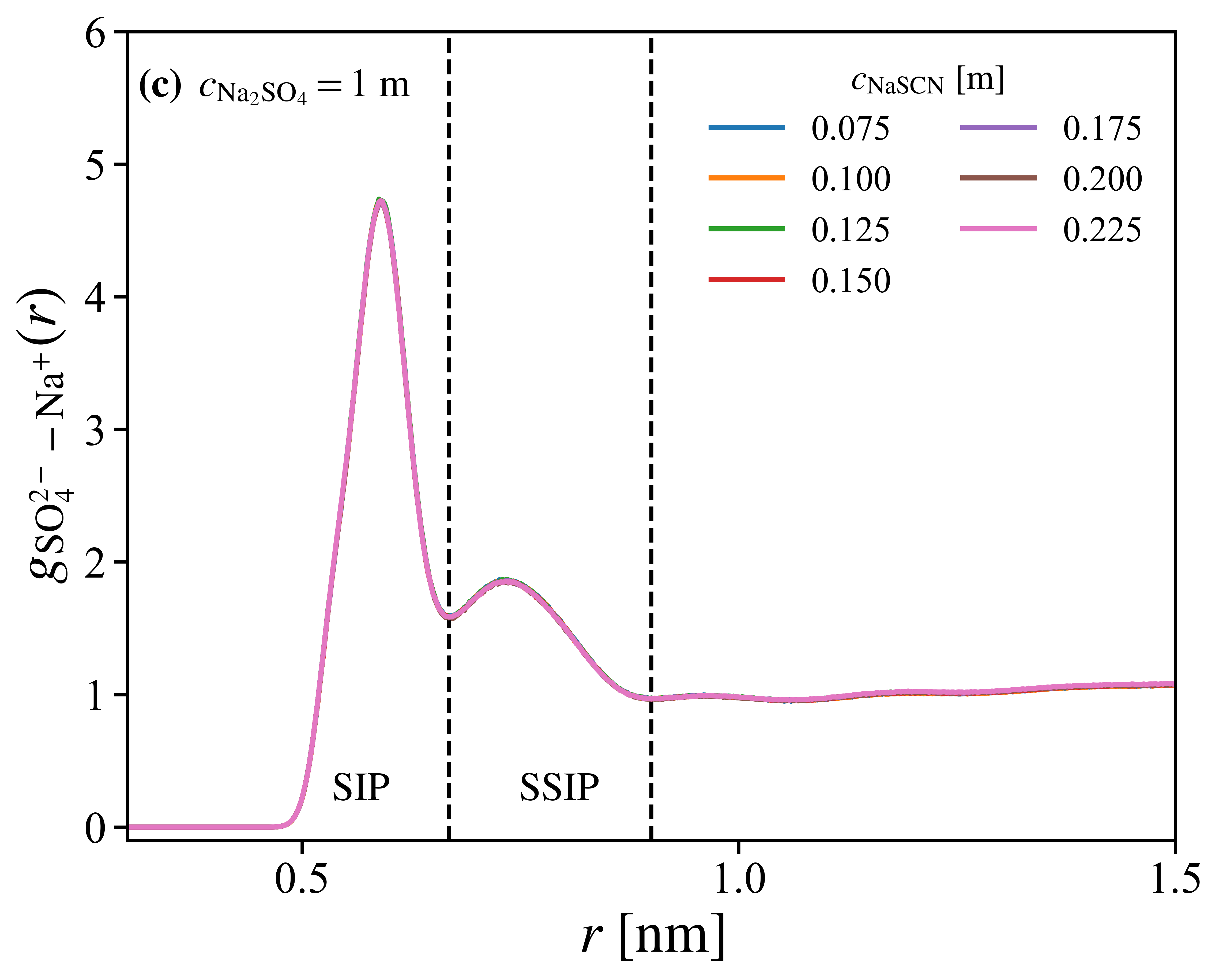}
\includegraphics[scale=0.35]{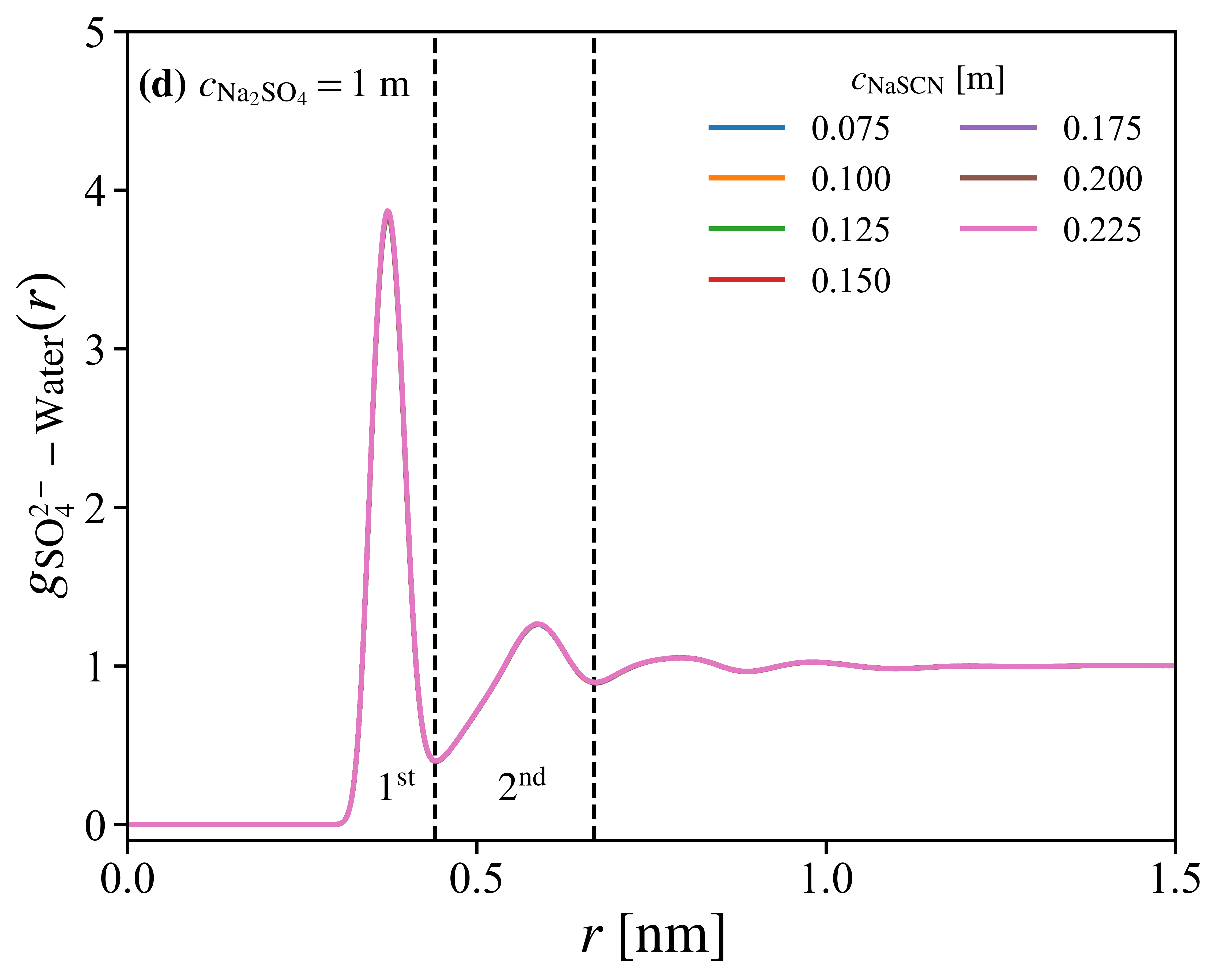}
\caption{
Radial distribution functions for mixed salt solutions at a
1~m Na$_2$SO$_4$ background salt concentration with varying weakly
hydrated salt concentrations: (a) SCN$^{-}$--water,
(b) SCN$^{-}$--Na$^{+}$, (c) SO$_4^{2-}$--Na$^{+}$, and
(d) SO$_4^{2-}$--water RDFs. The vertical lines in panels (a) and (d)
indicate the regions used to compute the thiocyanate--water and
sulfate--water affinities for the first and second hydration shells. The
vertical lines in panels (b) and (c) indicate the regions used to compute
the excess ion pairing corresponding to contact ion pairs (CIP),
solvent-shared ion pairs (SIP), and solvent-separated ion pairs (SSIP).}
\label{rdfg2}
\end{figure}
%%%%%%%%%%%%%%%%%%%%%%%%%%%%%%%%%%%%%%%%%%%%%%%%%%%%%%%%%%%%%%%%

\section{Ion pairing and Ion hydration}
The excess ion-pairing for the different systems are presented across Figures~\ref{rdfh}--\ref{rdfk}. Specifically, Figure~\ref{rdfh} displays the SCN$^{-}$--water affinity for the first and second hydration shells in the pure and mixed salt solution, while Figure~\ref{rdfi} shows the CIP and SIP states for SCN$^{-}$--Na$^{+}$. For the sulfate species, the SIP and SSIP excess ion-pairing of SO$_4^{2-}$--Na$^{+}$ are illustrated in Figure~\ref{rdfj}, and the corresponding SO$_4^{2-}$--water hydration behavior is shown in Figure~\ref{rdfk}.

\begin{table}[tbp]
\centering
\caption{Distance criteria used to define CIP, SIP, and SSIP states along with hydration shell boundaries.}
\label{tab:ion_pairs}
\renewcommand{\arraystretch}{1}
\begin{tabular}{ll cc|cc|cc}
\hline
\hline
 & & \multicolumn{2}{c}{CIP} & \multicolumn{2}{c}{SIP} & \multicolumn{2}{c}{SSIP} \\
\hline
 & & $r_{\mathrm{in}}$ & $r_{\mathrm{out}}$ & $r_{\mathrm{in}}$ & $r_{\mathrm{out}}$ & $r_{\mathrm{in}}$ & $r_{\mathrm{out}}$ \\
\hline
SCN$^{-}$ & Na$^{+}$ & 0 & 0.45 nm & 0.45 nm & 0.65 nm & - & - \\
SO$_4^{2-}$ & Na$^{+}$ & - & - & 0 & 0.668 nm & 0.668 nm & 0.90 nm \\
\hline
 & & \multicolumn{2}{c}{1st hydration shell} & \multicolumn{2}{c}{2nd hydration shell} & \multicolumn{2}{c}{} \\
\hline
SCN$^{-}$ & OW ($c_{\mathrm{Na_2SO_4}} = 0$ m) & 0 & 0.50 nm & 0.50 nm & 0.67 nm & & \\
SCN$^{-}$ & OW ($c_{\mathrm{Na_2SO_4}} = 0.5$ m) & 0 & 0.50 nm & 0.50 nm & 0.69 nm & & \\
SCN$^{-}$ & OW ($c_{\mathrm{Na_2SO_4}} = 1$ m) & 0 & 0.50 nm & 0.50 nm & 0.70 nm & & \\
SO$_4^{2-}$ & OW & 0 & 0.44 nm & 0.44 nm & 0.668 nm & & \\
\hline
\hline
\end{tabular}
\end{table}

\begin{figure}[tbp]
\centering
\includegraphics[scale=0.35]{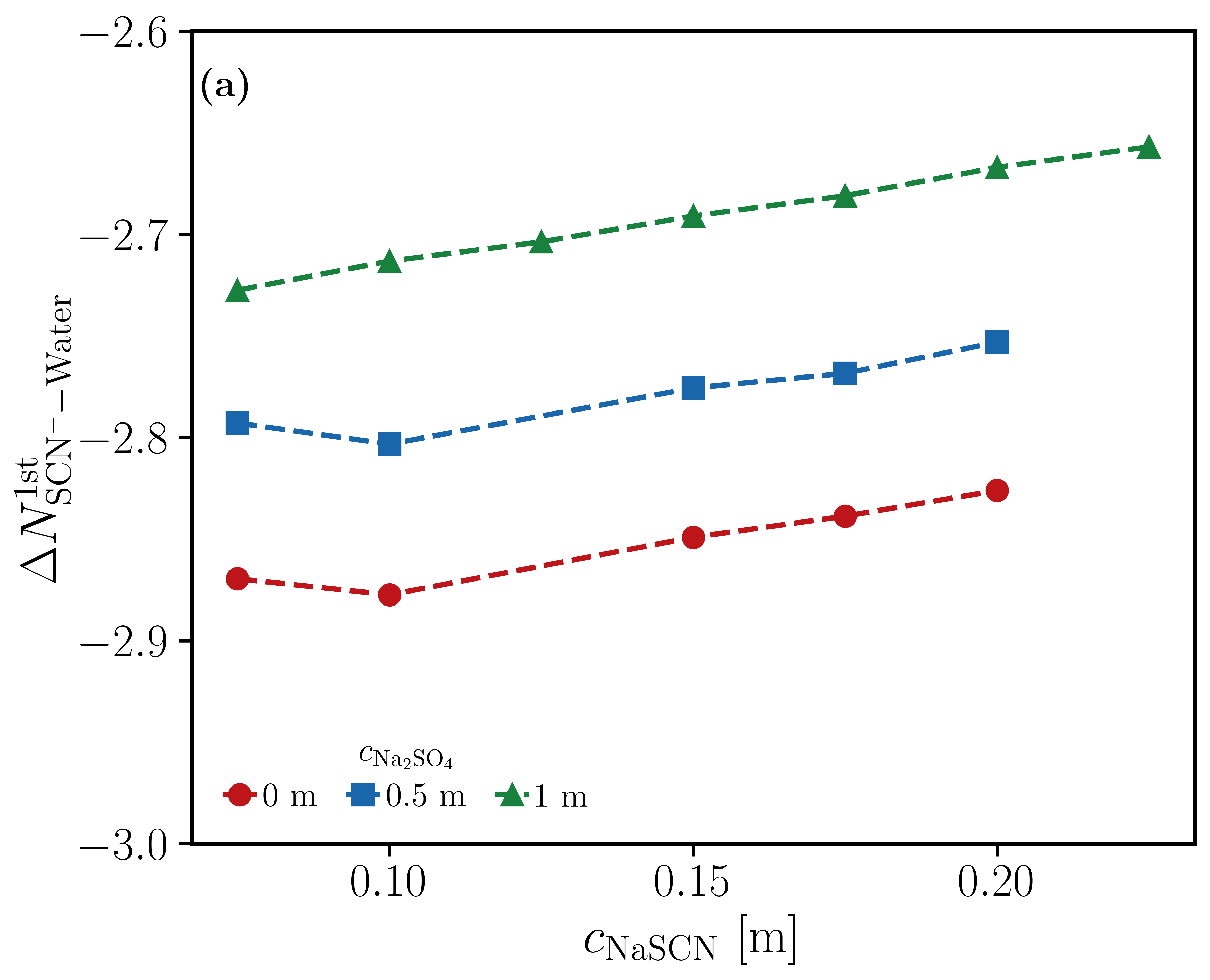}
\includegraphics[scale=0.35]{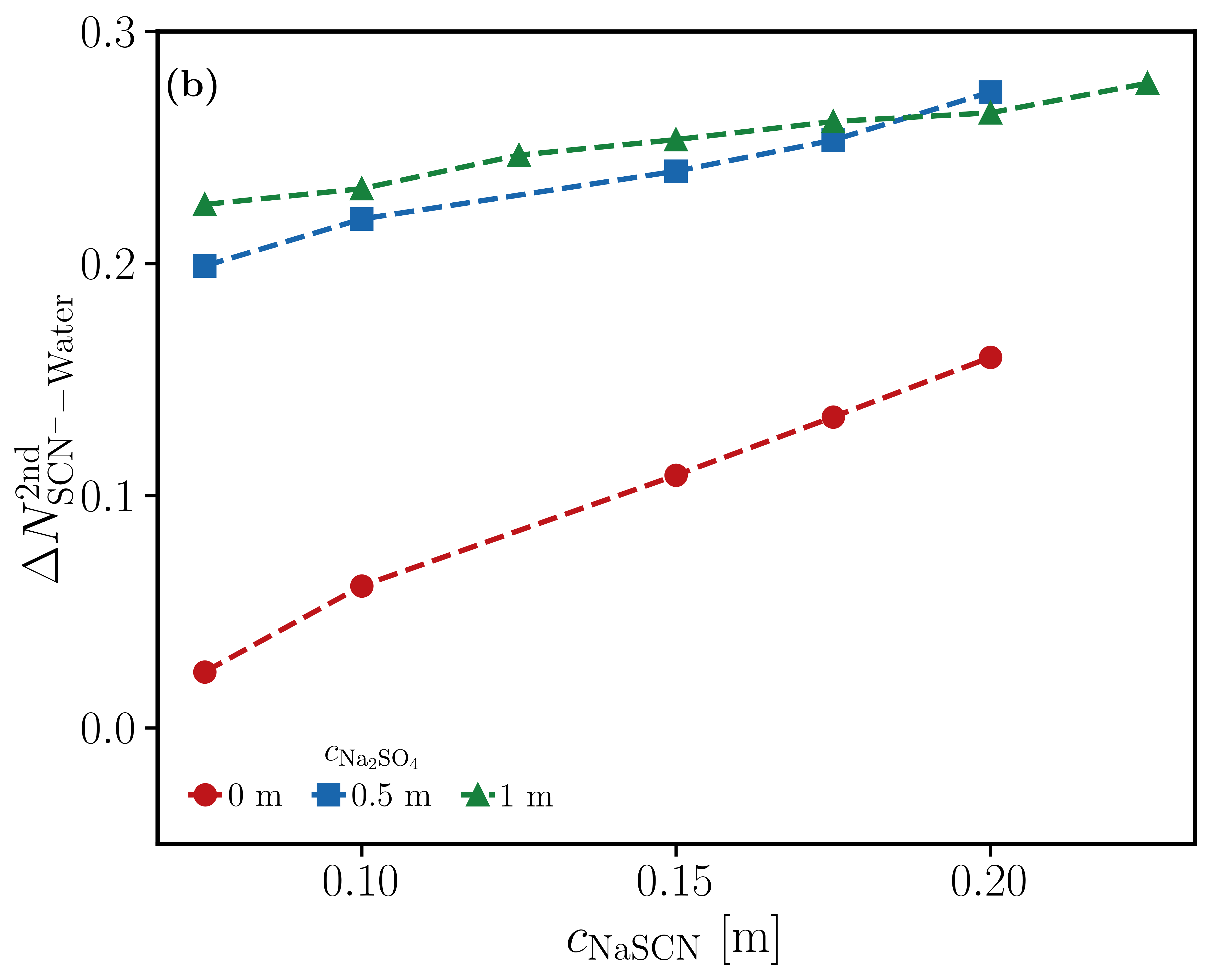} 
\caption{Dependence of the thiocyanate--water affinity on $c_{\rm NaSCN}$ at varying
Na$_2$SO$_4$ background salt concentrations:
(a) first hydration shell and (b) second hydration shell. }
\label{rdfh}
\end{figure}

\begin{figure}[tbp]
\centering
\includegraphics[scale=0.35]{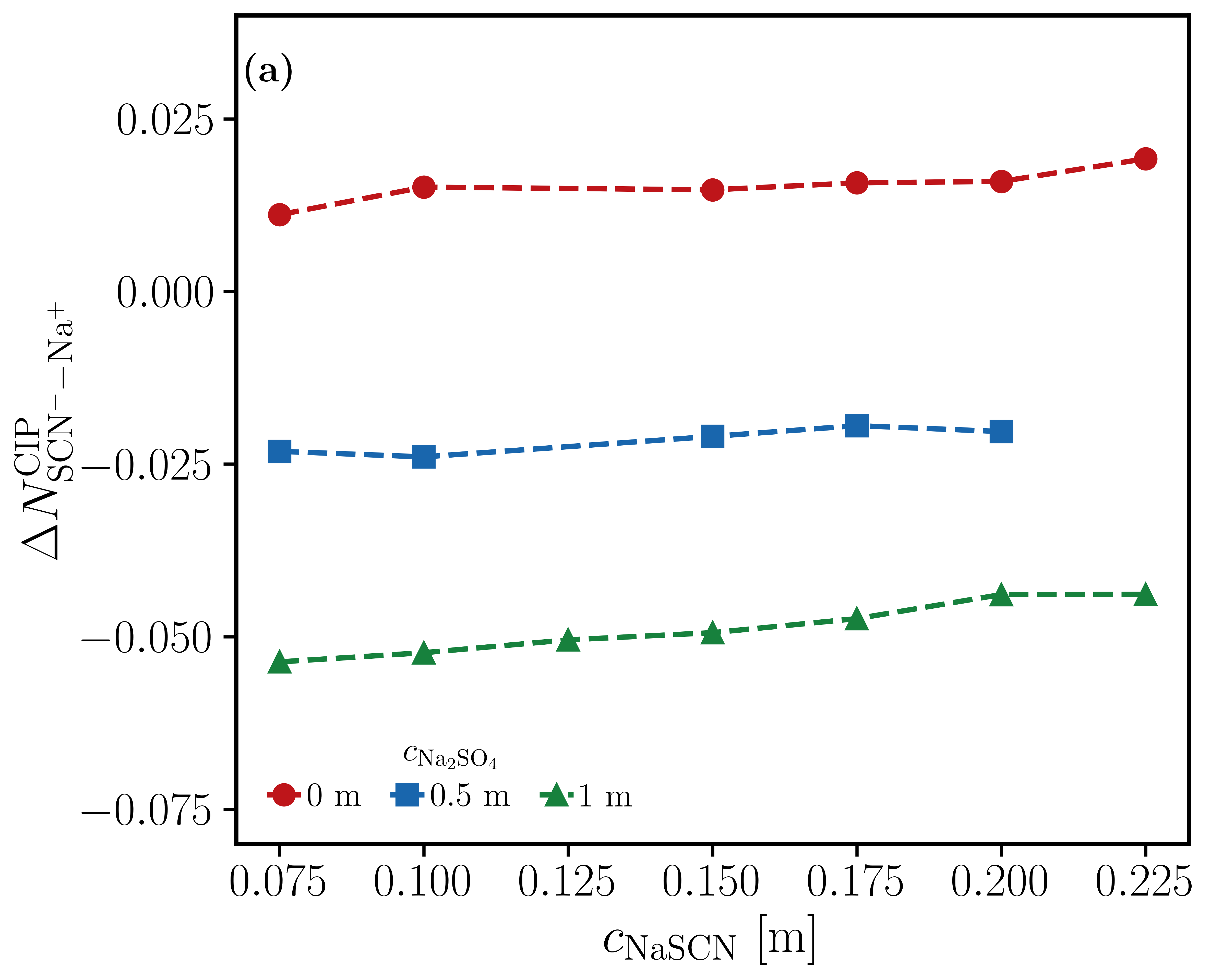}
\includegraphics[scale=0.35]{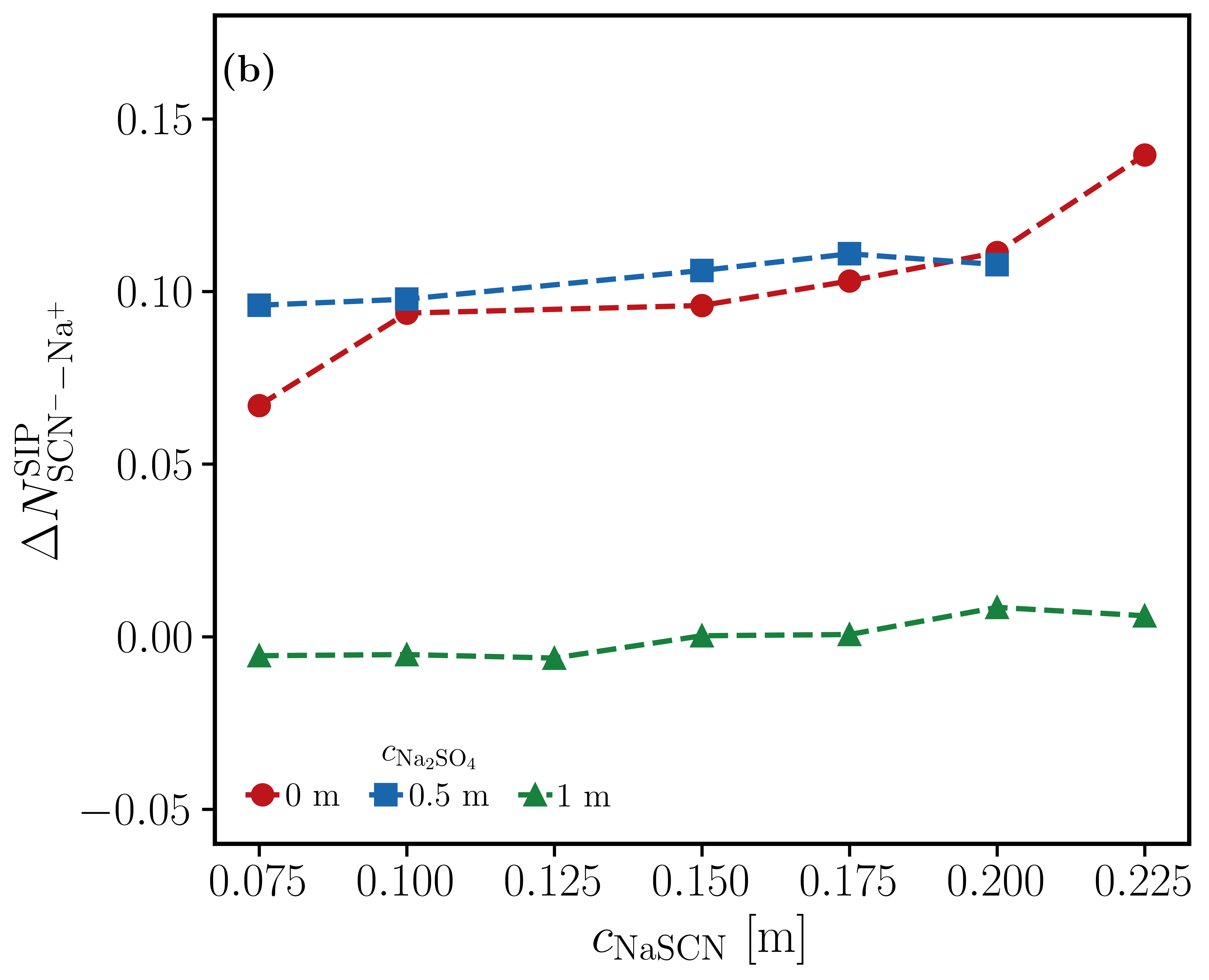}  
\caption{Dependence of the thiocyanate--sodium affinity on $c_{\rm NaSCN}$ at varying
Na$_2$SO$_4$ background salt concentrations:
(a) contact ion pairs and (b) solvent-shared ion pairs. }
\label{rdfi}
\end{figure}

\begin{figure}[tbp]
\centering
\includegraphics[scale=0.35]{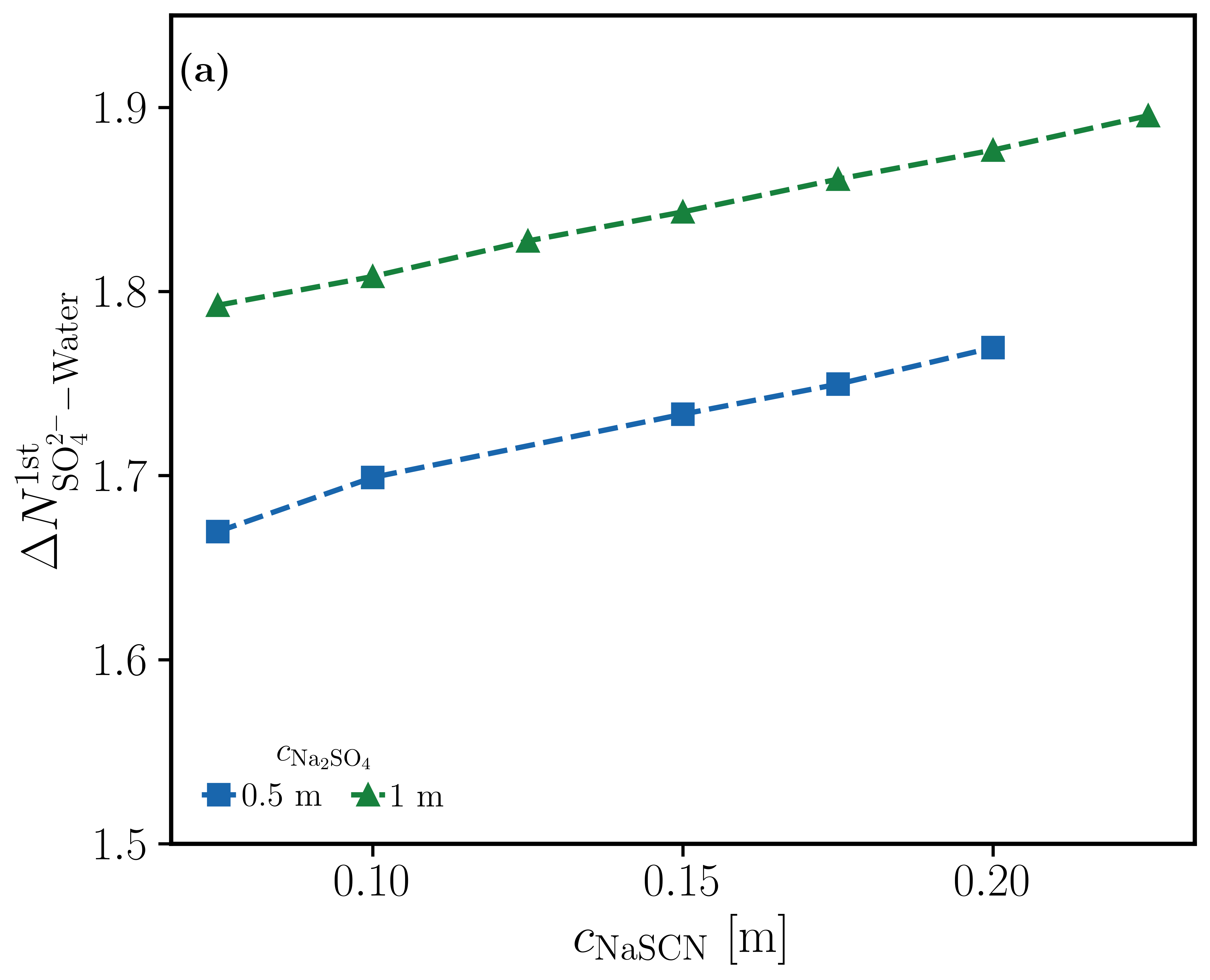}
\includegraphics[scale=0.35]{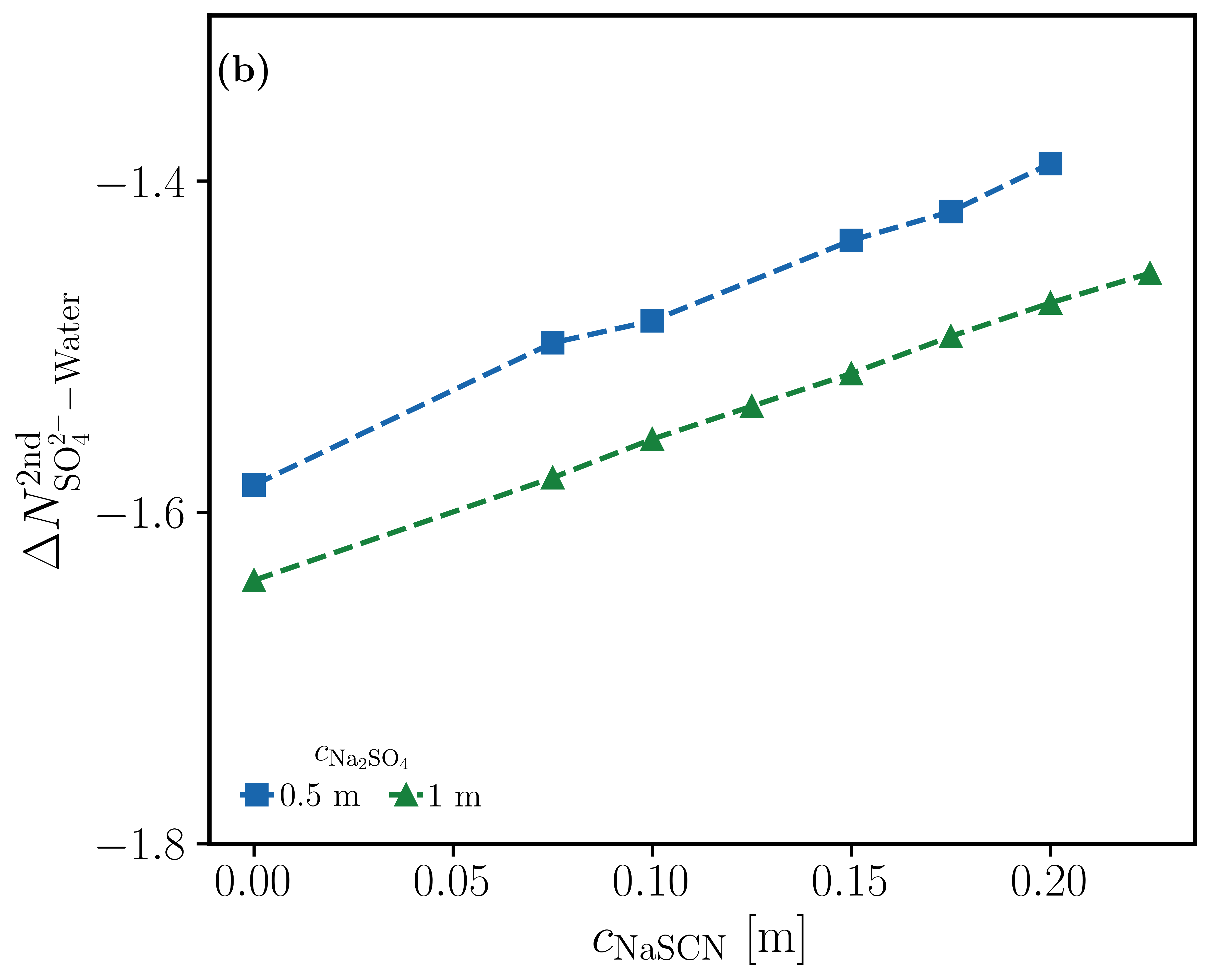}
\caption{Dependence of the sulfate--water affinity on $c_{\rm NaSCN}$ at
varying Na$_2$SO$_4$ background salt concentrations: (a) first hydration shell
and (b) second hydration shell.}
\label{rdfj}
\end{figure}

\begin{figure}[tbp]
\centering
\includegraphics[scale=0.35]{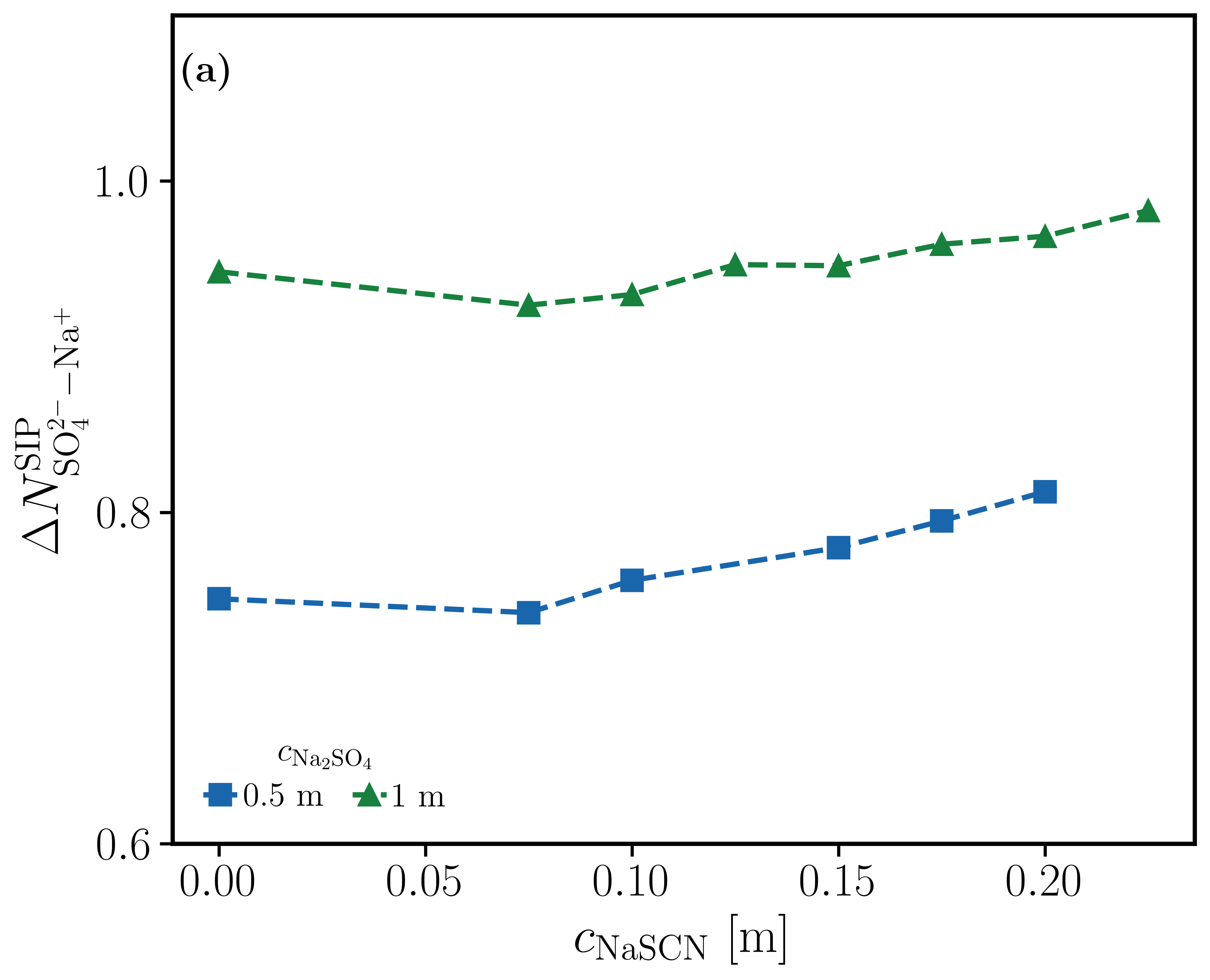}
\includegraphics[scale=0.35]{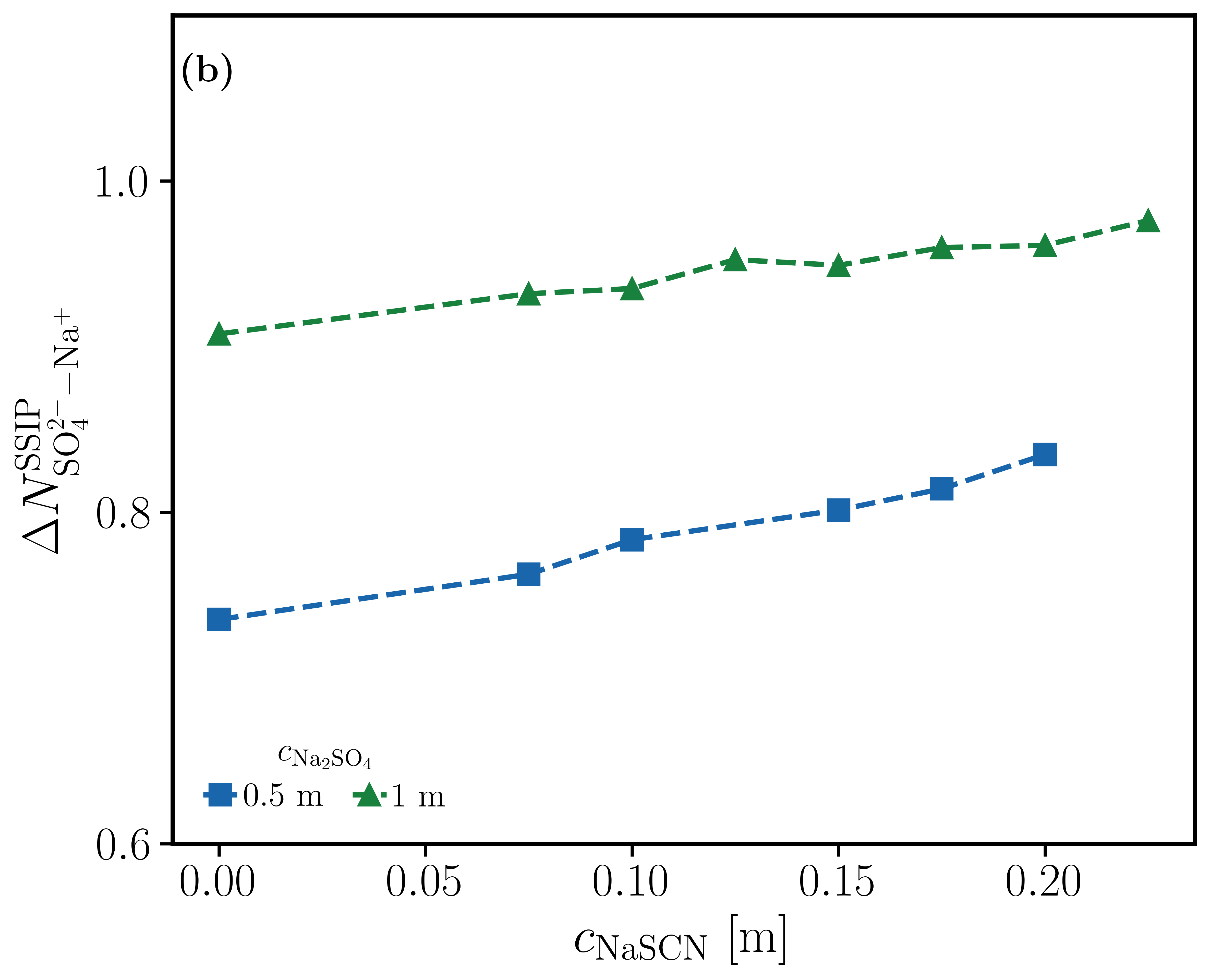}  
\caption{Dependence of the sulfate--sodium affinity on $c_{\rm NaSCN}$ at varying
Na$_2$SO$_4$ background salt concentrations:
(a) solvent-shared ion pairs (SIP) and
(b) solvent-separated ion pairs (SSIP).}
\label{rdfk}
\end{figure}
\section{Radius of gyration}
The radius of gyration ($R_g$) was calculated as the mean of the averaged $R_g$
values obtained from different time intervals of the equilibrated portion of
the 110 ns trajectory. The uncertainty in the mean was estimated as the
standard deviation of these averaged values.  Figure.~\ref{S3} shows the
dependence of $R_{\rm g}$ on the salt concentration in pure NaSCN and
Na$_{2}$SO$_{4}$ solutions. Figure.~\ref{fig:rg_mixed}(a) shows the dependence
of $R_{\rm g}$ on $c_{\rm NaSCN}$ for different background salt concentrations.
Figure.~\ref{fig:rg_mixed}(b) shows the dependence of $R_{\rm g}$ on $c_{\rm
NaSCN}/c_{\rm NaI}$ at 0.5 m background salt.

\begin{figure}[tbp]

\centering
\includegraphics[scale=0.35]{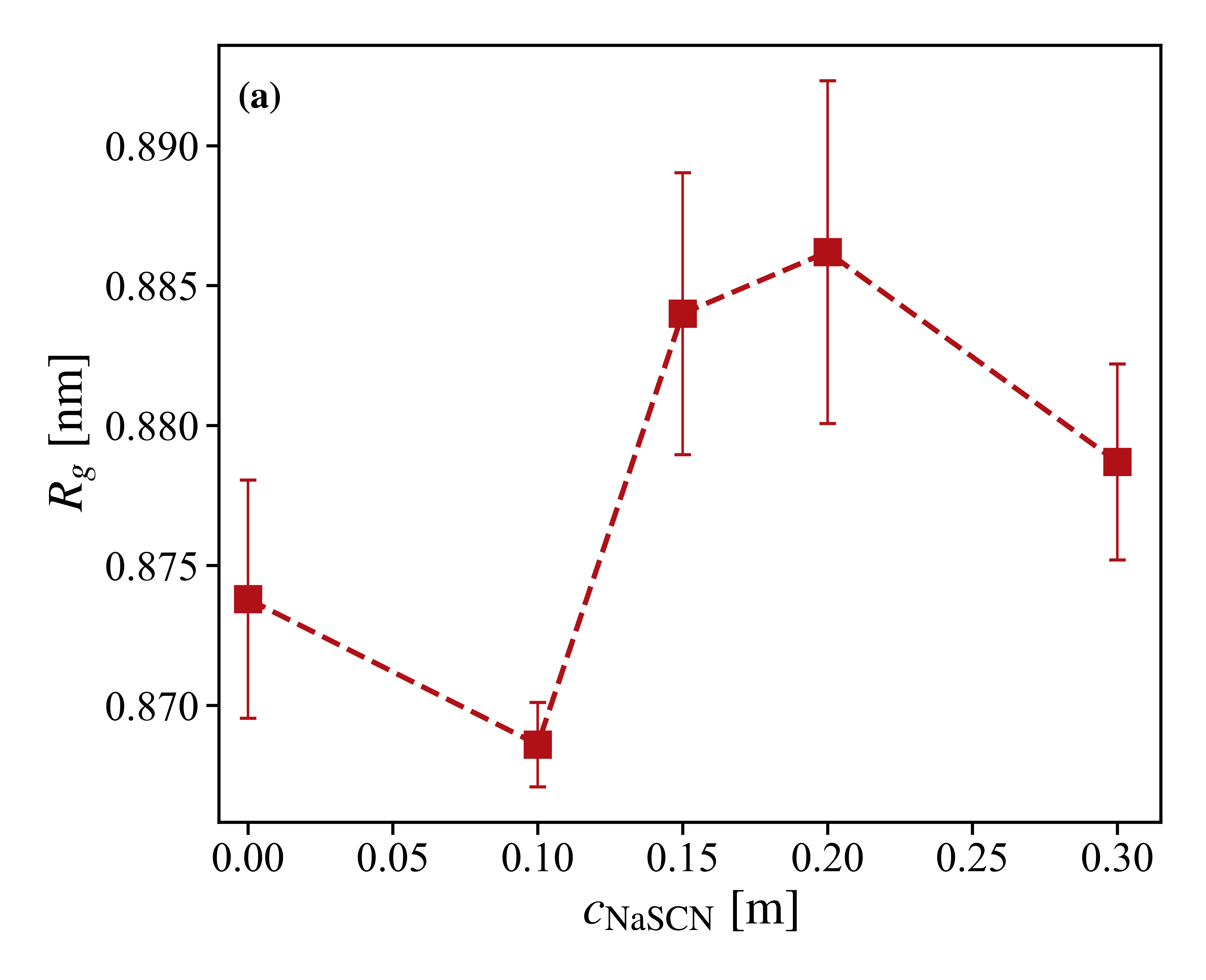}
\includegraphics[scale=0.35]{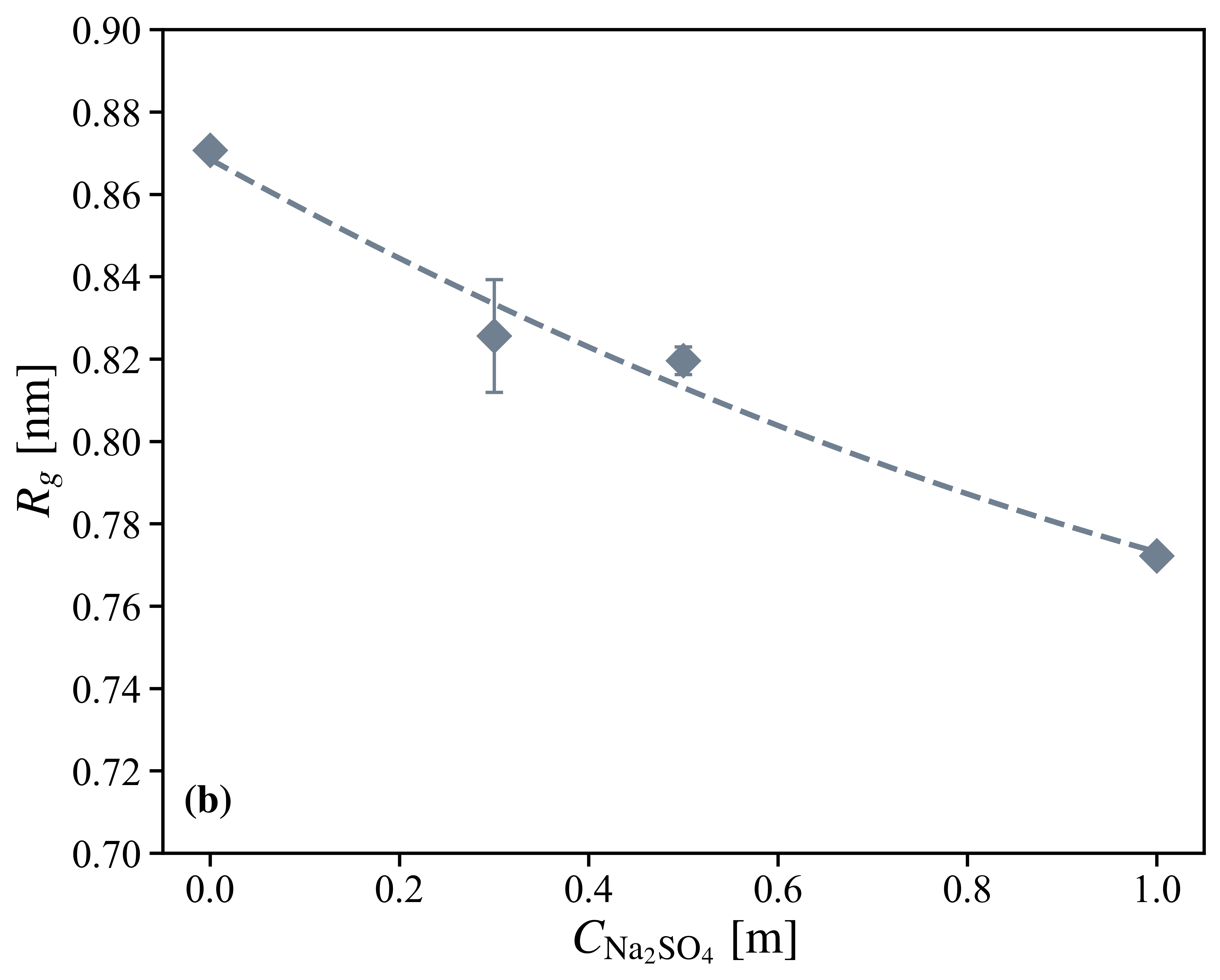} 
\caption{
(a) Dependence of the radius of gyration, $R_g$, on $c_{\rm NaSCN}$ for pure
NaSCN solutions.
(b) Dependence of the radius of gyration, $R_g$, on $c_{\rm Na_2SO_4}$ for
pure \ce{Na2SO4} solutions.}
\label{S3}
\end{figure}

\begin{figure}[tbp]
\centering
\includegraphics[scale=0.35]{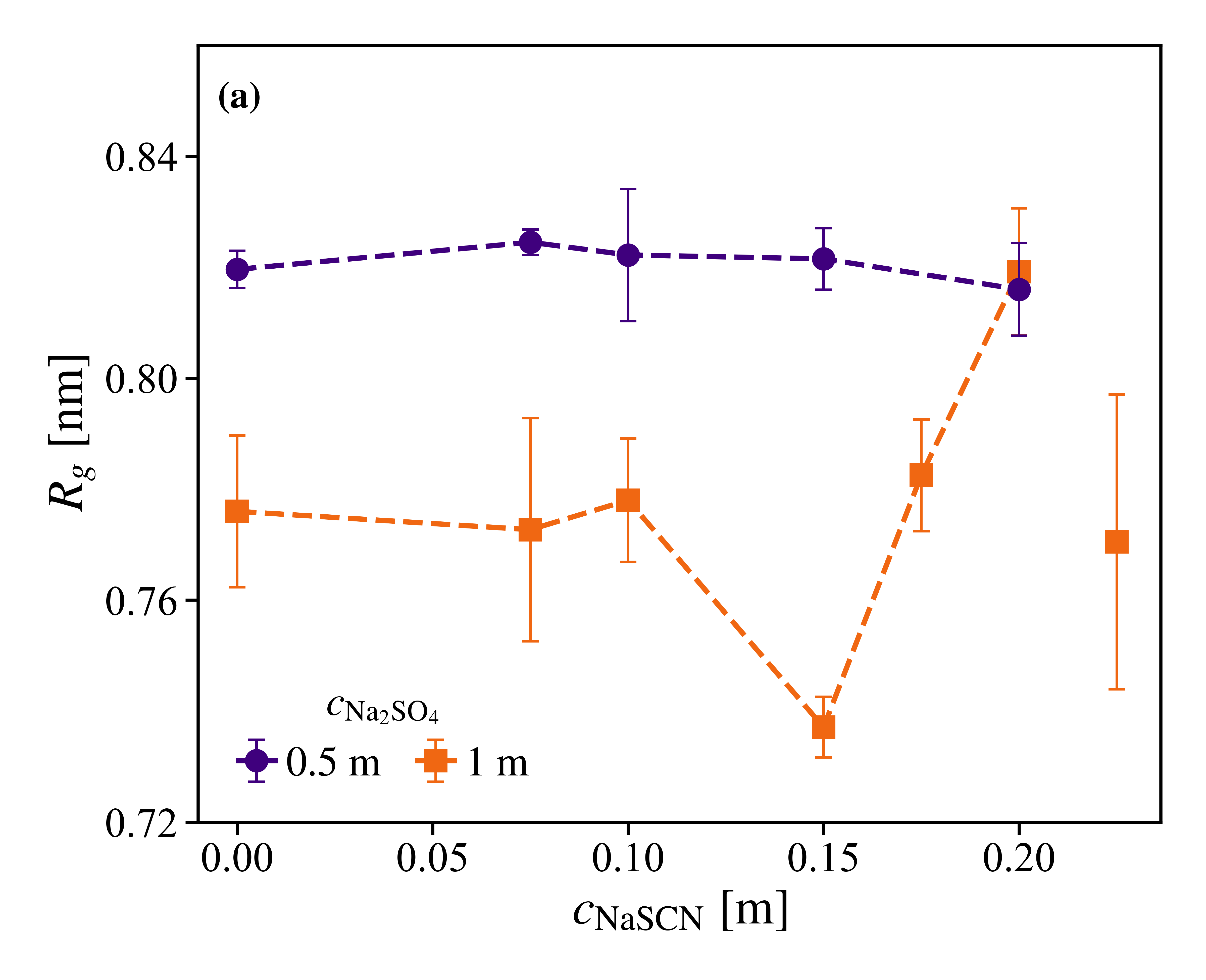}
\includegraphics[scale=0.35]{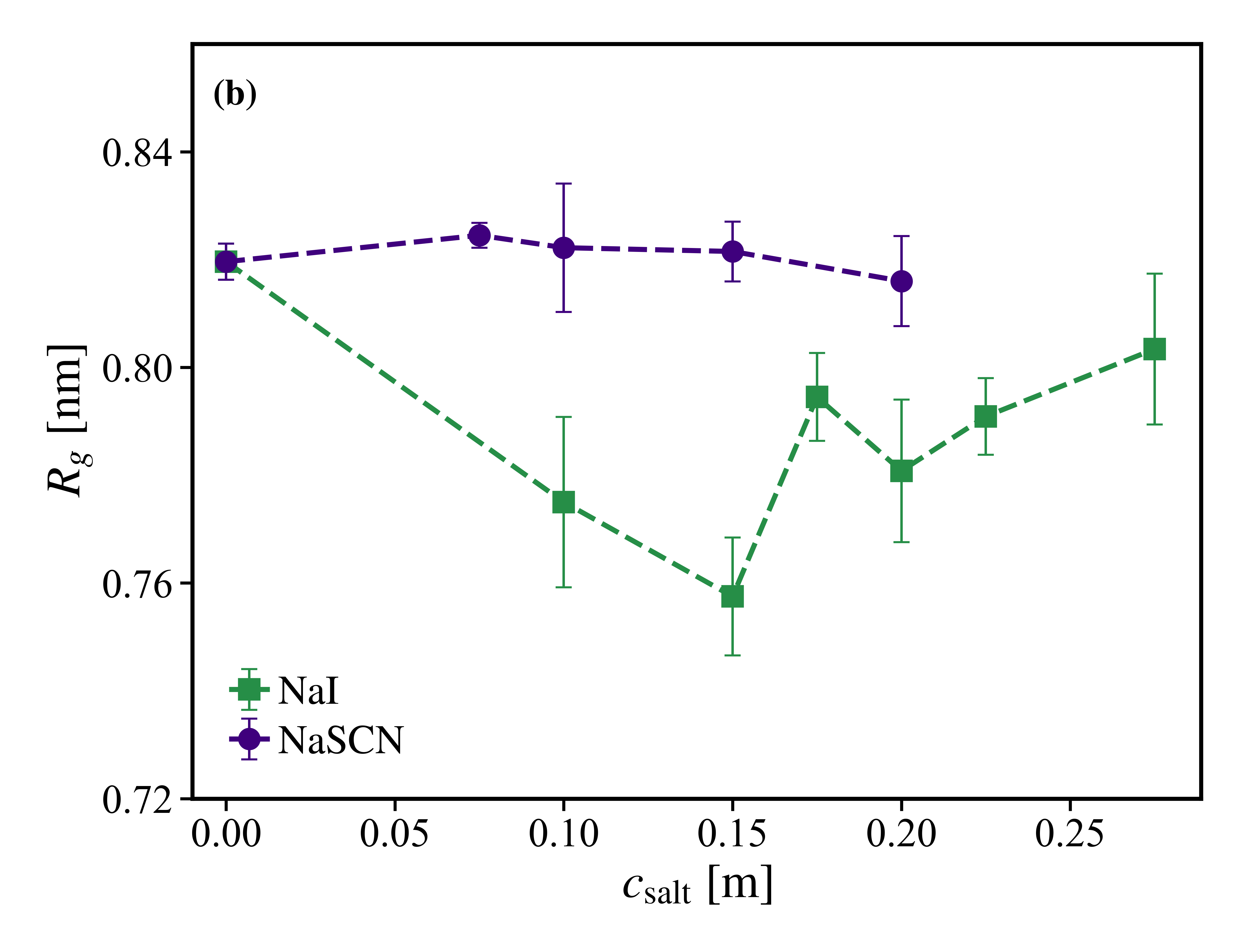}  
\caption{
(a) Dependence of the radius of gyration, $R_g$, on $c_{\rm NaSCN}$ for
mixed salt solutions with 0.5~m and 1~m \ce{Na2SO4} background salt
concentrations.
(b) Dependence of the radius of gyration, $R_g$, on weakly hydrated salt
concentration for NaSCN--Na$_2$SO$_4$ and NaI--Na$_2$SO$_4$ mixed salt
solutions with a 0.5~m \ce{Na2SO4} background salt concentration.}
\label{fig:rg_mixed}
\end{figure}

\end{document}